\documentclass[11pt,USenglish]{article}
\pdfoutput=1

\usepackage{setspace}
\usepackage{cite}

\renewcommand{\Phi}{\phi}

\usepackage[utf8]{inputenc}
\usepackage{geometry}
\usepackage[USenglish]{babel}
\usepackage{verbatim}
\usepackage{prettyref}
\usepackage{amsmath}
\usepackage[normalem]{ulem}
\usepackage{amssymb}
\usepackage{graphicx}
\usepackage{setspace}
\usepackage{esint}
\usepackage{dsfont}

\usepackage{color}
\definecolor{darkgreen}{rgb}{0,0.5,0}
\definecolor{darkblue}{rgb}{0,0,0.6}
\definecolor{purple}{rgb}{0.4,.2,0.7}
\definecolor{orange}{rgb}{0.95, 0.5, 0.3}

\usepackage[colorlinks=true,citecolor=darkgreen,linkcolor=darkblue,urlcolor=blue]{hyperref}

\usepackage{cleveref}

\numberwithin{equation}{section}
\numberwithin{table}{section}

\def\be{\begin{equation}}
\def\ee{\end{equation}}
\def\bea{\begin{eqnarray}}
\def\eea{\end{eqnarray}}
\def\ba{\begin{align}}
\def\ea{\end{align}}
\def\d{\text{d}}

\def\d{\text{d}}

\makeatletter
\newcommand{\vast}{\bBigg@{4}}
\newcommand{\Vast}{\bBigg@{5}}
\makeatother


\begin{document}\spacing{1.2}

\vspace*{2.5cm}
\begin{center}
{  \Large \textsc{The Phases of Chaos}} \\ \vspace*{1.3cm}

\end{center}

\begin{center}
Tarek Anous$^{1,2}$ and Diego M. Hofman$^2$\\ 
\end{center}
\begin{center}
{
\footnotesize
$^1$ School of Mathematical Sciences, Queen Mary University of London, Mile End Road, London, E1 4NS, UK \\
$^2$ Institute for Theoretical Physics and $\Delta$-Institute for Theoretical Physics, University of Amsterdam, Science Park 904, 1098 XH Amsterdam, The Netherlands \\

}
\end{center}
\begin{center}
{\footnotesize\textsf{\href{mailto:t.anous@qmul.ac.uk}{t.anous@qmul.ac.uk}, \href{mailto:d.m.hofman@uva.nl}{d.m.hofman@uva.nl}} } 
\end{center}

\vspace*{0.5cm}

\vspace*{1.5cm}
\begin{abstract}
\noindent We develop a novel physical picture to understand certain universal properties of the GUE matrix model which are typically ascribed to quantum chaos, i.e.  the ramp and the plateau. We argue that these features should instead be associated with a pattern of spontaneous (or weak explicit) symmetry breaking. In this language, the GUE matrix model corresponds to an effective theory that describes the symmetry-broken phase, and where the Hermitian matrix of the GUE should be understood as a massive $\sigma$ field. The physics of this symmetry-broken phase governs certain particular features of the ramp such as its length and shape. However, the simple existence of a ramp is more universal and phase independent; it is related to sum rules obeyed by a large class of matrix models that constrain the interpolation to the plateau regime. Finally, the plateau is controlled by the symmetry-restored phase, which we call \textit{confined chaos}.

\end{abstract}

\newpage
\tableofcontents

\section{Introduction}
The SYK revolution of 2015 has led us to think about low-dimensional gravity in terms of ensemble averages, where the probability distribution one uses to compute these averages is over the space of Hamiltonians. In a precise sense, this line of reasoning was first highlighted by the discovery that the symmetries of AdS$_2$ are realized in the SYK model \emph{on average} rather than exactly \cite{KitaevTalks,Anninos:2013nra}; that is, not for any particular realization of the couplings. A second clue pointing to the fact that we should be considering random Hamiltonians is that black hole horizons are known to be maximally \emph{scrambling}, in other words, as strongly chaotic as any quantum system is allowed to be \cite{Maldacena:2015waa}---and what could be more chaotic than a completely random Hamiltonian sampled from a distribution? 

The final and most important piece of evidence for this perspective was provided in \cite{Saad:2019lba}, where the authors showed that the JT-gravity \cite{Jackiw:1984je,Teitelboim:1983ux} path integral, with fixed boundaries, admits an asymptotic expansion in genus that may be computed recursively, exactly analogously to the $1/N$ expansion of loop operators in a matrix model \cite{mirzakhani2007simple,Eynard:2007fi}  (see also \cite{Klebanov:1991qa,Ginsparg:1993is,Polchinski:1994mb,Maldacena:2004sn} and references therein for a historic perspective on two-dimensional holography). 
Thus, our ``AdS/CFT'' dictionary is of the sort summarized by the following equation:\footnote{This schematic picture should be understood in the infinite-$N$ limit, where higher genus contributions are suppressed.}
\begin{equation}
\langle Z(\beta)\rangle_{\rm MM}=\begin{gathered}\includegraphics[height=2.8cm]{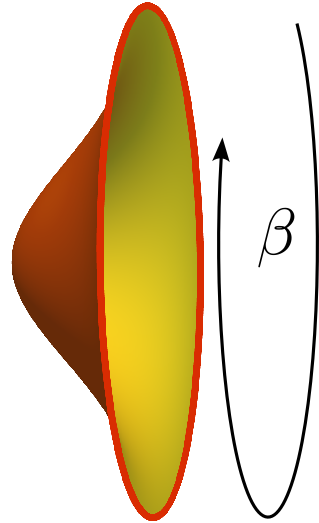}\end{gathered}\equiv Z_{\rm grav}(\beta)
\end{equation}
where the expectation value of the operator $Z(\beta)\equiv\text{Tr}\,e^{-\beta H}$ on the left-hand-side is computed in some matrix model
\begin{equation}
    \langle \cdot\rangle_{\rm MM}=\frac{1}{\mathcal{N}}\int [\d H]\, (\cdot)e^{-\text{Tr} \,V(H)}~,
 \end{equation}
while the diagram on the right-hand-side is obtained via a two-dimensional Euclidean-gravity path integral, e.g. the JT gravity path integral, with boundary conditions such that the (renormalized) geodesic length of the boundary circle is proportional to $\beta$.

We will push this perspective in a new way. What we will advocate for is that we should think of the operator $Z$ as \emph{charged under some symmetry}, and we will argue that the expectation values of the operator $Z(\beta)$ are consistent with the spontaneous (or weak explicit) breaking of that symmetry. This idea is not new, and has been advocated previously in \cite{Altland:2020ccq}, although the symmetry breaking principle (and symmetry group) we will discuss is different. 

If $Z$ were indeed charged under some continuous global symmetry, we would expect:
\begin{equation}
\langle Z(\beta)\rangle=0 \qquad\qquad \text{(symmetry preserving phase)}~.
\end{equation}
This is a simple equation with important consequences in the context of holography. To wit, if $Z(\beta)$ is indeed computed by a gravity path integral with fixed boundary conditions, then the above equality would suggest that the very notion of semiclassical geometry \emph{breaks down} in such a symmetry preserving phase. 
On the other hand, if we allow the underlying symmetry to be spontaneously broken, then we would expect $\langle Z \rangle \neq0$ in the spontaneous-symmetry-broken (SSB) phase. Hence if $Z$ is a charged operator in the underlying theory, then a standard gravitational description may only emerge if the continuous symmetry it is charged under is broken.

To be slightly more concrete, we will consider a compact $U(1)$ (more comments on the nature of this symmetry below and in the conclusions) acting by multiplication by a phase on $Z(\beta)$ (or more precisely its analytic continuation $Z(it)$). We can think of this symmetry as the dimensional reduction of the well known center symmetry that acts on Wilson loop operators, sometimes referred to as the single plaquette approximation in lattice gauge theory. Our analysis will be grounded in the standard understanding of generalized global symmetries \cite{Gaiotto:2014kfa}, and we will treat the operator $Z(\beta)$  using the language of SSB for \emph{Wilson-loop} operators $W(C)$ tracing out some curve $C$. For line operators charged under a one-form symmetry, the following general behavior is expected \cite{Hofman:2018lfz} depending on the phase of the theory we are in (whether the higher form symmetry is preserved or broken):
\begin{equation}\label{eq:gensymmetry}
\langle W(C)\rangle\sim\begin{cases}e^{-T_2\, \text{Area}(C)}~, &\text{symmetry preserved} \\e^{-T_1\, \text{Perimeter}(C)}~, &\text{symmetry spontaneously broken}
\end{cases}~,
\end{equation}
where $T_1$ and $T_2$ are dimensionful constants. 

A clear separation between these phases only happens in the large volume limit, where a phase transition becomes possible. In practice, all we really need is a large parameter available in our theory that quantifies the size of configuration space.  So, even at finite volume, a large $N$ (or $L$ in the notation we use below) limit for vector or matrix models can induce such a phase transition. We will call this regime the large volume regime irrespective of the details. It is important to remember that when our observables are made to scale with this large parameter, starting from the symmetry broken phase, it is possible to probe the physics of symmetry restoration.  In this sense one would expect to see a transition from a perimeter- to an area-law when the size of $C$ starts to probe the volume scale. This is another way to probe the different phases of the model within the symmetry broken ground state. We will argue for a similar effect that can be observed in $Z(\beta)$ at large $\beta$.

\begin{figure}[t!]
\begin{center}
\includegraphics[width=0.4\paperwidth]{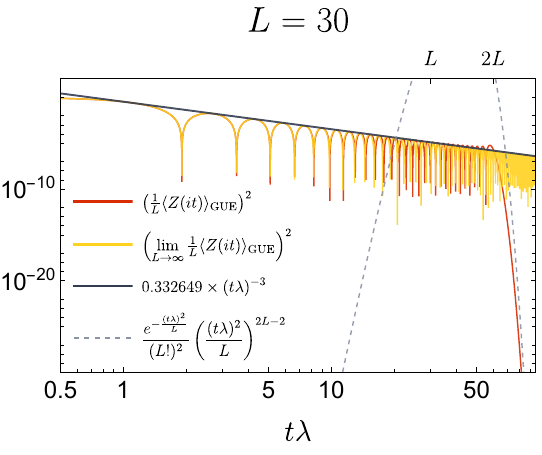}
\end{center}
\caption{Log-log plot of the exact $L=30$ \emph{(red)} and infinite-$L$ \emph{(yellow)} squared disk amplitude $\tfrac{1}{L}\langle Z(it)\rangle_{\rm GUE}$ as a function of complex temperature $\beta=it$. At small $t\lambda$, the disk amplitude oscillates around zero with a $(t\lambda)^{-3/2}$ envelope. In this range, it is clear that the exact and infinite-$L$ answers agree to very high accuracy. Noticeable differences begin to accumulate around $t\lambda\sim L$. Once $t\lambda\sim 2L$, the exact amplitude decays exponentially to zero, indicative of a phase transition. We plot the square of the observable because $\left\langle Z(it)\right\rangle_{\rm GUE}$ is not sign definite, meaning that, unless squared, its log will generally be complex. }\label{fig:symbreaking1pt}
\end{figure}

\paragraph{Why symmetry breaking is important in matrix models: } The first hint that there may be a notion symmetry breaking at play can be seen in the plot in figure \ref{fig:symbreaking1pt}. 
To make this plot we consider a particular matrix model known as the Gaussian-Unitary-Ensemble (GUE) where expectation values are computed as an integral over the entries of an $L\times L$ Hermitian Matrix $H$. Precisely, we are interested in a particular set of expectation values of so-called \emph{loop operators}: 
\begin{equation}\label{eq:GUEint}
 		\langle Z(\beta)\rangle_{\rm GUE}=\frac{1}{\mathcal{N}_{\rm GUE}}\int [\d H]\, \left[\text{Tr}\, e^{-\beta H}\right]e^{-\frac{L}{2\lambda^2 }\text{Tr} \,H^2}
 \end{equation}
and the normalization $\mathcal{N}_{\rm GUE}$ ensures that $\langle 1\rangle_{\rm GUE}=1$. GUE expectation values can be computed exactly. We review several ways to do this in \cref{ap:GUE}. In this case, we find: 
\begin{equation}\label{eq:onepointexacta}
     {\langle Z(\beta)\rangle_{\rm GUE}}={e^{\frac{(\beta \lambda)^2}{2L}}}L_{L-1}^1\left(-\frac{(\beta\lambda)^2}{L}\right),
 \end{equation}
where $L_\alpha^\beta(x)$ is a Laguerre polynomial.\footnote{Readers wishing to learn how to compute this expression are invited to consult \cref{ap:GUE1pt}.} Let us now analyze this expression. First note that, in the thermodynamic limit ($L\rightarrow\infty$, with $t\lambda$ fixed), we have: 
\begin{equation}\label{eq:powerlawdecay}
   \lim_{L\rightarrow\infty}\frac{\langle Z(it)\rangle_{\rm GUE}}{L}=\frac{J_1(2t\lambda)}{t \lambda}\underset{t\lambda\gg1}{\approx}\frac{\pi\,\sin\left(2 t\lambda -\frac{\pi}{4}\right)}{(\pi t\lambda)^{3/2}}~. 
\end{equation}
This is a power-law decay with oscillations on top. Using the language of \eqref{eq:gensymmetry}, we would like to interpret this as the perimeter-law phase, since the functional behavior is polynomial, and hence sub-exponential, in $t$. We plot this large-$L$ behavior in yellow in figure \ref{fig:symbreaking1pt}, whereas the exact answer (given in \eqref{eq:onepointexacta}) is plotted in red for $L=30$ in the same figure. Notice that the large-$L$ approximation agrees with the exact answer so long as $t\lambda< L$. Once we reach the volume scale $t\lambda\sim L$, errors between the exact and approximate answers start to build up, and at $t\lambda\sim 2 L$ something drastic happens: the exact expectation value of $\langle Z(it)\rangle_{\rm GUE}$ decays exponentially to zero, marking a stark difference between the large-$L$ and the exact answer. We would like to interpret this exponential decay as an area-law phase.

What we are advocating for is that the transition between power-law behavior and exponential decay in $t\lambda$ is evidence for a transition from a symmetry-broken to a symmetry-restored phase. In this and further examples we see that $L$ acts as the {volume scale}.  In this interpretation, once our dimensionless parameter $t\lambda$ begins to probe the volume scale, the theory has explored the entire phase space and there are no longer any effective degrees of freedom at play. In this case, emergent degeneracies are lifted and whatever symmetry was broken must be restored. 

But what symmetry could the operator $Z(\beta)$ be charged under? From the perspective of the GUE, this is not obvious. In what follows, we will argue that the GUE (and its perturbative deformations) should be understood as an effective theory for a symmetry broken phase. Here there is an important distinction between matrix models and $d \neq 0$ quantum field theories. In the latter case we would expect the low energy physics to be controlled by a Goldstone massless boson and a massive field $\sigma$ related to radial excitations.\footnote{This is the common nomenclature when studying symmetry breaking in linear $\sigma$-models.} Concretely, a simple $U(1)$ example can be described by the following action:
\be\label{eq:linearcomplex}
S_{U(1)} = \int \d^d x\,\left\lbrace \frac{1}{2} \left| \partial \Phi \right|^2 + V\left(|\Phi|^2\right)\right\rbrace = \int \d^d x\, \left\lbrace\frac{e^{2\sigma}}{2}\left[ (\partial \varphi)^2 + (\partial \sigma)^2 \right] + V\left(e^{2\sigma}\right) \right\rbrace~,
\ee
\noindent where we have parameterized $\Phi \equiv e^{i \varphi + \sigma}$. Near the minimum of $V$, the $\sigma$ field describes a massive excitation whereas $\varphi$ corresponds to a massless Goldstone boson.
 Evidence of the original $U(1)$ symmetry has been washed away once we flow to the effective theory where the symmetry acts non-linearly. In a matrix model (corresponding to $d=0$ above),  there is no spatial modulation of the fields and thus there is little functional difference between a spontaneously-broken global symmetry and a gauge symmetry. What we mean by this is the following: for any compact symmetry we can explicitly integrate over the degenerate vacua of the theory, leaving no light degrees of freedom behind. This means that the massive $\sigma$-field describes the full effective theory in a matrix model.

Hence we advocate for interpreting the GUE matrix model as the effective  $\sigma$-field ($H$ in the GUE) theory. As there is no Goldstone, it is not clear how to identify the underlying symmetry from within the GUE theory. But we do seem to get an important hint:  the exponential decay of $\langle Z(it)\rangle_{\rm GUE}$ signals that there \emph{exists} a symmetry preserving phase, without revealing what symmetry is being restored. What happens concretely is that, at large $t\lambda$ the potential term $\text{Tr} \,H^2$ in the GUE action mainly plays  the role of an IR regulator which suppresses large $H$ excursions in the matrix integral. But most of the physics is not sensitive to this IR regulator. A useful analogy is that of a quantum mechanical system in the WKB approximation. At large energies, all wave-functions become plane waves with small, potential-dependent, modulations; the potential acts as an IR regulator that determines the spread of this wave-function.  From this perspective we can say that at \emph{infinite} $t \lambda$, there exists a $U(1)$ symmetry which acts by translation: $H \rightarrow H +a$, with $a$ a constant, that is restored. However, at large but finite $t \lambda$, there are exponential tails in the eigenvalue distributions that weakly break the symmetry through the IR regulating scale $\lambda$.

In order to proceed, we need a UV theory, the analog of $S_{U(1)}$---the linear complex boson theory above in \eqref{eq:linearcomplex}---which preferably has a continuous symmetry that can be spontaneously broken, as well as a remaining massive $\sigma$ field. We will demand that the symmetry at stake be compact. This will ensure IR finiteness of the system  without invoking the scale $\lambda$, meaning the symmetry can be realized exactly. Whether this happens in practice as opposed to situations where the symmetry is weakly broken will not be crucial for the arguments below. The main difference between these cases is the explicit form of the exponential approach to the symmetry restored phase.

Once we have our UV theory, we would like to show that it reduces to the GUE in the symmetry broken phase. If the GUE describes the universal physics of a phase that what we call (deconfined) \textit{quantum chaos}, then the symmetry preserved phase represents an \emph{altogether different} effective theory which we henceforth  call \textit{confined chaos}. We will study this physics in the remainder of this work and show that universal features like the ramp or the plateau usually discussed in the context of universal quantum systems can be associated to the different possible phases of chaos that these systems can manifest.  

\subsection*{Summary of results}

The summary of our paper is as follows: In \cref{sec:sumrules} we define a new set of `sum rules' for the one- and two-point functions of loop operators $\left(Z(it)\equiv \text{Tr}\, e^{-i H t}\right)$, connecting the early- and late-time behaviors of these correlation functions. The conclusion of this section is that the very existence of a ramp in the spectral form factor is ensured by these sum rules. In other words, ramps are \emph{generic} in a broad class of matrix models. Our goal in this section is to highlight that, while ramps are generic, their shapes depend on the particular IR-phase of the system. In \cref{sec:unitary_matrix_models} we introduce the UV-model of interest to us: the Yang-Mills matrix model, which is invariant under a $U(1)$ symmetry. We review the exact solution of this model in \cref{sec:phases_of_YM} and its (symmetry broken or preserved) phases  in section \ref{sub:symmetry_breaking_YM}. In these sections we review the relationship between the Yang-Mills matrix model and the Gross-Wadia-Witten (GWW) matrix model, which is particularly useful for computing observables. In \cref{sec:gwwobservables} we review how to compute observables in the GWW model (although equation \eqref{eq:fullyconnectedGWW} is new). We assemble these reviewed facts in \cref{sec:observables_YM}, where we numerically demonstrate that, in the Yang-Mills matrix model, the behavior of the ramp is fully governed by the particular IR-phase (symmetry broken or preserved), and where the particular phase of the system has imprints on macroscopic aspects of the ramp, including its shape and duration. In this section, we also numerically demonstrate, by directly comparing the GUE and the Yang-Mills matrix models in the symmetry broken phase, that the plateau region (the exit from the ramp) depends sensitively on the UV completion, even if the ramps are qualitatively indistinguishable. We present our conclusions in \cref{sec:conclusions}. In the appendices we collect and review various exact results pertaining to the GUE and GWW matrix models. 

\section{General rules for single-trace Hermitian matrix models}\label{sec:sumrules}

Before getting into details about the specific UV theory we will consider, we first need to set a few expectations. We refer the reader to appendix \ref{ap:singletracemm} for a self-contained (for our purposes) review of single-trace Hermitian matrix models and their exact solutions via the method of orthogonal polynomials. In this section we use these known results to present certain behaviors that must be obeyed by the expectation values of interest to us, in any theory. We will call these behaviors \emph{sum rules}. These rules are theory-independent and rely on the finiteness of these models and the fact that we can compute integrals over a finite number of degrees of freedom. 

\subsection{Sum rule for the one-point function} \label{sec:1ptsumrule}
In a general Hermitian matrix model, the one-point function of $Z(it)\equiv \text{Tr}\, e^{-i H t}$ must satisfy certain constraints. The most obvious of which is that 
\begin{equation}\label{eq:t0onept}
\langle Z(0)\rangle_{\rm MM} = \langle \text{Tr}\, \mathds{1}\rangle_{\rm MM} = L~,
\end{equation} 
since this quantity computes the dimension of the Hilbert space of an $L\times L$ Hamiltonian. As reviewed in appendix \ref{ap:singletracemm}, for a generic matrix ensemble defined by a particular potential $V(H)$
\begin{equation}
    \langle \cdot\rangle_{\rm MM}=\frac{1}{\mathcal{N}}\int [\d H]\, (\cdot)e^{-\text{Tr} \,V(H)}~,
 \end{equation}
 the expectation value of $Z(it)$ must take the following general form 
 \begin{equation}
 \langle Z(it)\rangle_{\rm MM}=\int_{-\infty}^{\infty}\d E\, e^{-i E t}\sum_{j=0}^{L-1}\psi_j(E)^2~.
 \end{equation}
The quantities  $\psi_j(E)$ are called the `wavefunctions' of the ensemble. These wavefunctions are related to the orthogonal polynomials $P_j$ of the matrix ensemble  as follows (see \eqref{eq:orthpoldef}, \eqref{eq:wavefunctiondef}, and \eqref{eq:onepointgeneral}  for definitions):
\begin{equation}
\psi_j(E)\equiv\frac{P_j(E)}{\sqrt{h_j}}e^{-\frac{V(E)}{2}}~,
\end{equation}
and are orthonormal:
\begin{equation}\label{eq:orthonormality}
\int \d E\, \psi_i(E)\psi_j(E)=\delta_{ij}~.
\end{equation}

 We will shortly be interested in taking a large-$L$ limit, so let us define $V(E)\equiv L\, v(E)$ and let us moreover assume that our ensemble is such that the polynomial coefficients in the rescaled potential $v(E)$ are all $\mathcal{O}(1)$. Then at late times, $ t \sim L$, let us define $t\equiv L \tilde t$, and write 
 \begin{equation}
 \langle Z(it)\rangle_{\rm MM}=\int_{-\infty}^{\infty}\d E\, e^{-L(i E \tilde t+v(E))}\times (\text{polynomial in $E$})~.
 \end{equation}
 We can approximate this integral using the saddle point method. In this regime, the integral is dominated by a complex energy $E^*$ which satisfies: 
 \begin{equation}
 v'(E^*)=-i\tilde{t}~.
 \end{equation}
 Thus at late times we find: 
 \begin{equation}
 \langle Z\left(iL\tilde{t}\right)\rangle_{\rm MM}\propto e^{-L(i E^*\tilde{t}+v(E^*))}=e^{-L(v(E^*)-E^* v'(E^*))}~.
 \end{equation}
 If the quantity $(v(E^*)-E^* v'(E^*))$ has a positive real part, we note that this expectation value becomes exponentially small at late times. For example, taking $v(E)$ to be dominated by its Gaussian piece, that is $v(E)=\frac{E^2}{2\lambda^2}+\text{subleading}\dots$, we find: 
\begin{equation}\label{eq:latetime1ptsumrul}
 \langle Z\left(it\right)\rangle_{\rm MM}\underset{t\sim L}{\propto} e^{-\frac{(t\lambda)^2}{2L}+\text{subleading}\dots}~
 \end{equation}
 where we have reinstated $t\sim \mathcal{O}(L)$ in favor of $\tilde{t}$. Hence we have derived a particular general expection for the behavior of $\langle Z(it)\rangle_{\rm MM}$: it must start at $L$ at $t=0$ and eventually decay to zero, for a general class of Hermitian matrix models. While this rule does not tell us anything about the functional form of $\langle Z(it)\rangle_{\rm MM}$ save for its early- and late-time asymptotic values, we can distinguish \emph{qualitatively} between different matrix models by how quickly or slowly this transition from $L$ to zero happens.  
 
 In general, we can even estimate how long this interpolating region is. If we consider imaginary time $t= i \beta$ we can see that the result above comes from a saddle dominated by energies $E^*= - \frac{\beta \lambda}{L} \lambda$. On the other hand the calculation can only be valid when it is this exponentially suppressed region of the eigenvalue distribution that dominates. For energies in the range $ E \in [-2\lambda,2\lambda]$ it is well known that the large $L$ behavior is controlled by the Wigner semi-circle law, see appendix (\ref{largeLGUEsec}). Putting these results together we expect the exponentially suppressed regime to kick in at $\beta \lambda \sim 2L$, in agreement with the discussion in the introduction.

 How to interpret this result? One could write the exponential decay above in a suggestive fashion:
 
\begin{equation}
 \langle Z\left(it\right)\rangle_{\rm MM}\underset{t\sim L}{\propto} e^{- t \Lambda_{IR}(t)}
 \end{equation}
 \noindent where $\Lambda_{IR}$ is a $\mathcal{O}(L^0)$ (time dependent) IR scale. Looking at this we are tempted to interpret the exponential decay as a sort of  symmetry restored phase emerging where we can shift $H \rightarrow H + a$, where $a$ is a constant. The specific time dependence of $\Lambda_{IR}$ captures the fact that the potential actually weakly breaks the symmetry by exponentially small effects.

\subsection{Sum rule for the spectral form factor } \label{sec:2ptsumrule}

Here we derive a similar sum rule for the connected two-point function and particularly the spectral form factor, similar to the sum rules obtained in \cite{Winer:2023btb}. The spectral form factor is generally defined as follows (see e.g. \cite{Haake2018} for a standard reference):
\begin{equation}
\text{S}(t)\equiv\langle Z(it)Z(-it)\rangle_{\rm MM}
\end{equation}
although we will sometimes use it to mean simply the \emph{connected part}  of the spectral form factor, $\text{S}^c(t)\equiv\langle Z(it)Z(-it)\rangle_{\rm MM}^c$~.

The method of orthogonal polynomials allows us to write the following general formula for the connected two-point function in a general class of matrix models (see equations \eqref{eq:twopointconnecteddef} and \eqref{eq:exact2ptdensity}): 
\begin{equation}
\langle Z(\beta_1)Z(\beta_2)\rangle^c_{\rm MM} =-\sum_{j,k=0}^{L-1}\int \d E_1 \d E_2\, e^{-\beta_1 E_1-\beta_2 E_2}\psi_j(E_1)\psi_j(E_2)\psi_k(E_1)\psi_k(E_2)~,
\end{equation}
The orthonormality relation \eqref{eq:orthonormality} immediately implies that, as in \eqref{eq:t0onept}:
\begin{equation}\label{eq:t0twopt}
\text{S}^c(0)=\langle Z(0)Z(0)\rangle^c_{\rm MM}=-L~.
\end{equation}
Moreover, by following the same steps that led to \eqref{eq:latetime1ptsumrul},  then we can similarly conclude that, at late times the connected SFF goes as: 
\begin{equation}\label{eq:latetime2ptsumrul}
 \text{S}^c(t)\underset{t\sim L}{\propto}  -e^{-2(V(E^*)-E^* V'(E^*))}~,
 \end{equation}
 and thus is exponentially small if $V(E^*)-E^* V'(E^*)$ has a positive real part, as before. 
 For ensembles governed by GUE universality, that is, if the matrix ensemble is such that $V(E)=L\frac{E^2}{2\lambda^2}+\text{subleading}\dots$, this means $\text{S}^c(t)\propto-e^{-\frac{(t\lambda)^2}{L}+\text{subleading}\dots}$ at late times. This late behavior is called the \emph{plateau}.
 
 Once again it is tempting to write this as:
\begin{equation}
 \text{S}^c(t)\underset{t\sim L}{\propto}  -e^{-2 t \Lambda_{IR}(t)}~,
 \end{equation}
 which is the familiar behavior we expect for correlators of charged operators in a theory that linearly realizes the symmetry these operators are charged under, and where $\Lambda_{IR}(t)$ takes the place of the mass of the relevant degree of freedom. 

 Hence, it is a generic expectation in a large class of matrix models that the (connected) SFF interpolate between $-L$ and zero in a time proportional to $L$. This interpolating behavior is often called the \emph{ramp}. In essence, we are concluding that ramps are generic in a large class of matrix models, \emph{but their shapes are not}. The initial shape of the ramp and the time required to reach the plateau depend crucially on the realized symmetries of the model.

 In the next section we will introduce a particular UV model that will exhibit the physics that we've so far been alluding to. The symmetry of interest will be realized exactly but can be spontaneously broken. In this UV model the `partition function' operator $Z(it)$ is naturally charged under a continuous $U(1)$ symmetry. What we will show is that, while the sum rules we derived above are generic, the shapes of the various correlation functions, will depend in important ways on whether this $U(1)$ symmetry is preserved or spontaneously broken. For the two-point function in particular, we will show that, while the existence of a ramp is generic across a large class of matrix models,  the \emph{shape of the ramp}---how the sum rule and the eventual approach to the plateau is fulfilled---depends heavily on the ensemble of interest and whether the underlying $U(1)$ symmetry is preserved or broken.

\section{Unitary matrix models}\label{sec:unitary_matrix_models}
We will now discuss \emph{unitary} matrix models.  Unitary matrix models were first introduced by Dyson \cite{Dyson:1962es,Dyson:1962es2,Dyson:1962oir,Dyson:1970tza} and are sometimes referred to as the ``circular ensembles'' \cite{Meh2004}. Like in the GUE example \eqref{eq:GUEint}, we will define expectation values in a general unitary matrix model as follows:
\begin{equation}\label{eq:defofUnitarymatrixensemble}
   \langle\cdot\rangle_{\rm UMM}=\frac{1}{\mathcal{N}}\int  [\d U]\, (\cdot)\exp\left\lbrace- V(U)\right\rbrace~
 \end{equation}
 where now $U$ is an $L\times L$ unitary matrix, meaning its eigenvalues can be expressed as $\lambda_i=e^{i\theta_i}$ with $-\pi\leq \theta_i\leq \pi$, $[\d U]$ is the standard Haar measure, and $\mathcal{N}$ is a normalization that ensures $\langle 1\rangle_{\rm UMM}=1$. 

 A unitary matrix $U$ satisfies 
 \begin{equation}
 U U^\dagger = \mathds{1}~, 
 \end{equation}
and this condition is invariant under 
 \begin{equation}\label{eq:centersymmetry}
    U\rightarrow e^{i\varphi}U~,\qquad \qquad \varphi\in (0,2\pi)~.
 \end{equation} 
 We would like to build our effective theory around this $U(1)$ symmetry, and to explore the consequences of its spontaneous breaking. To start, let us discuss how to build matrix potentials $V(U)$ that are invariant under this $U(1)$ symmetry (see e.g. section 3 of \cite{Alvarez-Gaume:2005dvb}). 

\begin{figure}[t!]
\begin{center}
\includegraphics[width=0.4\paperwidth]{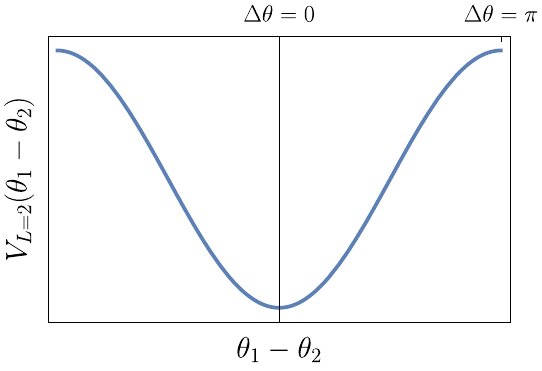}
\end{center}
\caption{Unitary Matrix Model potential \eqref{eq:potentialSYM} for $L=2$. The potential is minimized when the eigenvalues are equal and maximized when they are diametrically opposed. }\label{fig:unitarypotential}
\end{figure}

 Naively, one might imagine starting with a single-trace term, for example $\text{Tr} U+\text{Tr} U^\dagger$ or perhaps  $\text{Tr} U^2+\text{Tr} U^{-2}$ as in the GUE,\footnote{We add the complex conjugates so that our potentials are real.} but neither choice is invariant under the $U(1)$ symmetry of interest, so do not suit our purposes. We might also be compelled to consider a term such as $\text{Tr}\left(U U^\dagger\right)$, but this is obviously trivial since $\text{Tr}\left(U U^\dagger\right)=\text{Tr}\,\mathds{1}=L$. Instead, the basic building blocks for such an invariant $V(U)$ turn out to be powers of multi-trace objects: 
 \begin{equation}
 \left(\text{Tr}U^{n_1}\text{Tr}U^{n_2}\dots\text{Tr}U^{n_k}\right)^{t_k}~,\quad\quad \sum_{i=1}^k n_k=0~, \quad\quad t_k>0~.
 \end{equation}
Thus the lowest-order non-trivial $U(1)$-symmetric potential is of the following form:
\begin{equation}\label{eq:potentialSYM}
   V(U)= -a\text{Tr} U \text{Tr} \, U^\dagger~, \qquad\qquad a>0~.
\end{equation}
Note our choice of sign for this potential, as it will be important in what follows. Unlike the Hermitian matrix models described earlier, this is a double trace potential, and is negative semi-definite. As we will show, this model has all the features of symmetry breaking that we seek. This will be our UV theory. We call this particular unitary matrix model, with the potential given by (\ref{eq:potentialSYM}), the \emph{Yang-Mills Matrix Model}. We will explain the origin of this name further below. We denote expectation values in this model as follows:
\begin{equation}\label{eq:defofYMensemble}
   \langle\cdot\rangle_{\rm YM}=\frac{1}{Z(a)}\int  [\d U]\, (\cdot)\exp\left\lbrace a\text{Tr} U \text{Tr} \, U^\dagger\right\rbrace~.
 \end{equation}
To get some intuition on the potential \eqref{eq:potentialSYM}, let us study the case $L=2$, for which: 
\begin{equation}V_{L=2}=-4a\cos^2\left(\frac{\theta_1-\theta_2}{2}\right)~.
\end{equation} 
In the $U(1)$ preserving phase, by definition, we must have: 
\begin{equation}
\left\langle \text{Tr} \, U^k\right\rangle_{\rm YM}=0 ~, \quad \forall k\in \mathbb{Z}\backslash \{0\}~.
\end{equation}
Let us consider what this means for $k=1$. If $\left\langle\text{Tr} U\right\rangle_{\rm YM} = 0$ then
\begin{equation}
\left\langle\text{Tr}\,U\right\rangle_{\rm YM}=\left\langle e^{i\theta_1}+e^{i\theta_2}\right\rangle=\left\langle e^{i\theta_2}\left(1+e^{i(\theta_1-\theta_2)}\right)\right\rangle=0 \quad \implies\quad \left\langle\theta_1-\theta_2\right\rangle\approx\pi~. 
\end{equation}  Plugging this into the potential above, we notice that this corresponds to a local \emph{maximum} (see figure \ref{fig:unitarypotential}).  Hence, we have chosen the sign of our potential in \eqref{eq:potentialSYM} such that it is energetically disfavored to preserve the $U(1)$ symmetry. In other words, the chosen sign of the potential favors symmetry breaking, with the coupling constant $a$ governing the depth of the potential. This intuition will continue to hold for higher values of $L$. Thus, in the $L\rightarrow\infty$ limit, we expect the $U(1)$ symmetry to be spontaneously broken for large enough $a$. This is precisely the type of physics we're after. 

To make the connection with the GUE and other Hermitian matrix models more precise,  consider writing the unitary matrix $U$ as the exponential of a Hermitian matrix $H$: $U\equiv e^{-i H}$,\footnote{It is well known that for every unitary matrix model, there exists and equivalent Hermitian matrix model, see \cite{Mizoguchi:2004ne} and references therein.} then 
\begin{equation}
\text{Tr}\, U^k=\text{Tr}\,e^{-ik H}\equiv Z(ik)~,
\end{equation}
hence Hermitian matrix model expectation values, as computed in \eqref{eq:GUEint}---for example, $\left\langle Z(ik)\right\rangle_{\rm GUE}$, where $k$ acts as a discrete time variable---can be related to unitary matrix model expectation values. Since our unitary matrix models are designed to obey a continuous global $U(1)$ symmetry, this translates to a particular shift symmetry of the energy variables in the related Hermitian matrix model. Thus we have given a precise sense in which the operator $Z(it)$ can be charged in the underlying theory, although we have yet to make a direct connection to the GUE.

\subsection{Four-dimensional origin}
 Incidentally, this exact matrix model \eqref{eq:defofUnitarymatrixensemble} with potential \eqref{eq:potentialSYM} has a four-dimensional origin. It can be obtained starting from $SU(L)$ Yang-Mills theory on a  $S^1_\beta\times S^3$ at large-$L$, after integrating out all the massive modes on the $S^3$  \cite{Aharony:2003sx,Liu:2004vy} (see also \cite{Alvarez-Gaume:2005dvb} for deformations of this model), which explains our \emph{Yang-Mills} nomenclature. The $S^1_\beta$ of size $\beta$ is interpreted as the Euclidean-time circle. After following this procedure, one finds an effective theory of the Polyakov loop---a Wilson loop wrapping the Euclidean time circle. From the gauge theory perspective, the $k^{\rm th}$ power of the Polyakov loop: 
\begin{equation}
   U^k=P\exp\left(i\int_0^{k\beta} \d\tau\,A\right)~,
\end{equation}
winds the Euclidean-time circle $k$ times. Thus the power $k$ is a stand-in for the perimeter of the loop. This gauge theory has a $\mathbb{Z}_L$ center symmetry, which gets promoted to a $U(1)$ symmetry at large-$L$, which is the $U(1)$ symmetry of interest to us. If the gauge group is $U(L)$ the center symmetry is $U(1)$ for all values of $L$. 

 From the four-dimensional perspective, the coupling constant $a$ is a theory-dependent function of temperature. In $\mathcal{N}=4$ SYM, with gauge group $SU(L)$, for example, it is given by the following monotonically increasing function of $T \equiv \frac{1}{\beta}$ \cite{Aharony:2003sx}:
\begin{equation}
   a(T)=\frac{2e^{-\beta}\left(3-e^{-\frac{\beta}{2}}\right)}{\left(1-e^{-\frac{\beta}{2}}\right)^3}~. 
\end{equation}
The theory is known to have a Hagedorn phase transition when $a(T)=1$, characterized by whether the $U(1)$ symmetry is preserved or broken \cite{Aharony:2003sx,Liu:2004vy}. Thus there is a direct line between symmetry breaking in this matrix model, and the behavior of loop operators in gauge theory, whose expectation values follow the general equation \eqref{eq:gensymmetry} across the various phases.  As we saw in figure \ref{fig:symbreaking1pt}, starting from a $U(1)$-symmetry broken phase, we expect that as $k$ begins to probe the volume scale, the behavior will begin to show signs of symmetry restoration, by morphing from a sub-exponential, to an exponential decay. We will demonstrate this behavior precisely below.

\subsection{Phases of the Yang-Mills matrix model}\label{sec:phases_of_YM}

We will now review this model's solution \cite{Liu:2004vy} (see also \cite{Klebanov:1994pv,Klebanov:1994kv} for a similar analysis), and explain how to compute the various correlation functions of interest. First, remark that we can turn the double trace potential \eqref{eq:potentialSYM} into a single-trace one, via a Hubbard-Stratonovitch transformation: 
\begin{equation}\label{eq:lambdab4g}
   \int  [\d U]\, (\cdot)\exp\left\lbrace a\text{Tr} U \text{Tr} \, U^\dagger\right\rbrace=\frac{1}{2\pi a}\int [\d U] \d\lambda \d \bar{\lambda}\, (\cdot)\exp\left\lbrace -\frac{\lambda\bar{\lambda}}{a}+\lambda\text{Tr} U +\bar{\lambda}\text{Tr} \, U^\dagger\right\rbrace.
\end{equation}
At this stage, the $U(1)$ symmetry is still manifest as the action and measure are invariant under 
 \begin{equation}\label{eq:manifestshiftymmetry}
    U\rightarrow e^{i\varphi}U~,\qquad \qquad \lambda\rightarrow e^{-i\varphi}\lambda~.
 \end{equation} 
This means that, if we parametrize 
\begin{equation}\label{eq:lambdaparam}
   \lambda=\frac{L g}{2}e^{i\alpha}~,\qquad\qquad \bar{\lambda}=\frac{L g}{2}e^{-i\alpha}~,
\end{equation}
then the variable $\alpha$ can be integrated out at will, since its dependence in the action can be removed by a suitable transformation in $U$ which leaves the measure $[\d U]$ invariant.\footnote{One technical caveat worth mentioning is that integrating out the $U(1)$ mode $\alpha$ can be done so long as we commit to computing expectation values of neutral operators. However, in practice we will show that the expectation values of charged operators can be computed in the SSB phase once we've selected a vacuum where the symmetry is spontaneously broken.} In the end, we will be interested in computing: 
\begin{equation}\label{eq:fullexpectationSYM}
  \langle\cdot\rangle_{\rm YM}= \frac{\frac{L^2}{2a}\int_0^\infty g\d g\,e^{-\frac{L^2g^2}{4a}}\int [\d U]\,(\cdot)\exp\left\lbrace\frac{L g}{2}\text{Tr}\left(U+U^\dagger\right)\right\rbrace}{\mathcal{Z}(a)}~,
\end{equation}
where the partition function in the denominator is defined as: 
\begin{equation}\label{eq:PartitionFuncSYM}
   \mathcal{Z}(a)=\frac{L^2}{2a}\int_0^\infty g\d g\,e^{-\frac{L^2g^2}{4a}}\int [\d U]\,\exp\left\lbrace\frac{L g}{2}\text{Tr}\left(U+U^\dagger\right)\right\rbrace~,
\end{equation}
and the coefficient out front comes from the Jacobian determinant from the parametrization \eqref{eq:lambdaparam}, and the $2\pi$ in the denominator of \eqref{eq:lambdab4g} cancels with the integral over the variable $\alpha$.

For all practical purposes, we have reduced our Yang-Mills matrix model calculations to first computing \emph{un-normalized}  expectation values in the famed Gross-Wadia-Witten (GWW) model \cite{Gross:1980he,Wadia:2012fr} at coupling $g$ (see \cite{Marino:2008ya} for a review), that is, if we define:
\begin{equation}\label{eq:GWWdef}
   \langle\cdot\rangle_{\rm GWW}\equiv\frac{1}{\mathcal{N}_{\rm GWW}}\int [\d U]\,(\cdot)\exp\left\lbrace\frac{L g}{2}\text{Tr}\left(U+U^\dagger\right)\right\rbrace~,
\end{equation}
with $\mathcal{N}_{\rm GWW}$ tuned such that $\langle1\rangle_{\rm GWW}=1$,  then we can obtain results in the Yang-Mills model by using the following formula:
\begin{equation}\label{eq:YMexpectationvalues}
 \langle\cdot\rangle_{\rm YM}= \frac{\frac{L^2}{2a}\int_0^\infty g\d g\,e^{-\frac{L^2g^2}{4a}}\mathcal{N}_{\rm GWW}\, \langle\cdot\rangle_{\rm GWW}}{\mathcal{Z}(a)}~,
\end{equation}
where $\mathcal{Z}(a)$ is included in the denominator to ensure $ \langle1\rangle_{\rm YM}=1$. Explicitly this means: 
\begin{equation}\label{eq:ZaYM}
\mathcal{Z}(a)=\frac{L^2}{2a}\int_0^\infty g\d g\,e^{-\frac{L^2g^2}{4a}}\mathcal{N}_{\rm GWW}~.
\end{equation}

One comment is in order: If we stare at how expectation values in the Yang-Mills model are computed, starting from \eqref{eq:GWWdef}, it would appear that the original $U(1)$ symmetry of interest has disappeared, since the action \eqref{eq:GWWdef} now transforms non-trivially under shifts $ U\rightarrow e^{i\varphi}U$. This was not the case in \eqref{eq:lambdab4g}, since the field $\lambda$ also transforms, but this symmetry becomes hidden once we integrate out $\alpha$. The resolution is that the $U(1)$ symmetry will ultimately be restored upon integrating over the coupling $g$ as in \eqref{eq:fullexpectationSYM}.

\subsection{Symmetry breaking at large-\texorpdfstring{$L$}{L}}\label{sub:symmetry_breaking_YM}
\subsubsection{Summary}
Before explaining how we derive these results, let us briefly summarize the phase structure of the model defined in \eqref{eq:PartitionFuncSYM}. As previously mentioned, at large-$L$, there is a Hagedorn phase transtion across $a=1$ \cite{Aharony:2003sx,Liu:2004vy}. Consider the free energy as a function of $a$:
\begin{equation}
   F(a)\equiv\log \mathcal{Z}(a)~.
\end{equation}
The result of a simple large-$L$ analysis (which we will explain below) gives \cite{Liu:2004vy}: 
\begin{equation}\label{eq:SYMfreeE}
    F(a)\approx\log(L!)+\begin{cases}
   -\log(1-a) &a<1~,\\
   L^2 \mathcal{F}_{0}(a) &a>1,
   \end{cases}
\end{equation}
where 
\begin{equation}
   \mathcal{F}_0(a)\equiv-\frac{g_*^2}{4a}+g_*-\frac{1}{2}\log g_*-\frac{3}{4}~,\qquad\qquad g_*\equiv a+\sqrt{a(a-1)}~.
\end{equation}
First notice that across $a=1$ we see a first-order confinement/deconfinement transition where the free energy (after subtracting the $\log (L!)$ piece) goes from scaling like $\mathcal{O}(1)$ for $a<1$ to scaling like $L^2$ beyond $a>1$. Furthermore, as we will show, for $a<1$, the theory is gapped and the expectation values of most interesting operators, e.g. $\langle\text{Tr} U^k\rangle\approx 0$ in the large-$L$ limit, as expected in a $U(1)$ preserving phase. Finally, the $a>1$ free energy matches that of the GWW model at a particular value of the coupling given by $g_*$, see equation \eqref{eq:GWWfreeE}. So for $a>1$, we have an effective theory governed by GWW universality, where the $U(1)$ symmetry is spontaneously broken. We will later demonstrate, in a precise sense, that this GWW universality is the same as GUE universality.

\subsubsection{Derivation of the Large-\texorpdfstring{$L$}{L} free energy}\label{sec:freeederivation}

Now let us explain how we derive this result. We start with the GWW partition function, which we have labeled as $\mathcal{N}_{\rm GWW}$ in \eqref{eq:GWWdef}. Since the potential involves traces of $U$, we may integrate out the off-diagonal components of $U$ and write the remainder of the Haar measure in terms of the eigenvalues. Thus:
\begin{equation}
   \mathcal{N}_{\rm GWW}=\int_{-\pi}^\pi \left[\prod_{i=1}^L \frac{\d\theta_i}{2\pi}\right]\, \left|\Delta\left(e^{i\theta}\right)\right|^2 \exp\left\lbrace L g\sum_{i=1}^L \cos \theta_i\right\rbrace~,
\end{equation}
where 
\begin{equation}
   \Delta\left(e^{i\theta}\right)\equiv \prod_{k<j}^L \left(e^{i\theta_j}-e^{i\theta_k}\right)=\text{det}\left[e^{i(j-1)\theta_k}\right]~
 \end{equation}
 is the standard Vandermonde determinant. We may bring the Vandermonde determinant into to the exponent as follows:
\begin{equation}
   \mathcal{N}_{\rm GWW}=\int_{-\pi}^\pi \left[\prod_{i=1}^L \frac{\d\theta_i}{2\pi}\right]\,  \exp\left\lbrace L g\sum_{i=1}^L \cos \theta_i+\sum_{i<j}^L\log\left[4{}\sin^2\left(\frac{\theta_i-\theta_j}{2}\right)\right]\right\rbrace~,
\end{equation}
which will allow us to perform a Coulomb-gas analysis (see \cite{Eynard:2015aea} for a review).
 First let us introduce the ``equilibrium'' eigenvalue density: 
\begin{equation}\label{eq:equildensityGWW}
\rho_{\rm eq}(\theta)\equiv \frac{1}{L}\sum_{i=1}^L\delta(\theta-\theta_i)~.
\end{equation}
To establish some nomenclature: We will use the term \textbf{equilibrium} to denote quantities whose limits exist \emph{at infinite $L$}. For example, in \eqref{eq:equildensityGWW}, the multiplicative factor of $1/L$ is needed to ensure that this limit exists. When we denote an equilibrium quantity in an expectation value, e.g. $\left\langle\rho_{\rm eq}(\theta)\right\rangle_{\rm GWW}$, as we have in \eqref{eq:smallgdensity} and \eqref{eq:largegedensity} below, we specifically mean these quantities are to be understood strictly at infinite-$L$. 

The equilibrium density allows us to replace the sums with integrals in the following way:
\begin{equation}
   L g\sum_{i=1}^L \cos \theta_i\rightarrow L^2 g\int_{-\pi}^\pi\frac{\d \theta}{2\pi} \rho_{\rm eq}(\theta)\cos\theta
\end{equation}
and
\begin{equation}
   \sum_{i< j}^L\log\left[4\sin^2\left(\frac{\theta_i-\theta_j}{2}\right)\right]\rightarrow \frac{L^2}{2}\fint_{-\pi}^\pi\frac{\d\theta}{2\pi}\fint_{-\pi}^\pi\frac{\d\gamma}{2\pi}\,\rho_{\rm eq}(\theta)\rho_{\rm eq}(\gamma)\log\left[4\sin^2\left(\frac{\theta-\gamma}{2}\right)\right]~,
\end{equation}
where the cuts in the integral instruct us to compute the principal value with the divergence at $\theta=\gamma$ removed.  In the large-$L$ limit, we will work with the following approximation: 
\begin{equation}\label{eq:approxngww}
   \mathcal{N}_{\rm GWW}\approx L!\int [D\rho_{\rm eq}(\theta)]\d\xi \,e^{L^2F_{\rm eq}[\rho_{\rm eq}]}~,
\end{equation}
where the $L!$ incorporates all the permutations of the $L$ eigenvalues and the measure $[D\rho_{\rm eq}(\theta)]$ instructs us to extremize over the equilibrium eigenvalue density $\rho_{\rm eq}(\theta)$, weighted according to the free energy functional:
\begin{align}\label{eq:freeenergy}
   F_{\rm eq}[\rho_{\rm eq}]\equiv&  ~g\int_{-\pi}^\pi\frac{\d \theta}{2\pi} \rho_{\rm eq}(\theta)\cos\theta+\frac{1}{2}\fint_{-\pi}^\pi\frac{\d\theta}{2\pi}\fint_{-\pi}^\pi\frac{\d\gamma}{2\pi}\,\rho_{\rm eq}(\theta)\rho_{\rm eq}(\gamma)\log\left[4\sin^2\left(\frac{\theta-\gamma}{2}\right)\right]\nonumber\\& +\xi\left(\int_{-\pi}^\pi\frac{\d \theta}{2\pi}\,\rho_{\rm eq}(\theta)-1\right)+\text{const.}
\end{align}
In the above expression, we have introduced a Lagrange multiplier $\xi$ to ensure that $\rho_{\rm eq}$ is appropriately normalized. Now, given the expression \eqref{eq:approxngww}, we would conclude that at large $L$
\begin{equation}\label{eq:gwwpartitionfunctionapprox} 
   \mathcal{N}_{\rm GWW}\approx L!\, e^{L^2 \langle F_{\rm eq}(g)\rangle_{\rm GWW}}~, 
\end{equation}
Thus, the additive constant in \eqref{eq:freeenergy} is there to ensure that $\langle F_{\rm eq}(g=0)\rangle_{\rm GWW}=0$, matching the finite $L$ case. 

We now proceed to evaluate \eqref{eq:freeenergy} via a saddle-point approximation. The saddle point equation is
\begin{equation}
   g\cos\theta+\int_{-\pi}^\pi\frac{\d\gamma}{2\pi}\,\rho_{\rm eq}(\gamma)\log\left[4\sin^2\left(\frac{\theta-\gamma}{2}\right)+4\epsilon^2\right]+\xi=0~,
\end{equation}
where we have introduced a parameter $\epsilon^2$ to regulate the integral, which we will take to zero at the very end. Instead of working with the above expression, it will be more convenient to take a derivative with respect to $\theta$, which gives: 
\begin{equation}
   g\sin\theta=\lim_{\epsilon\rightarrow0}\int_{-\pi}^\pi \frac{\d\gamma}{2\pi}\,\rho_{\rm eq}(\gamma)\frac{\cos\left(\frac{\theta-\gamma}{2}\right)\sin\left(\frac{\theta-\gamma}{2}\right)}{\epsilon^2+\sin^2\left(\frac{\theta-\gamma}{2}\right)}
\end{equation}
For $g<1$ the unique solution to the above equation is \cite{Gross:1980he,Wadia:2012fr}: 
\begin{equation}\label{eq:smallgdensity}
   \left\langle\rho_{\rm eq}(\theta)\right\rangle_{\rm GWW}=1+g\cos\theta~,  \qquad\qquad g<1~,
\end{equation}
as is easily checked. Notably, for $g=0$, the density is uniform. 

This solution stops being a reasonable density for $g>1$. This is because, for $g>1$ the density of states \eqref{eq:smallgdensity} becomes negative beyond some value of $\theta$, which is clearly not allowed. To remedy this, we must allow for densities with compact support $\theta\in[-\theta_c,\theta_c]$. The edges of the distribution are to be determined dynamically. The analysis can be found in \cite{Gross:1980he,Wadia:2012fr}, and the answer is: 
\begin{equation}\label{eq:largegedensity}
   \left\langle\rho_{\rm eq}(\theta)\right\rangle_{\rm GWW}=2 g\cos\left(\frac{\theta}{2}\right)\sqrt{\frac{1}{g}-\sin^2\left(\frac{\theta}{2}\right)}~,  \qquad\qquad g>1~,
\end{equation}
and $\theta_c= 2\sin^{-1}\left(g^{-1/2}\right)$~. We will discuss this more below, but for now, the initiated should already realize that the above solution, with its square-root cut and compact support of eigenvalues, is reminiscent of the GUE.  

To summarize, we have
\begin{equation}\label{eq:GWWdensity}
   \left\langle\rho_{\rm eq}(\theta)\right\rangle_{\rm GWW}=\begin{cases}1+g\cos\theta~, & g<1~,~~~ |\theta|\leq\pi\\
2 g\cos\left(\frac{\theta}{2}\right)\sqrt{\frac{1}{g}-\sin^2\left(\frac{\theta}{2}\right)}~,& g>1~,~~~ |\theta|\leq2\sin^{-1}\left(g^{-1/2}\right)
   \end{cases}~,
\end{equation}
while
\begin{equation}\label{eq:GWWfreeE}
   \langle F_{\rm eq}(g)\rangle_{\rm GWW} =\begin{cases}\frac{g^2}{4}~, & g<1~,\\
g-\frac{1}{2}\log g-\frac{3}{4}~,& g>1~.
   \end{cases}~.
\end{equation}
Now we are in a position to evaluate the free energy $F(a)$ in the Yang-Mills model, at least at large $L$, by direct integration. We must evaluate
\begin{equation}
   F(a)\approx\log L!+ \log\left[\frac{L^2}{2a}\int_0^\infty g\d g\,e^{-\frac{L^2g^2}{4a}+L^2\langle F_{\rm eq}(g)\rangle_{\rm GWW}}\right]~.
\end{equation}
Since we are taking $L$ large, the integral above can now be computed via saddle-point. 
\begin{figure}[t!]
\begin{center}
\includegraphics[width=12cm]{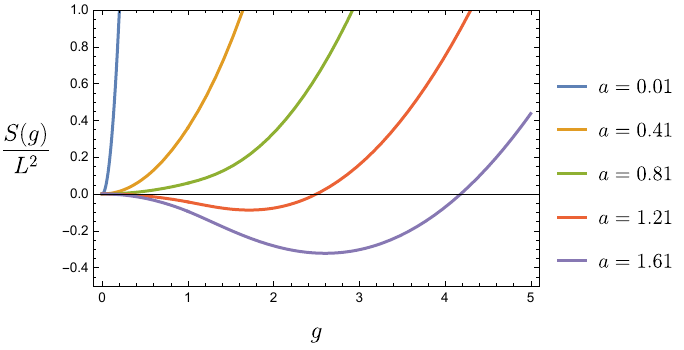}
\end{center}
\caption{Effective action over the coupling $g$ for the Yang-Mills model at various values of $a$}\label{fig:YMeffac}
\end{figure}
The analysis can be summarized as follows . For $a<1$ the effective action 
\begin{equation}
S(g)=\frac{L^2g^2}{4a}-L^2\langle F_{\rm eq}(g)\rangle_{\rm GWW}
\end{equation}
has a single minimum at $g=0$. However, as we increase $a$, this saddle becomes unstable, and for $a>1$ a new stable saddle point arises at $g=g_*\equiv a+\sqrt{a(a-1)}$ (see \cref{fig:YMeffac}). Incorporating the one-loop determinants around these saddles, we find the promised behavior in \eqref{eq:SYMfreeE}:\footnote{For $a>1$ a more refined analysis produces a power series in $L$ corresponding to a genus expansion in the matrix model \cite{Liu:2004vy,Okuyama:2017pil,Goldschmidt:1979hq,Periwal:1990gf}, giving: 
\begin{equation}
F(a)=\log(L!)+\sum_{n=0}^\infty L^{2-2n}\mathcal{F}_n(a)~.
\end{equation}
The first correction in the Yang-Mills model can be found in \cite{Okuyama:2017pil} and reads
\begin{equation}
\mathcal{F}_1(a)\equiv\zeta'(1)+\frac{11}{12}\log L-\frac{1}{8}\log\left(1-\frac{1}{g_*}\right)-\frac{1}{2}\log\left(\frac{a}{\pi g_*^2}\left(1-\frac{a}{g_*^2}\right)\right)~.
\end{equation}
}
\begin{equation}\label{eq:SYMfreeEnumber2}
   F(a)\approx\log(L!)+\begin{cases}
   -\log(1-a) &a<1~,\\
   L^2 \mathcal{F}_{0}(a) &a>1,
   \end{cases}
\end{equation}
with 
\begin{equation}\label{eq:gstarderive}
   \mathcal{F}_0(a)\equiv-\frac{g_*^2}{4a}+g_*-\frac{1}{2}\log g_*-\frac{3}{4}~,\qquad\qquad g_*\equiv a+\sqrt{a(a-1)}~.
\end{equation}
The upshot of this analysis is two-fold. First we see that for $a<1$, the theory is governed by a GWW model with $g=0$. The density of eigenvalues is uniform as in \eqref{eq:GWWdensity} evaluated at $g=0$. A GWW model at $g=0$ is called \emph{Haar random}. Thus the original $U(1)$ symmetry is preserved in this phase. Past $a>1$, the story changes, and we learn that we can  compute YM-correlation functions simply by borrowing formulas from the GWW model, evaluated at the effective coupling $g_*=a+\sqrt{a(a-1)}>1$. The non-trivial GWW potential demonstrates that the $U(1)$ symmetry has been spontaneously broken in this regime.

\subsection{Exact observables in the Haar random model}\label{sec:Haarobservables}
We have established that the phases of the model are controlled by the GWW model on either side of the transition. For $a<1$ at large-$L$, we are dealing with a GWW model at $g=0$, in other words a Haar-random model. Since they are much simpler to compute than GWW expectation values, we will begin by presenting computations of the one- and two-point function in the Haar random ensemble in this section.

The simplicity boils down to the fact that, in this ensemble, we have exact expressions for the one and two-point densities \cite{Meh2004}: 
\begin{equation}
\langle\rho(\theta)\rangle_{\rm Haar}=L~, \qquad \langle\rho(\theta)\rho(\phi)\rangle_{\rm Haar}^c=-\left[\frac{\sin\left(L\frac{\theta-\phi}{2}\right)}{\sin\left(\frac{\theta-\phi}{2}\right)}\right]^2~.
\end{equation}
A simple computation gives
\begin{equation}
\left\langle\text{Tr} U^k\right\rangle_{\rm Haar}=\int_{-\pi}^{\pi}\frac{\d\theta}{2\pi}e^{i k\theta}\langle\rho(\theta)\rangle_{\rm Haar}=L\,\delta_{k0}~,\label{eq:onepointhaar}
\end{equation}
revealing a very sharp example of the sum rule described generally in section \ref{sec:1ptsumrule}, where this function starts its life at $L$ and decays immediately to zero. 
For the two-point function we find:  
\begin{align}
\text{S}_{\rm Haar}^c(k)&=\left\langle\text{Tr} U^k\text{Tr}U^{-k}\right\rangle^c_{\rm Haar}=\int_{-\pi}^{\pi}\frac{\d\phi}{2\pi}\int_{-\pi}^{\pi}\frac{\d\theta}{2\pi}e^{i k(\theta-\phi)}\langle\rho(\theta)\rho(\phi)\rangle^c_{\rm Haar}\nonumber\\
&=\begin{cases}-L+k & k\leq L\\0 &k>L\end{cases}~,\label{eq:twopointhaar}
\end{align}
revealing an exact linear ramp. This again obeys the sum rule described in section \ref{sec:2ptsumrule}, where this two-point function goes to zero in exactly $L$ steps. One important observation of this paper is that, the \emph{shape} of this ramp will change in the symmetry broken phase. Precisely, the ramp in the GUE phase is \emph{not} linear (see \eqref{eq:nonlinearamp}). Moreover, note that  in the confined phase we observe a ramp of length $L$. This changes drastically both in the GUE and the symmetry broken phase of this model, which agree with each other, as we will show.

Also notice that there is no exponential behavior for $ k \geq L$---the ramp simply ends. This is a consequence of the preserved symmetry in this phase. The effective potential is exactly flat and once different eigenvalues decorrelate, the spectral form factor is exactly zero. In other words there is no $\Lambda_{IR}$ scale that enters the problem, so the two-point function must vanish exactly once the symmetry is restored.

\subsection{Exact observables in the GWW model}\label{sec:gwwobservables}

 As we tune past $a>1$, YM expectation values can be computed from observables in the GWW model evaluated at effective coupling $g=g_*$ as we derived in \eqref{eq:gstarderive}. Thus it will be important of us to describe how to compute exact expectation values in the GWW model for arbitrary $g$.

We will adapt a calculation in \cite{Okuyama:2017pil} to compute the exact free energy, as well as the one and two-point functions in the GWW model. This will differ from other approaches that rely on orthogonal polynomials. However, we provide a review of the orthogonal polynomials for the GWW model in \cref{ap:GWWorthogonalpolynomials}.  

Let us begin by considering the following integral 
\begin{equation}
   A^{f,g}\equiv\int [\d U]\,\text{det}[f(U)]\,\text{det}[g(U)]\exp\left\lbrace\frac{L g}{2}\text{Tr}\left(U+U^\dagger\right)\right\rbrace~, 
\end{equation}
where, for now, $f$ and $g$ are arbitrary scalar functions. To start, we rewrite the above expression as an integral over the eigenvalues of the unitary matrix $U$:
\begin{equation}
   A^{f,g}=\int_{-\pi}^\pi\left|\Delta\left(e^{i\theta}\right)\right|^2 \prod_{j=1}^L \frac{\d\theta_j}{2\pi}\,  e^{ L g \cos \theta_j}f\left(e^{i\theta_j}\right)g\left(e^{i\theta_j}\right)~. 
\end{equation}
Now, recall that the Vandermonde can be expressed as follows: 
\begin{equation}
   \Delta\left(e^{i\theta}\right)=\text{det}\left[e^{i(j-1)\theta_k}\right]=\sum_{\sigma \in S_L} (-1)^\sigma\prod_{k=1}^L e^{i(\sigma_k-1)\theta_k}=\sum_{\sigma \in S_L} (-1)^\sigma\prod_{k=1}^L e^{i(k-1)\theta_{\sigma_k}}~,
\end{equation}
where $S_L$ is the permutation group of $L$ elements.
Since the Vandermonde determinant is squared, plugging in the above expression leads to two separate sums over $S_L$. However, we can exploit the fact that the integrand is permutation symmetric to collapse the double sum into a single sum, at the expense of a factor of $L!$. Thus we are tasked with evaluating:
\begin{equation}\label{eq:permutationexpAfg}
   A^{f,g}=L!\sum_{\sigma \in S_L} (-1)^\sigma\prod_{j=1}^L\int_{-\pi}^\pi  \frac{\d\theta_j}{2\pi}\,e^{i\left(\sigma_j-j\right)\theta_j}  e^{ L g \cos \theta_j}f\left(e^{i\theta_j}\right)g\left(e^{i\theta_j}\right)~.  
\end{equation}
If we denote by $A_k^{f,g}$ the following $k$-th moment: 
\begin{equation}
   A_k^{f,g}\equiv\int_{-\pi}^\pi  \frac{\d\theta}{2\pi}\,e^{ik\theta}  e^{ L g \cos \theta}f\left(e^{i\theta}\right)g\left(e^{i\theta}\right)~, 
\end{equation}
then we see that the expression \eqref{eq:permutationexpAfg} can be cast as a determinant over a matrix whose entries are the above moments: 
\begin{equation}
   A^{f,g}=L!\,\text{det} \left[A_{i-j}^{f,g}\right]_{i,j=1,\dots,L}~. 
\end{equation}
This observation will allow us to derive exact expressions for the free-energy, as well as the one- and two-point functions of Polyakov loop operators in the GWW model.

To do so, let us now make the following choice for the scalar functions $f$ and $g$:
\begin{equation}\label{eq:fgchoice}
   f(U)=\mathds{1}+x\, U^n~, \qquad\qquad g(U)=\mathds{1}+y\, U^m~.
\end{equation}
This will be helpful because:  
\begin{equation}
   \text{det}[f(U)]=1+x\, \text{Tr}\, U^n+\mathcal{O}\left(x^2\right)\dots
\end{equation}
so we will be able to extract the correlation functions of interest by taking derivatives of this function with respect to $x$ and $y$ and evaluating it at $x=y=0$.

As a final step, let us introduce the following integral representation of the modified Bessel function:
\begin{equation}\label{eq:besselintegral}
   I_n(Lg)=\int_{-\pi}^\pi \frac{\d \theta}{2\pi}e^{L g\cos\theta}e^{in\theta}~.
\end{equation}
 Using \eqref{eq:besselintegral}, we find that, for the choice of $f$ and $g$ given in \eqref{eq:fgchoice}:\footnote{To avoid confusion, for $f$ and $g$ as in \eqref{eq:fgchoice}, we denote $A^{f,g}=A(x,y)$~.} 
\begin{equation}
   A({x,y})=L!\,\text{det}\left[I_{i-j}(Lg)+x\,I_{n+i-j}(Lg)+y\,I_{m+i-j}(Lg)+xy\,I_{n+m+i-j}(Lg)\right]_{i,j=1,\dots,L}~. 
\end{equation}
Our next step is to define matrices $M_k$ such that their components are given by:
\begin{equation}
   (M_k)_{ij}=I_{k+i-j}(Lg)~, \qquad i,j=1,\dots,L~.
\end{equation}
Using this definition, we can rewrite: 
\begin{equation}
   A({x,y})=L!\,\text{det}\left[M_0+x\,M_n+y\,M_m+xy\,M_{n+m}\right]~. 
\end{equation}
Now we are ready to write down exact expressions for zero-, one-, and two-point functions. We can compute the normalization (a.k.a the partition function) of the GWW matrix integral by evaluating the above expression at $x=y=0$:
\begin{equation}\label{eq:ngww}
\boxed{   \mathcal{N}_{\rm GWW}=A(0,0)=L!\,\text{det}[M_0]\,~.}
\end{equation}
Moreover, for the one-point function, we find
\begin{equation}\label{eq:onepointGWWexact}
  \boxed{ \left\langle \text{Tr}\, U^n\right\rangle_{\rm GWW}=\frac{1}{\mathcal{N}_{\rm GWW}}\left.\partial_x A({x,y})\right\rvert_{x,y=0}=\text{Tr}\left(M_0^{-1}M_n\right)~. }
\end{equation}
So far we have reproduced the results of \cite{Okuyama:2017pil}.
We may go one step further and compute the two-point function, where we find: 
\begin{multline}\label{eq:twopointGWWexact}
   \left\langle \text{Tr}\, U^n\text{Tr}\, U^m\right\rangle_{\rm GWW}=\frac{1}{\mathcal{N}_{\rm GWW}}\left.\partial_x\partial_y A({x,y})\right\rvert_{x,y=0}\\=\text{Tr}\left(M_0^{-1}M_{n+m}\right)+\text{Tr}\left(M_0^{-1}M_n\right)\,\text{Tr}\left(M_0^{-1}M_m\right)-\text{Tr}\left(M_0^{-1}M_nM_0^{-1}M_m\right)~. 
\end{multline}
We can group terms in the above expression as follows: 
\begin{equation}\label{eq:GWW2pt}
   \left\langle \text{Tr}\, U^n\text{Tr}\, U^m\right\rangle_{\rm GWW}\equiv \left\langle \text{Tr}\, U^{n+m}\right\rangle_{\rm GWW}+ \left\langle \text{Tr}\, U^n\right\rangle_{\rm GWW} \left\langle \text{Tr}\, U^m\right\rangle_{\rm GWW}+\left\langle \text{Tr}\, U^n\text{Tr}\, U^m\right\rangle_{\rm GWW}^c
\end{equation}
which should be compared with \eqref{eq:2ptgeneral2} and where we have labeled the `fully-connected' part of the two-point function as:
\begin{equation}\label{eq:fullyconnectedGWW}
\boxed{\left\langle \text{Tr}\, U^n\text{Tr}\, U^m\right\rangle_{\rm GWW}^c\equiv-\text{Tr}\left(M_0^{-1}M_nM_0^{-1}M_m\right)~,}
\end{equation}
 which we call the SFF of the GWW model below.

Parsing these formulas is a difficult task, especially since they involve traces of products of large-$L$ matrices whose entries are Bessel functions. So for the sake of clarity, let us provide the equilibrium answers as well, at least for the one-point function \cite{Bars:1979xb,Rossi:1996hs}:
\begin{align}
\langle \text{Tr}\, U_{\rm eq}\rangle_{\rm GWW}&\equiv\lim_{L\rightarrow\infty}\frac{1}{L}\left\langle\text{Tr}\, U\right\rangle_{\rm GWW}~,\nonumber\\
&=\int_{-\pi}^\pi\frac{\d\theta}{2\pi}e^{i\theta}\left\langle\rho_{\rm eq}(\theta)\right\rangle_{\rm GWW}=\begin{cases}
   \frac{g}{2}~, &g<1\\
   \left(1-\frac{1}{2g}\right)~, & g>1
    \end{cases}   
\end{align} 
and for $|n|\geq2$ 
\begin{align}
  \langle \text{Tr}\, U^n_{\rm eq}\rangle_{\rm GWW}&\equiv \lim_{L\rightarrow\infty}\frac{1}{L}\left\langle\text{Tr}\, U^n\right\rangle_{\rm GWW}~,\nonumber\\
  &=\int_{-\pi}^\pi\frac{\d\theta}{2\pi}e^{in\theta}\left\langle\rho_{\rm eq}(\theta)\right\rangle_{\rm GWW}=\begin{cases}
   0~, &g<1\\
  \frac{1}{|n|-1}\left(1-\frac{1}{g}\right)^2 P_{|n|-2}^{(1,2)}\left(1-\frac{2}{g}\right)~, & g>1
    \end{cases}   
\end{align}
where $P_n^{(\alpha,\beta)}(x)$ is a Jacobi polynomial.  The symmetry under $n\rightarrow -n$ comes from the fact that $\rho_{\rm eq}(\theta)=\rho_{\rm eq}(-\theta)$.  Note that $\langle \text{Tr}\, U_{\rm eq}\rangle_{\rm GWW}$ is nonzero unless we take the strict $g=0$ limit. This is reasonable, since for $g\neq0$ the GWW model breaks the underlying $U(1)$ symmetry of the Haar random ensemble. 

\subsection{Observables in the Yang-Mills matrix model} \label{sec:observables_YM}
Now that we have derived exact expressions for the observables of interest in the GWW model, we can convert them to observables in the Yang-Mills model using \eqref{eq:YMexpectationvalues}. We will analyze the free energy, one- and two-point functions each in turn below. 

\begin{figure}[t!]
\begin{center}
\includegraphics[height=6.1cm]{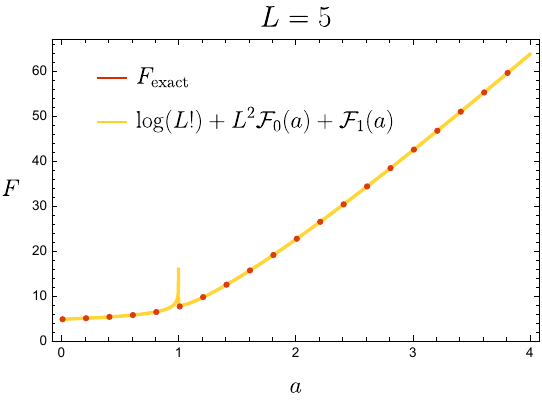}
\includegraphics[height=6.1cm]{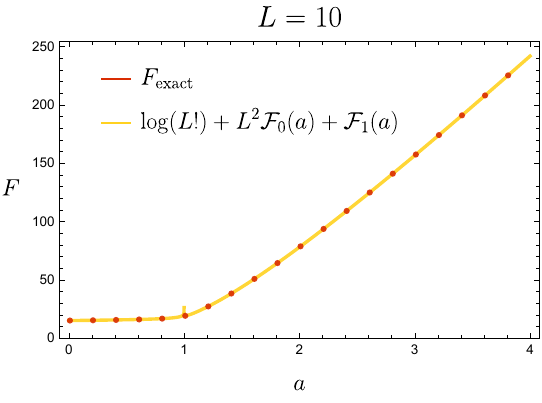}
\end{center}
\caption{Comparison between the numerically-evaluated exact (\emph{red}) and large-$L$ (\emph{yellow}) (including the first non-planar order) free energies in the Yang-Mills matrix model as a function of $a$. The spike in the large-$L$ free energy at $a=1$ is the tell-tale signal of a first-order phase transition. As expected, the first-order phase transition in the large-$L$ expression gets smoothed out at finite-$L$.}\label{fig:freeEcompare}
\end{figure}
\subsubsection{Free energy}
We start with the expression for the free energy $F(a)=\log \mathcal{Z}(a)$ with $\mathcal{Z}(a)$ given in \eqref{eq:ZaYM} and where we have derived the expression for $\mathcal{N}_{\rm GWW}$ in \eqref{eq:ngww}. Given the complicated form of $\mathcal{N}_{\rm GWW}$, an explicit and exact finite-$L$ expression for $F(a)$, would be hopeless. Luckily it is not crucial for our needs. We already discussed in \cref{sec:freeederivation} that as $L\rightarrow\infty$ the expected behavior of the free energy tends to \eqref{eq:SYMfreeEnumber2}. We can compare this expectation with the numerically-computed finite-$L$ result. Our results are displayed  in figure \ref{fig:freeEcompare}.

\subsubsection{One-point function in the Yang-Mills matrix model }

\begin{figure}[t]
\begin{center}
\includegraphics[height=5cm]{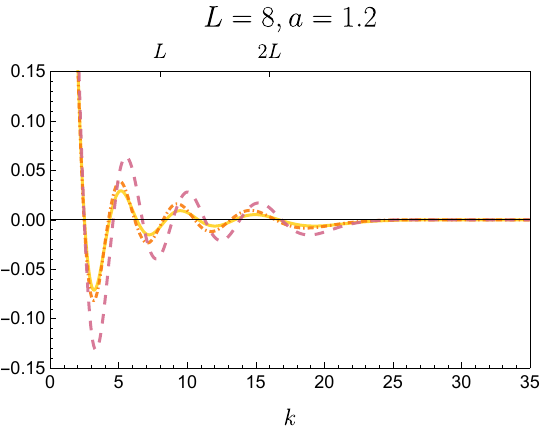}\qquad\includegraphics[height=5cm]{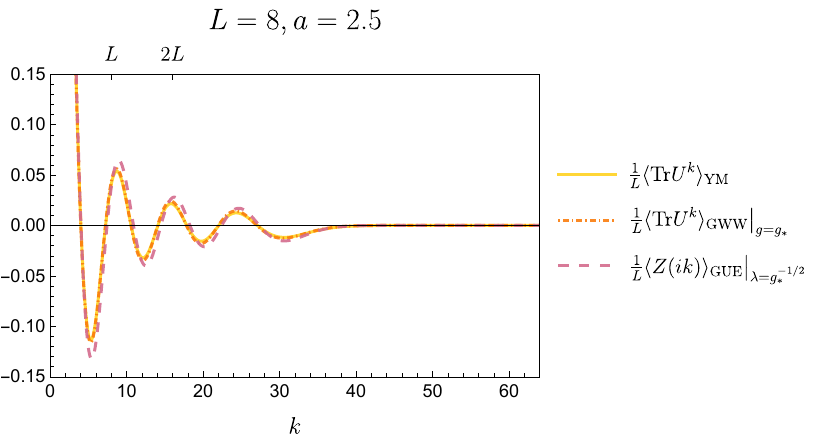}
\end{center}
\caption{We compare the numerically-evaluated one-point function in the Yang-Mills matrix model $\tfrac{1}{L}\langle\text{Tr}\,U^k\rangle_{\rm YM}$ with the equivalent observable in the GUE and the GWW ensembles in the deconfined phase ($a>1$).  For $a$ slightly larger than 1 (at least for small values of $L$),  YM and GWW expectation values do not match with the GUE. However, for $a\gg1$, we observe a nice agreement between these functions in the oscillatory phase. These plots are made using discrete data in {\tt Mathematica}'s {\tt ListPlot} command with options \emph{Joined$\rightarrow$True, InterpolationOrder$\rightarrow$3}.}  \label{fig:oneptYM}
\end{figure}

We are now ready to discuss our first non-trivial observable in the Yang-Mills matrix model, the one-point function: $\frac{1}{L}\left\langle \text{Tr}\, U^k\right\rangle_{\rm YM}$. Before showing any plots, let us remind the reader of our expectations. For $a<1$, the theory is confined and the $U(1)$ symmetry \eqref{eq:centersymmetry} is unbroken, hence the one-point function is zero on the nose for all $k>0$, matching the Haar-random calculation \eqref{eq:onepointhaar}:
\begin{equation}
\frac{1}{L}\left\langle \text{Tr}\, U^k\right\rangle_{\rm YM}=0~~\forall k>0\qquad \text{if } a<1~. 
\end{equation}

When we crank $a$ above the transition, the theory deconfines and the $U(1)$ symmetry is spontaneously broken. We now expect to find GUE universality. Let us remind the reader of what was discussed in the introduction regarding the GUE. The one-point function of interest in the GUE can be expressed as follows: 
\begin{equation}
   \frac{\langle Z(it)\rangle_{\rm GUE}}{L}=\int_{-\infty}^\infty\d E\, \frac{\langle\rho(E)\rangle_{\rm GUE}}{L}\, e^{-i E t}~.
\end{equation}
with the exact expression for $\langle\rho(E)\rangle_{\rm GUE}$ given in \eqref{eq:GUEdensity}. This can be thought of as the expectation value of a partition function at complex temperature $\beta=it$ (normalized by the dimension of the Hilbert space). We provided an example plot of this function for $L=30$ in figure \ref{fig:symbreaking1pt}. 

At large-$L$ and for $t\lambda\ll L$, we may use the famous result that the equilibrium density approaches the Wigner semi-circle:
\begin{equation}\label{eq:GUEdensityequilibrium}
\langle\rho_{\rm eq}(E)\rangle_{\rm GUE}\equiv\lim_{L\rightarrow\infty}\frac{\langle\rho(E)\rangle_{\rm GUE}}{L} =\frac{1}{\pi\lambda}\sqrt{1-\left(\frac{E}{2\lambda}\right)^2}~,
\end{equation}
which we review around equation \eqref{eq:rhoinfinitydef}. Thus, at least at early times: 
\begin{align}
   \frac{\langle Z(it)\rangle_{\rm GUE}}{L}&\approx\int_{-2\lambda}^{2\lambda}\d E \,\frac{e^{-i E t}}{\pi\lambda}\sqrt{1-\left(\frac{E}{2\lambda}\right)^2}=\frac{J_1(2t\lambda)}{t \lambda}~,\nonumber\\
   &\underset{t\lambda\gg1}{\approx}\frac{\pi\,\sin\left(2 t\lambda -\frac{\pi}{4}\right)}{(\pi t\lambda)^{3/2}}~.\label{eq:t32decay}
\end{align}
This analysis is what gives the oscillatory behavior with the $(t\lambda)^{-3/2}$ envelope which we showed in figure \ref{fig:symbreaking1pt}.

\paragraph{Early-time behavior in the YM model:}
\begin{figure}[t!]
\begin{center}
\begin{minipage}{0.52\paperwidth}\begin{tabular}{c c}
\includegraphics[width=0.245\paperwidth]{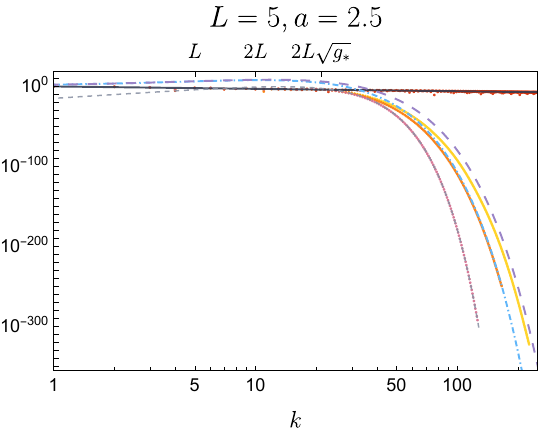}
&\includegraphics[width=0.245\paperwidth]{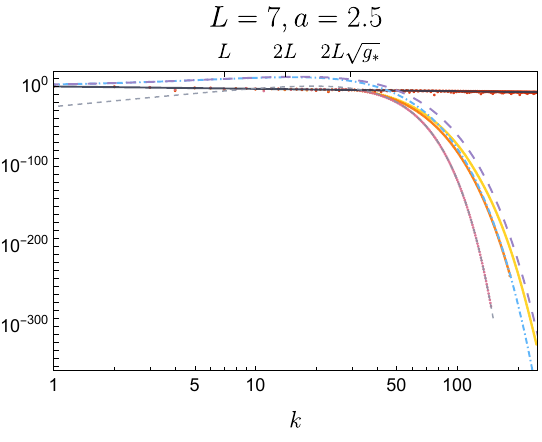}\\
\includegraphics[width=0.245\paperwidth]{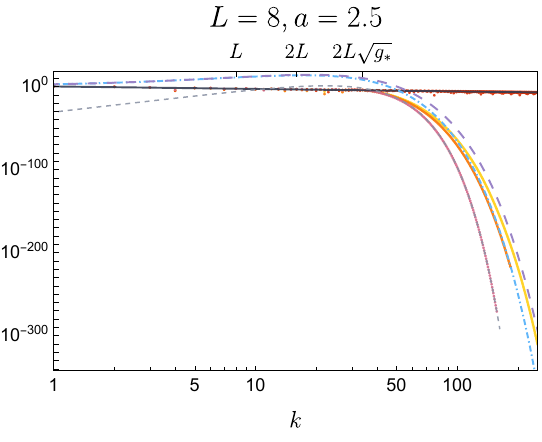}
&\includegraphics[width=0.245\paperwidth]{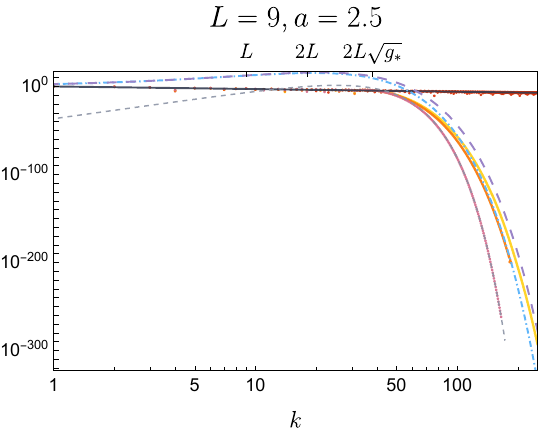}
\end{tabular}\end{minipage}
\begin{minipage}{0.245\paperwidth}\includegraphics[width=0.24\paperwidth]{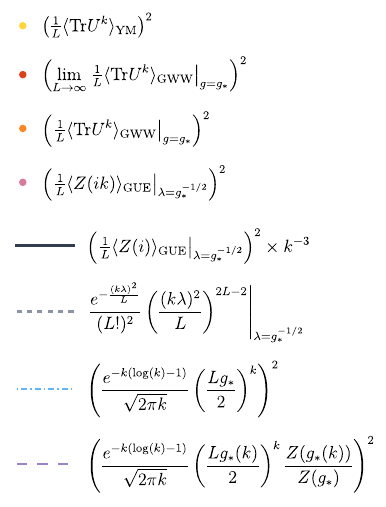}\end{minipage}
\end{center}
\caption{Log-log plot of the numerically-computed exact one-point function $\tfrac{1}{L}\left\langle\text{Tr}\, U^k\right\rangle_{\rm YM}$ (squared) (\emph{yellow}) compared with the infinite-$L$ result (\emph{red}) for various values of $L$. For reference, we also plot $\left.\tfrac{1}{L}\left\langle\text{Tr}\, U^k\right\rangle_{\rm GWW}\right\rvert_{g=g_*}$ (\emph{orange}) and $\left.\tfrac{1}{L}\langle Z(ik)\rangle_{\rm GUE}\right\rvert_{\lambda=g_*^{-1/2}}$ (\emph{pink}). Notice that for $k<L$ all of these functions behave as expected: oscillating around a $k^{-3/2}$ polynomial decay (\emph{black line}). For $k>2L$ we note that the functional behavior changes drastically in favor of characteristic exponential decay towards the symmetry preserving phase. Here we notice a stark difference between the GUE late-time behavior, which decays like $e^{-\frac{(k\lambda)^2}{2L}}$ (\emph{dashed gray}), and the Yang-Mills behavior, which goes like $e^{-k\log \left(\frac{2k}{L g_*}\right)}$ (\emph{dot-dashed blue}). We have also verified that $\tfrac{1}{L}\left\langle\text{Tr}\, U^k\right\rangle_{\rm YM}$ and $\left.\tfrac{1}{L}\left\langle\text{Tr}\, U^k\right\rangle_{\rm GWW}\right\rvert_{g=g_*}$ (\emph{yellow} and \emph{orange} curves) approach each other as we increase $L$, even in the exponentially decaying part of the plot. 
} \label{fig:loglogoneptYM}
\end{figure}

 Let us now demonstrate how to derive a similar early-time behavior in the Yang-Mills matrix model at large-$L$ and in the symmetry broken phase when $a>1$. We begin by making a similar approximation as in the GUE, assuming the exact density of states can be approximated by its large-$L$ equilibrium value found in the previous section (see figure \ref{fig:densitycompare} for numerical proof of this equivalence): 
\begin{align}
  \frac{1}{L}\left\langle\text{Tr}\, U^k\right\rangle_{\rm YM}&\approx\int_{-\theta_c}^{\theta_c}\frac{\d\theta}{2\pi}\left.\left\langle\rho_{\rm eq}(\theta)\right\rangle_{\rm GWW}\right\rvert_{g=g_*}\,e^{ik\theta}~,\nonumber\\
&=\int_{-\theta_c}^{\theta_c}\frac{\d\theta}{2\pi}2 g_*\cos\left(\frac{\theta}{2}\right)\sqrt{\frac{1}{g_*}-\sin^2\left(\frac{\theta}{2}\right)}\,e^{ik\theta}~,\nonumber\\
&=\int_{-2\lambda}^{2\lambda}\d E\frac{e^{2i k\sin^{-1}\left(\frac{E}{2\lambda\sqrt{g_*}}\right)}}{\pi\lambda}\sqrt{1-\left(\frac{E}{2\lambda}\right)^2}~.
\end{align}
In the last line we performed a change of variables $\theta= 2\sin^{-1}\left(\frac{E}{2\lambda\sqrt{g_*}}\right)$ and we have introduced the width $\lambda$ by hand. Now, note that for $g_*\gg 1$ we can approximate $\theta\approx \frac{E}{\lambda\sqrt{g_*}}$. This allows us to write
\begin{equation}\label{eq:YMGUEapprox}
\frac{1}{L}\left\langle\text{Tr}\, U^k\right\rangle_{\rm YM}\approx\int_{-2\lambda}^{2\lambda}\d E\frac{e^{i E t_k}}{\pi\lambda}\sqrt{1-\left(\frac{E}{2\lambda}\right)^2}
\end{equation}
where we have identified an emergent continuous-time variable $t_k\equiv \frac{k}{\lambda\sqrt{g*}}$~. The emergence of this continuous-time variable is intricately linked with the breaking of the original $U(1)$-symmetry, since the compactness of the original `energy' variable $\theta$ is washed out in the symmetry-broken phase. Thus, we have derived the approximate equivalence
\begin{equation}
\frac{\langle Z(it_k)\rangle_{\rm GUE}}{L}\approx\frac{1}{L}\left\langle\text{Tr}\, U^k\right\rangle_{\rm YM}
\end{equation}
for $k< L$ and $g_*=a+\sqrt{a(a-1)}\gg1$ and we remind the reader that $t_k\lambda\equiv \frac{k}{\sqrt{g_*}}$~. The large $a$ limit corresponds to the effective theory being weakly coupled and, therefore, matching the GUE predictions. For $k<L$ we expect to see the characteristic $k^{-3/2}$ decay (with oscillations on top) of the GUE one-point function as demonstrated in \eqref{eq:t32decay}. We demonstrate this approximate equality explicitly in figures \ref{fig:oneptYM} and \ref{fig:loglogoneptYM}.

Let us, for a moment, step back and make a particular observation about the GUE. The GUE is an ensemble of random Hermitian Hamiltonians where the distribution favors states with energies near $E=0$. Hence this ensemble has a preferred energy and any potential shift symmetry $E\rightarrow E+a$ is explicitly broken by the ensemble. We would expect such a shift symmetry only in an ensemble where the distribution over Hamiltonians is flat. What we have done here is show that behavior of the GUE, with its \emph{explicit} symmetry breaking, can be recovered in a model where the $E\rightarrow E+a$ shift symmetry is obeyed by the ensemble but \emph{spontaneously} broken. The operative difference is that we can now think of the GUE as a spontanous symmetry broken phase of an altogether different ensemble of matrix models. We will explore the consequences of this for the spectral form factor, but let us first touch upon symmetry restoration as diagnosed by the late-time behavior of the one-point function $\langle\text{Tr}\, U^k\rangle_{\rm YM}$ as $k$ approaches the volume scale $L$.

\paragraph{Late-time behavior:} To get a crude estimate of the late-time behavior of the one-point function in the Yang-Mills model, and to illustrate our logic, let us remind the reader how we estimated this behavior in the GUE in section \ref{sec:1ptsumrule}. 
The important observation is that for $t$ large with respect to some characteristic scale set by the potential $V(E)$, we may approximate the density of states as being dominated by the exponential part of these wavefunctions. 
For the GUE, this analysis gives:\footnote{For a more refined anayslis, capable of extracting the subleading time dependence at late times, see \cref{ap:saddleanalysisonepoint}. } 
\begin{equation}
\langle Z(it)\rangle_{\rm GUE} \underset{t\to \infty}{\approx}\int_{-\infty}^{\infty}\d E\, e^{-iEt}e^{-\frac{L}{2\lambda^2}E^2}\propto e^{-\frac{(t\lambda)^2}{2L}}~,
\end{equation}
which agrees with the late-time behavior of the GUE one-point function as can be seen from the gray dashed curves in figures \ref{fig:symbreaking1pt} and \ref{fig:loglogoneptYM}. 

In the GWW matrix model, the analysis follows similarly to the GUE. Here we have: 
\begin{align}
\left\langle\text{Tr}\, U^k\right\rangle_{\rm GWW}&\underset{k\sim L}{\approx}\int_{-\pi}^\pi\frac{\d\theta}{2\pi}e^{ik\theta}e^{-V(\theta)}\nonumber\\
&=\int_{-\pi}^\pi\frac{\d\theta}{2\pi}e^{ik\theta}e^{L g \cos\theta}=I_k(Lg)~.
\end{align}
For large order $k\sim L\gg 1$ we can use the asymptotic expression for the modified Bessel function \cite{nistDLMFxA71041}: 
\begin{equation}
\left\langle\text{Tr}\, U^k\right\rangle_{\rm GWW}\propto I_k(Lg)\underset{k\sim L\gg1}{\approx}\frac{e^{-k(\log k-1)}}{\sqrt{2\pi k}}\left(\frac{L g}{2}\right)^k~.
\end{equation}
Having derived this result, let us recall how to convert this to an expectation value in the Yang-Mills matrix model using equation \eqref{eq:YMexpectationvalues}: 
\begin{equation}
   \left\langle\text{Tr}\, U^k\right\rangle_{\rm YM}\underset{k\sim L\gg1}{\propto} \frac{e^{-k(\log k-1)}}{\sqrt{2\pi k}}\frac{L!\times \frac{L^2}{2a}\int_0^\infty g\d g\,e^{-\frac{L^2g^2}{4a}+L^2\langle F_{\rm eq}(g)\rangle_{\rm GWW}}
   \left(\frac{L g}{2}\right)^k}{Z(a)}~,\label{eq:YMasymptotics}
\end{equation}
where $\langle F_{\rm eq}(g)\rangle_{\rm GWW}$ can be found in \eqref{eq:GWWfreeE}. We can now use this to compute the asymptotics behavior of the YM model in the deconfined phase. To do this, we evaluate the integral in \eqref{eq:YMasymptotics} by saddle point. A naive calculation yields:
\begin{equation}
 \left\langle\text{Tr}\, U^k\right\rangle_{\rm YM}\underset{L^2\gg k\sim L\gg1}{\propto}\frac{e^{-k(\log k-1)}}{\sqrt{2\pi k}}\left(\frac{L g_*}{2}\right)^k~,
\end{equation}
with $g_*$ given in \eqref{eq:SYMfreeEnumber2}.\footnote{
The  complexity of numerically evaluating the expectation values in the YM model scale polynomially in $L$ and are generally onerous. We were only able to perform the numerical integrals up to $L=9$ within a reasonable time-frame. Hence our numerical plots probe regions where $k\sim L^2$. To get a better estimate for the exponential behavior in these regions we needed to account for the fact that, for small enough $L$, the insertion of the operators shift the saddle. We are able to calculate the shifted saddle point systematically, yielding: 
\begin{equation}
 \left\langle\text{Tr}\, U^k\right\rangle_{\rm YM}\underset{k\sim L^2\gg1}{\propto}\frac{e^{-k(\log k-1)}}{\sqrt{2\pi k}}\left(\frac{L g_*(k)}{2}\right)^k \frac{Z(g_*(k))}{Z(g_*)}~,
\end{equation}
where $g_*(k)$ is the shifted saddle-point value of $g$ (with $g_*=g_*(k=0)$):
\begin{equation}
g_*(k)\equiv a+\sqrt{a\left(a-1+\frac{2k}{L^2}\right)}~.
\end{equation}}
In figure \ref{fig:loglogoneptYM} we observe that this function upper-bounds the behavior of the late-time Yang-Mills one-point function.

\paragraph{Discussion:} In figure \ref{fig:oneptYM}, we have shown that the early-time behavior in the GUE and YM matrix models agree, displaying the tell-tale $t^{-3/2}$ behavior that is universal to random matrix theories and 2d gravity alike. We have understood this behavior to arise in the Yang-Mills model due to a particular breaking of the underlying $U(1)$ symmetry present in the model.\footnote{It is certainly possible to alter the $t^{-3/2}$ behavior in the matrix model. The universal nature this exponent stems from the infinite-$L$ square-root edge of the density of states in $\langle \rho_{\rm eq}\rangle$ (see equations \eqref{eq:GUEdensityequilibrium} and \eqref{eq:equildensityGWW}). Achieving a different type of cut or edge of the distribution can be achived by fine-tuning the potential, as reviewed, for example, in \cite{Anninos:2020ccj} and discussed for unitary matrix models in \cite{Alvarez-Gaume:2005dvb}.} In figure \ref{fig:loglogoneptYM} we showed that, once we reach volume scale $L$, the exponential approach to symmetry restoration distinguishes between the various possible ensembles, meaning the approach depends sensitively on the exact form of the potential. Starting in the preserved phase $a<1$, the expectation values are completely different, we find $\left\langle \text{Tr} U^k\right\rangle_{\rm YM}=0$ for all $k$.

Let us tie this observation to holography. We typically understand one-point functions, such as $\langle Z(it)\rangle_{\rm GUE}$ or $\left\langle \text{Tr}\, U^k\right\rangle_{\rm YM}$ as arising holographically from a low-dimensional gravity path integral with certain boundary conditions. What we would like to advocate for here is that \emph{the geometric description only makes sense in such a symmetry broken phase}. Of course, since expectation values vanish in the preserving phase, there can be no geometric picture. More to the point: the exit from the preserving phase, diagnosed by the exponential parts of the plots in figures \ref{fig:oneptYM} and \ref{fig:loglogoneptYM}, are sensitive to the UV completion of the model. Here we see how the GUE and YM results differ in important ways, while agreeing in the `geometric' part of the curve.

Now that we have this out of the way, we will continue and analyze the spectral form factor using the same logic. Here we will also uncover interesting results. 

\subsubsection{Spectral form factor in the Yang-Mills matrix model }
\begin{figure}[t]
\begin{center}
\includegraphics[height=6cm]{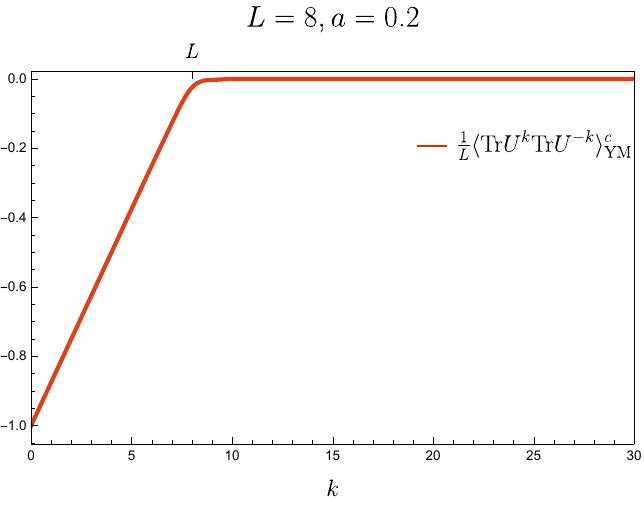}\includegraphics[height=6cm]{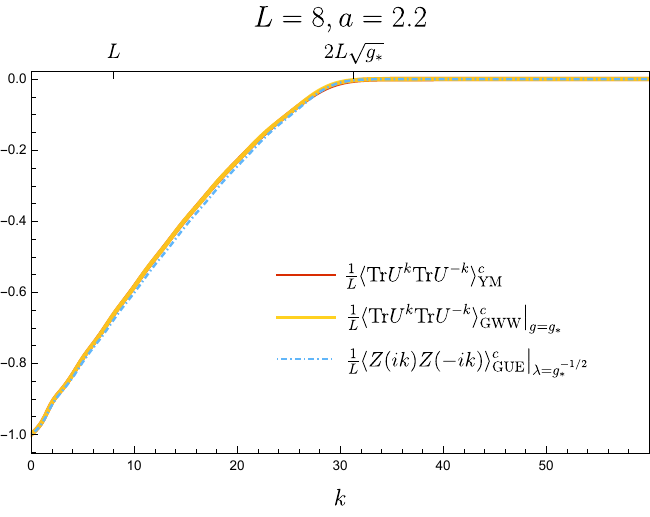}
\end{center}
\caption{\emph{Left}: We compute the numerically-evaluated SFF in the Yang-Mills matrix model $\tfrac{1}{L}\langle\text{Tr}\,U^k\text{Tr}\,U^{-k}\rangle^c_{\rm YM}$ in the confined phase where $a<1$. We see an exactly linear ramp that hits zero in $L$ steps, indicative of Haar-random universality. \emph{Right}: We compare the YM two-point function with the GWW answer at $g=g_*$ and the GUE answer at $\lambda=g_*^{-1/2}$ in the deconfined phase ($a>1$). These plots follow each other very closely. We have also verified that the ramp closely follows the behavior in \eqref{eq:largeLramp} and hits zero in a time of order $k\sim 2L\sqrt{g_*}$. These plots are made using discrete data in {\tt Mathematica}'s {\tt ListPlot} command with options \emph{Joined$\rightarrow$True, InterpolationOrder$\rightarrow$3}.}  \label{fig:twoptYM}
\end{figure}

We are now ready to apply the previous analysis to the SFF in the Yang-Mills matrix model. This is a more refined observable than the simple one-point function that we treated in the previous section, since the SFF is non-zero on either side of the confinement/deconfinement transition across $a=1$. By now we have everything we need to set expectations. In the symmetry preserving phase $(a<1)$, we are effectively dealing with a Haar random model.  In section \ref{sec:Haarobservables}, (equation \eqref{eq:twopointhaar}) we derived the following:
\begin{equation}
\text{S}_{\rm Haar}^c(k)=\left\langle\text{Tr} U^k\text{Tr}U^{-k}\right\rangle^c_{\rm Haar}=\begin{cases}-L+k & k\leq L\\0 &k>L\end{cases}~,
\end{equation}
that is an exactly linear ramp which obeys the sum rule derived in section \ref{sec:2ptsumrule}. 

\begin{figure}[t!]
\begin{center}
\includegraphics[width=0.5\paperwidth]{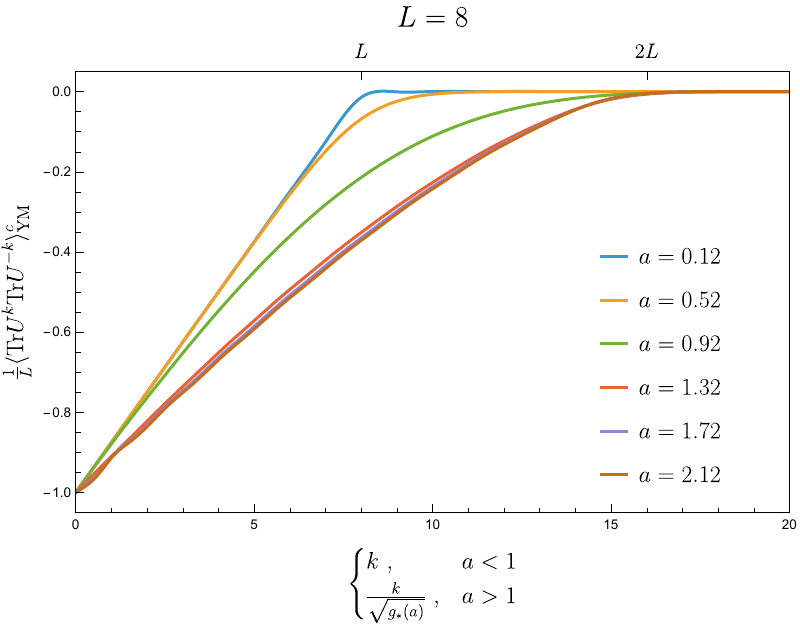}
\end{center}
\caption{Demonstration that the ramp lasts for $k\sim L$ in the symmetry preserving phase $a<1$, whereas it lasts for $t_k\lambda\sim 2L$ in the spontaneously broken phase $a>1$ in terms of the emergent continuous time variable $t_k\lambda=\frac{k}{\sqrt{g_*(a)}}$.  One straightforward diagnostic that distinguishes between the two families of curves is their slope near $k=0$: in the symmetry preserving phase ($a<1$), this slope (w.r.t. $k$) is $\frac{1}{L}$, whereas in the symmetry broken phase, this slope (w.r.t. $t_k\lambda$) is $\frac{2}{\pi L}$. While the green curve for $a=0.92$ (in the preserving phase) looks to be smoothly interpolating between the two sets of behaviors, this is a finite-$L$ effect that gets washed out if we scale $L$ larger. Note however that this curve has the tell-tale $\frac{1}{L}$ slope near $k=0$, clearly indicating what phase it should be associated with. }\label{fig:phasetransitionSFF}
\end{figure}

We would like to set our sights on how this exactly linear behavior changes when we cross into the symmetry broken phase $(a>1)$. Is the SFF able to distinguish whether it is in a preserving or broken phase? Recall from the discussion around the one-point function, that in the SSB phase, for all intents and purposes, we are now dealing with an effective GWW model at $g=g_*=a+\sqrt{a(a-1)}$. Moreover, we showed in the case of the one-point function that for $g_*\gg1$, the effective dynamics is reproduced by a GUE model where we identify the dimentionless time: $t_k\lambda =\frac{k}{\sqrt{g_*}}$. 

As is well-known, although not necessarily well-advertised, the SFF in the GUE is \emph{not} exactly linear. In the appendix \ref{ap:firstderivation}, we review that the large-$L$ limit of the SFF behaves as follows \cite{Okuyama:2018yep,Liu:2018hlr}: 
\begin{equation}\label{eq:largeLramp}
\text{S}_{\rm GUE,\,eq}^c(t)\equiv\lim_{L\rightarrow\infty}\frac{\text{S}_{\rm GUE}^c(t)}{L}=-1+\frac{1}{\pi}\left(\frac{t\lambda}{L}\sqrt{1-\left(\frac{t\lambda}{2L}\right)^2}+2\,\text{arcsin}\left(\frac{t\lambda}{2L}\right)\right)~,\qquad \frac{t\lambda}{L}\text{  fixed.}
\end{equation}
This is interesting for two reasons. First, we notice that, although the SFF obeys the sum rule taking it from $-L$ to zero, the shape of this curve has changed across the transition, and is no longer exactly linear. That is, the functional form goes from exactly linear in the symmetry preserving phase, to a function with a bit more structure in the SSB phase.  
Putting everything together, we propose the following for the YM SFF at large-$L$:
\begin{equation}
\frac{1}{L}\left\langle\text{Tr} U^k\text{Tr}U^{-k}\right\rangle^c_{\rm YM}\approx\begin{cases}\frac{1}{L}\text{S}_{\rm Haar}^c(k)~, & a<1\\
\text{S}_{\rm GUE,\,eq}^c(t_k)~, &a>1\end{cases}~,
\end{equation}
where we remind the reader that we have defined the discrete time $t_k\lambda\equiv\frac{k}{\sqrt{g_*}}$ around equation \eqref{eq:YMGUEapprox}. We demonstrate this sharp change in behavior across the phase transition point $a=1$ in the plots in figures \ref{fig:twoptYM} and \ref{fig:phasetransitionSFF}. 

Secondly note that the \emph{duration} of the ramp differs across the phases. In the symmetry preserving phase, the SFF  vanishes at $k\sim L$, while in the spontaneously broken phase this is delayed to $k \sim 2L \sqrt{g_*}$, which is also displayed in figure \ref{fig:twoptYM}. In terms of the emergent continuous time variable in the broken phase we would say that the ramp ends at $t_k \lambda \sim 2 L$. We demonstrate this stark transition in figure \ref{fig:phasetransitionSFF}. In the case where the GUE model emerges as an SSB phase of the Yang Mills model, we have $g_* \gg 1$. In this regime the size of the ramp in the confined phase becomes parametrically shorter than the ramp in the deconfined phase. We propose a diagnostic that distinguishes between the two families of curves: their slope near $k=0$. In the symmetry preserving phase, this slope (with respect to $k$) is $\frac{1}{L}$, whereas in the symmetry broken phase, this slope (with respect to $t_k\lambda)$ is $\frac{2}{\pi L}$, this can be clearly seen in figure \ref{fig:phasetransitionSFF}.

As stated before, all this shows that the GUE can be thought of as the effective theory describing the broken phase of the of the Yang Mills matrix model. The one- and two-point functions agree to a very close approximation in the region where they are finite and $\mathcal{O}(1)$. What happens when we crank up $k$ up to the volume scale? In the introduction we described how this would allow us to probe the symmetry restored phase---and the approach to the symmetry restored phase will depend on the UV physics. We dedicate the last few lines of this section to studying the exit. 

\begin{figure}[t!]
\begin{center}
\includegraphics[height=8cm]{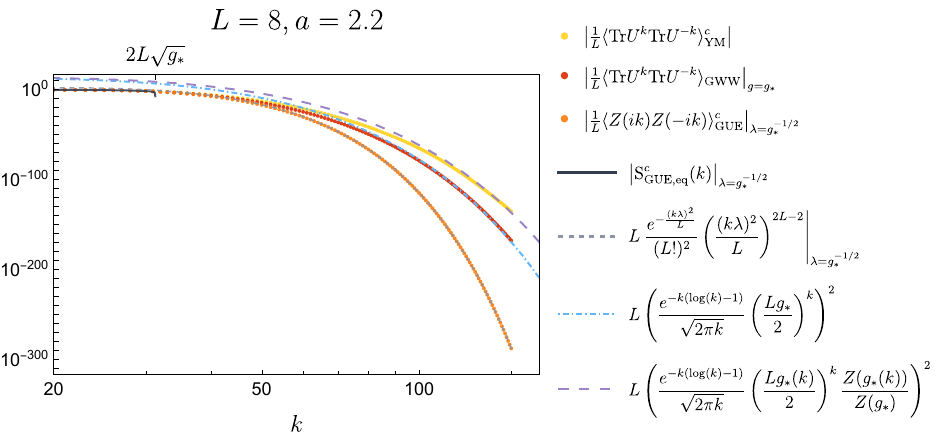}
\end{center}
\caption{We plot the late-time exit from the ramp in the deconfined phase $a>1$ for the YM, GWW and GUE matrix models. All three plots agree very closely in the ramp region (see figure \ref{fig:twoptYM}), but differ in how they exit from the ramp, demonstrating that the physics of symmetry restoration depends sensitively on the UV theory.}  \label{fig:twoptYMdeconfinedexit}
\end{figure}

One nice fact about the GUE is that it allows for an exact solution which can be studied to death. And a nice thing we can show (see the discussion around \eqref{eq:asymptoticratioGUE}) is that, at late times: 
\begin{equation}
\langle Z(it)Z(-it)\rangle_{\rm GUE}\underset{t\lambda\gg L}{\approx}-\left(\langle Z(it)\rangle_{\rm GUE}\right)^2~.
\end{equation}
Our presumption is that this equation holds to describe the exit from the ramp in the YM matrix model. That is we will assume that 
\begin{equation}
\langle\text{Tr} U^k\text{Tr}U^{-k}\rangle^c_{\rm YM}\underset{k\gg L}{\approx}-\left(\langle\text{Tr}\, U^k\rangle_{\rm YM}\right)^2
\end{equation}
and verify that this is true, at least in the deconfined phase. We verify this explicitly in figure \ref{fig:twoptYMdeconfinedexit}. We should also mention that this picture differs slightly from the symmetry-restoration description of the plateau provided in \cite{Altland:2022xqx}, where it is discussed in terms of whether the operators of interest are able to resolve the difference between a continuous density of eigenvalues (the one-cut phase) versus a sum of $\delta$-functions.

The exit from the ramp in the confined phase ($a<1$) is governed entirely by the fluctuations since the Haar caclculation is sharp. We do not have much to say about the exit from the ramp in this phase and do not show any plots of it.

\pagebreak
\section{Conclusions}\label{sec:conclusions}

In this work we have presented a novel perspective on the physics of random matrix models and their quantum-chaotic properties. It is often said that the GUE captures several universal features of quantum chaos, but in our view, the origin and implications of this universality are not often highlighted, particularly for the observables we care about.  We have advocated that the GUE is the universal effective action describing the symmetry-broken phase of a system enjoying a global $U(1)$ symmetry. It matters little for our results if this symmetry is actually present in the UV or if it is emergent. Even if we think of the GUE matrix model as descending from a UV theory with explicit symmetry breaking, it remains true that for observables that probe very large times, i.e. $ t \lambda \gg 1$, the $U(1)$ symmetry is emergent since the matrix model potential acts simply as an IR regulator, and hence contributes only to exponentially small effects in this regime. As a consequence, the plateau indeed corresponds to symmetry restoration. 

We conclude from the above analysis that there exist (at least) two phases of quantum chaos that describe \emph{different} universal behaviors. The first is the usual (deconfined) quantum chaos described by the GUE as an effective model. This phase is responsible for an extended ramp and a weak non-linear dependence on time during the ramp. The other phase is the \emph{confined chaos phase}, which controls the physics of the plateau.

Let us quickly summarize some of the results presented above and interpret them in the context of quantum mechanics:
\begin{enumerate}
\item Sum rules: once a system is described by a finite (but large) matrix model of size $L$ some general features are universal and related to the finiteness of configuration space. Observables involving $Z(i t)$ must be of order $L$ as $t \rightarrow 0$ as they encode the size of this space. Furthermore, if the potential has a typical scale $\lambda$, at $\lambda t \sim \alpha L$ (for some $\alpha$ of $\mathcal{O}(1)$)  the integral over matrices is controlled by a saddle point approximation and observables become exponentially suppressed in $L$. This gives a universal ramp, and observables must go from being order $L$ to exponentially suppressed within times of order $L$. The precise length and shape of the ramp depends on the particular phase that the model describes.

\item Once we consider times $t \lambda \gg L$ we enter a symmetry restored phase where the form of the potential contributes only exponentially small effects and there is a (possibly emergent) $U(1)$ symmetry. 

\item When the symmetry is unbroken the system becomes Haar random. One-point functions vanish exactly and two-point functions show linear ``short'' ramps that reach $0$ at finite discrete time $L$. This phase corresponds to \textit{confined chaos}.

\item If the symmetry is spontaneously broken the system  has an effective description in terms of a Hermitian matrix model. When this theory is weakly coupled it can be described by the GUE ensemble. In that case the ramp becomes parametrically longer than in the symmetry preserved phase (when $g_* \gg 1$). Therefore, long ramps are controlled by a \textit{deconfined} quantum chaos phase. The end of the ramp is nothing other than the restoration of the symmetry at late times. While the form of the exponential decay is not universal (as it depends on the effective $\Lambda_{IR}$ scale), the existence of the plateau is controlled by the \textit{confined} phase.

\end{enumerate}

From the point of view of reproducing the properties of quantum mechanical chaotic systems there is a very precise way in which we can characterize the physics of the phases described above. An abstract finite Hilbert space of size $L$ has a natural $U(L)$ gauge symmetry acting on it which corresponds to the conjugation by unitary matrices of all operators in the theory. There is also a $U(1)$ symmetry that acts as multiplication by a phase. This symmetry can be thought as global\footnote{In a matrix model, the difference between a compact global and gauge symmetry blurs for neutral observables.} as phase differences between quantum states are in principle observable (e.g. through particle statistics and interference experiments). The confined phase discussed before corresponds to an ensemble of quantum gates given by unitary operators acting on this space. We can think of these quantum gates as a form of discrete time evolution. The potential corresponding to the YM model described in this work corresponds to a probability distribution that favours unitary operators that on (Hilbert space) average act just by multiplying states by a phase. Even in this case (for $a<1$) the system can preserve the $U(1)$ symmetry because of configuration space effects encoded in the Haar measure which favor unitary operators that are random. When the potential becomes too strong the symmetry breaks and, instead, we end up choosing a unitary operator close the identity with a gaussian measure (for $a \gg 1$) for the effective Hamiltonian associated to the gate. Effectively we have morphed the problem from the choice of a random gate to the choice of a random Hamiltonian. This makes time evolution continuous in the effective description. Here we are in the deconfined phase associated to the GUE universality of quantum chaos. While this picture is satisfying from the quantum system perspective, it remains unexplored in the context of holography in quantum gravity. Interesting connections to the realization of bulk geometries through tensor networks \cite{Swingle:2009bg} are possible here.

It remains for us to state that we have yet to develop a fully-systematic characterization, in a Wilsonian sense, of the space of effective theories for matrix models. It would be interesting to formalize this in the future, potentially connecting to the hydrodynamic theory for the spectral form factor proposed in \cite{Winer:2020gdp}. Sadly,  little of our intuition from higher dimensions translates to the realm of matrix models. There is no base space notion of energy scales or an associated Hilbert space. The distinction between (compact) global and  gauge symmetries is blurred. The difference between spontaneous and explicit symmetry breaking is more subtle. But the results presented here point to a possible reorganization of the degrees of freedom of matrix models that might prove useful in understanding the various universal behaviors present in different phases of these models. Maybe holography is the appropriate language here.  A more complete and systematic understanding could have multiple applications, ranging from the study of quantum chaotic system to models of quantum gravity. As we hinted at in the introduction, it would seem that the very notion of gravity is deeply tied to symmetry breaking as the partition function can only acquire an expectation value in this phase. What, then, is the meaning of the symmetry preserving phase? Is there a non-geometric dual to \textit{confined chaos}? We leave these questions for future research.\footnote{The curious reader can take a look at the formulae in appendix \ref{qghint} for some interesting connections between finite $L$ matrix models and quantum gravity.}

\section*{Acknowledgements}
We would like to thank Dionysios Anninos, Sean Hartnoll, Jorge Kurchan, Raghu Mahajan, Julian Sonner and particularly Kazumi Okuyama, who walked us through his code on numerically computing expectation values in the YM matrix model. We would especially like to acknowledge and thank Antonio Rotundo for his collaboration during the (multi-year) initial stages of this project. T.A. is supported by the UKRI Future Leaders Fellowship ``The materials approach to quantum spacetime'' under reference MR/X034453/1.

\appendix
\section{Review: Hermitian matrix ensembles with single-trace potentials}\label{ap:singletracemm}

In this review section, we review for the reader how to compute exact expectation values in a random hermitian matrix model with an arbitrary single-trace potential. We adapt the techniques found in several reviews on the subject \cite{Ginsparg:1993is,forrester2010log,DiFrancesco:1993cyw,Anninos:2020geh,Haake2018}.
In a random matrix model (RMT), we consider an $L\times L$ random Hermitian matrix $H$ and compute expectation values with respect to the norm:
 \begin{equation}\label{eq:defofmatrixensemble}
   \langle\cdot\rangle=\frac{1}{\mathcal{N}}\int  [\d H]\, (\cdot)\exp\left\lbrace-\text{Tr} \,V(H)\right\rbrace~
 \end{equation}
 Here the normalization constant $\mathcal{N}$ is chosen such that $\langle1\rangle=1$, and $V(H)$ is some polynomial in the matrix $H$. 
The measure is explicitly:
\begin{equation}\label{eq:defofmeasure}
   [\d H]\equiv \prod_{i=1}^L \d H_{ii}\prod_{i<j}^L \d\text{Re}H_{ij}\,\d\text{Im}H_{ij}
\end{equation}
 We will focus mainly on the \emph{Gaussian Unitary Ensemble} (GUE), for which the explicit formula is: 
 \begin{equation}\label{eq:GUE}
   \langle\cdot\rangle_{\rm GUE}=2^{-L/2}\left(\frac{L}{\pi\lambda^2}\right)^{L^2/2}\int [\d H]\, (\cdot)\exp\left\lbrace-\frac{L}{2\lambda^2 }H_{ij}H_{ji}\right\rbrace~.
 \end{equation}
 Our main interest will be in computing correlation functions of \emph{loop operators}:
 \begin{equation}
 \langle Z(\beta_1) Z(\beta_2)\dots Z(\beta_k)\rangle=\frac{1}{\mathcal{N}}\int [\d H]\, \left[\text{Tr}\, e^{-\beta_1 H}\text{Tr}\, e^{-\beta_2 H}\dots \text{Tr}\, e^{-\beta_k H}\right]e^{-\text{Tr} \,V(H)}~.
 \end{equation}
 The loop operator is itself related to the \emph{resolvent operator} via a Laplace transform:
 \begin{equation}
   R(E)=\text{Tr}\frac{1}{E-H}=-\int_0^\infty \d\beta\, Z(\beta)\,e^{\beta E}~, 
  \end{equation} 
Which incidentally implies: 
\begin{equation}
   Z(\beta)=\frac{1}{2\pi i}\int_{c-i\infty}^{c+i\infty} \d E\, R(E)\, e^{-\beta E}. 
\end{equation}
From this definition we naturally see how the density of states giving rise to $\langle Z(\beta)\rangle$ can be thought of as arising from the residues of the resolvent, or, in case $\langle R(E)\rangle $ develops a branch cut, from the discontinuity across the cut. In any case, this motivates why we should study loop operators, as they contain the exact same information as the resolvent.

\subsection{General theory of orthogonal polynomials} \label{ap:orthpol}
Since we will only be interested in objects composed of traces of powers $H$, we can reduce the problem to the eigenvalues. This well known reduction leads to: 
\begin{equation}\label{eq:eigenvaluesnorm}
   \langle\cdot\rangle=\frac{1}{\mathcal{N}'}\int \left[\prod_{i=1}^L \d\lambda_i\right]\, \Delta(\lambda)^2 (\cdot)\exp\left\lbrace-\sum_{i=1}^L \,V(\lambda_i)\right\rbrace~,
 \end{equation}
 where again $\mathcal{N}'$ is chosen such that $\langle 1\rangle =1$ and we have introduced the Vandermonde determinant
 \begin{equation}
   \Delta(\lambda)\equiv \prod_{i<j}^L (\lambda_j-\lambda_i)=\text{det}\,\lambda_i^{j-1}~.
 \end{equation}
 The exact solution of our problem lies in assuming that $V(\lambda)$ is a function that admits a set of orthogonal polynomials as follows \cite{Ginsparg:1993is,forrester2010log,DiFrancesco:1993cyw,Anninos:2020geh}: 
 \begin{equation}\label{eq:orthpoldef}
   \int \d\lambda\, e^{-V(\lambda)}P_n(\lambda)P_m(\lambda)=h_n\,\delta_{nm}~. 
 \end{equation}
 where we have chosen to $P_n$ to be normalized such that its highest power of $\lambda$ comes with coefficient 1: 
 \begin{equation}
   P_n(\lambda)=\lambda^n+c_{n-1}\lambda^{n-1}+\dots+c_0~.
 \end{equation}
 This definition is convenient, since it allows{} us to write: 
 \begin{equation}
   \Delta(\lambda)=\text{det}\,\lambda_i^{j-1}=\det P_{j-1}(\lambda_i)~,
 \end{equation}
 which we can see by starting with the matrix $\lambda_i^{j-1}$ and constructing a new matrix by adding linear combinations of preceding columns, which leaves the determinant invariant, until we ultimately obtain the matrix $P_{j-1}(\lambda_i)$~. From the definition \eqref{eq:orthpoldef} we can equivalently define a set of `wavefunctions' $\psi_n$
 \begin{equation}\label{eq:wavefunctiondef}
   \psi_n(\lambda)\equiv\frac{P_n(\lambda)}{\sqrt{h_n}}e^{-\frac{V(\lambda)}{2}}
 \end{equation}
 which satisfy: 
 \begin{equation}
   \int \d\lambda\, \psi_n(\lambda)\psi_m(\lambda)=\delta_{nm}~.
 \end{equation}

\subsubsection{Normalization}\label{ap:norm}
 Now we can compute the normalization $\mathcal{N}'$. This quantity can be thought of as the partition function of the random matrix theory. Writing out the determinant: 
 \begin{equation}
 \det P_{j-1}(\lambda_i)=\sum_{\sigma \in S_L} (-1)^\sigma\prod_{i=1}^L P_{\sigma_i-1}(\lambda_i)~,
 \end{equation}
 where $S_L$ is the set of permutations of $L$ elements, we have: 
 \begin{align}
   \mathcal{N}'&=\int \left[\prod_{i}^L \d\lambda_i\right]\, \left(\sum_{\sigma \in S_L} (-1)^\sigma\prod_{i=1}^L P_{\sigma_i-1}(\lambda_i)\right)\left(\sum_{\gamma \in S_L} (-1)^\gamma\prod_{j=1}^L P_{\gamma_j-1}(\lambda_j)\right)\exp\left\lbrace-\sum_{k=1}^L \,V(\lambda_k)\right\rbrace~\\
   &=\sum_{\sigma \in S_L}\int \left[\prod_{i}^L \d\lambda_i\right]\, \prod_{i=1}^L P_{\sigma_i-1}(\lambda_i)^2\exp\left\lbrace-\sum_{k=1}^L \,V(\lambda_k)\right\rbrace~\\
   &=L! \prod_{n=0}^{L-1}h_n~.
 \end{align}
 In going from the first line to the second line we used orthogonality, and in going from the second line to the third line we used that all $L!$ permutations give the same result. 

 \subsubsection{One-point function}\label{sec:1pfctgene}
 The one point function is defined as follows
 \begin{align}
   \langle Z(\beta)\rangle&=\frac{1}{\mathcal{N}}\int [\d H]\, \text{Tr}\, e^{-\beta H}e^{-\text{Tr} \,V(H)}\\
   &=\frac{1}{\mathcal{N}}\sum_{n=0}^\infty\frac{(-\beta)^n}{n!}\int [\d H]\, (\text{Tr}H^n)\,e^{-\text{Tr} \,V(H)}
 \end{align}
 therefore it suffices to compute $\langle \text{Tr}\,H^n\rangle$ for arbitrary $n$ for us to obtain the full one-point function. To do so, we pass to the eigenvalue language: 
 \begin{equation}
   \langle \text{Tr}\,H^n\rangle=\frac{1}{L! \prod_{s=0}^{L-1}h_s}\int \prod_{i}^L \d\lambda_i\, \Delta(\lambda)^2 \left(\sum_{j=1}^L\lambda_j^n\right)\prod_{k=1}^Le^{-V(\lambda_k)}~.
 \end{equation}
 By relabeling, we can write this as: 
 \begin{align}
   \langle \text{Tr}\,H^n\rangle&=\frac{L}{L! \prod_{s=0}^{L-1}h_s}\int \prod_{i}^L \d\lambda_i\, \Delta(\lambda)^2 \lambda_1^n\prod_{k=1}^Le^{-V(\lambda_k)}\\
&=\frac{L}{L! \prod_{s=0}^{L-1}h_s}\int \prod_{i}^L \d\lambda_i\, \left(\sum_{\sigma \in S_L} (-1)^\sigma\prod_{i=1}^L P_{\sigma_i-1}(\lambda_i)\right)\left(\sum_{\gamma \in S_L} (-1)^\gamma\prod_{j=1}^L P_{\gamma_j-1}(\lambda_j)\right)\lambda_1^n\prod_{k=1}^Le^{-V(\lambda_k)}~.
 \end{align}
 Now for every $\lambda_i$ for $i>1$, this proceeds similarly as when we calculated the normalization factor $\mathcal{N}'$, but now we are only permuting $L-1$ elements. This leads to
 \begin{align}
 \langle \text{Tr}\,H^n\rangle=\frac{L}{L! \prod_{s=0}^{L-1}h_s}(L-1)!\sum_{j=0}^{L-1}\frac{\prod_{s=0}^{L-1}h_s}{h_j}\int \d\lambda_1\, P_{j}(\lambda_1)^2\lambda_1^n e^{-V(\lambda_1)}
 \end{align}
 which we may conveniently rewrite as
 \begin{equation}
   \langle \text{Tr}\,H^n\rangle=\int \d\lambda \,\lambda^n\sum_{j=0}^{L-1}\psi_j(\lambda)^2~.
 \end{equation}
 From this we conclude that 
 \begin{equation}\label{eq:onepointgeneral}
   \langle Z(\beta)\rangle =\int \d\lambda\, e^{-\beta \lambda}\sum_{j=0}^{L-1}\psi_j(\lambda)^2
 \end{equation}
 It is tempting to read this equation as defining an expectation value for the density of states of some theory: 
 \begin{equation}
   \langle Z(\beta)\rangle =\int \d E\, e^{-\beta E}\langle\rho(E)\rangle
 \end{equation}
 where 
 \begin{equation}\label{eq:exactdensityofstates}
   \langle\rho(E)\rangle=\sum_{j=0}^{L-1}\psi_j(E)^2~. 
 \end{equation}
 We will study this expression further in a specific example. 

 \subsubsection{Two-point function}
Let us now repeat the analysis of the previous section now for the two point function: 
 \begin{align}
    \langle Z(\beta_1) Z(\beta_2)\rangle&=\frac{1}{\mathcal{N}}\int [\d H]\, \text{Tr}\, e^{-\beta_1 H}\text{Tr}\, e^{-\beta_2 H}e^{-\text{Tr} \,V(H)}\\
    &=\frac{1}{\mathcal{N}}\sum_{n,m=0}^\infty\frac{(-\beta_1)^n}{n!}\frac{(-\beta_2)^m}{m!}\int [\d H]\, (\text{Tr}H^n)(\text{Tr}H^m)\,e^{-\text{Tr} \,V(H)}~,
 \end{align}
 meaning it suffices to compute $\langle \text{Tr}\,H^n\text{Tr}\,H^m\rangle$ for arbitrary $n,m$ for us to obtain the full two-point function. To do so, we pass to the eigenvalue language: 
 \begin{equation}
    \langle \text{Tr}\,H^n\,\text{Tr}\,H^m\rangle=\frac{1}{L! \prod_{s=0}^{L-1}h_s}\int \prod_{i}^L \d\lambda_i\, \Delta(\lambda)^2 \left(\sum_{j=1}^L\lambda_j^n\right)\left(\sum_{l=1}^L\lambda_l^m\right)\prod_{k=1}^Le^{-V(\lambda_k)}~.
 \end{equation}
 Now we have to be careful and treat the $j=l$ and $j\neq l$ terms separately in the above sum. Thus we have, after a relabeling:
  \begin{align}
    \langle \text{Tr}\,H^n\,\text{Tr}\,H^m\rangle=&\frac{L}{L! \prod_{s=0}^{L-1}h_s}\int \prod_{i}^L \d\lambda_i\, \Delta(\lambda)^2 \lambda_1^{n+m}\prod_{k=1}^Le^{-V(\lambda_k)}\nonumber\\
    &+\frac{L(L-1)}{L! \prod_{s=0}^{L-1}h_s}\int \prod_{i}^L \d\lambda_i\, \Delta(\lambda)^2 \lambda_1^n\lambda_2^m\prod_{k=1}^Le^{-V(\lambda_k)}~.
 \end{align} 
 The first line is the same as in the one-point function section, meaning it will give a contribution of $\langle Z(\beta_1+\beta_2)\rangle $ to the two-point function. It is not so difficult to follow the same steps as in the one-point function section and conclude that, once all the dust settles: 
\begin{equation}\label{eq:2ptgeneral}
   \langle Z(\beta_1)Z(\beta_2)\rangle =\langle Z(\beta_1+\beta_2)\rangle+\int \d E_1 \d E_2\, e^{-\beta_1 E_1-\beta_2 E_2}\langle\rho(E_1)\rho(E_2)\rangle
\end{equation}
where 
\begin{equation}
   \langle\rho(E_1)\rho(E_2)\rangle = \det \begin{pmatrix} K_L(E_1,E_1) & K_L(E_1,E_2)\\ K_L(E_2,E_1) & K_L(E_2,E_2)\end{pmatrix}
\end{equation}
and 
\begin{equation}
   K_L(x,y)=\sum_{j=0}^{L-1}\psi_j(x)\psi_j(y)~.
\end{equation}
Let us now expand out \eqref{eq:2ptgeneral}, we find
\begin{equation}
  \langle\rho(E_1)\rho(E_2)\rangle =  \langle\rho(E_1)\rangle\langle\rho(E_2)\rangle+\langle\rho(E_1)\rho(E_2)\rangle^c
\end{equation}
where $\langle\rho(E)\rangle$ is as in \eqref{eq:exactdensityofstates} and 
\begin{equation}\label{eq:exact2ptdensity}
    \langle\rho(E_1)\rho(E_2)\rangle^c\equiv-\sum_{j,k=0}^{L-1}\psi_j(E_1)\psi_j(E_2)\psi_k(E_1)\psi_k(E_2)
\end{equation}
is the connected contribution to the exact density of states. Given this structure, we now identify
\begin{equation}\label{eq:2ptgeneral2}
   \langle Z(\beta_1)Z(\beta_2)\rangle =\langle Z(\beta_1+\beta_2)\rangle+\langle Z(\beta_1)\rangle\langle Z(\beta_1)\rangle+\langle Z(\beta_1)Z(\beta_2)\rangle^c~,
\end{equation}
where 
\begin{equation}\label{eq:twopointconnecteddef}
\langle Z(\beta_1)Z(\beta_2)\rangle^c\equiv \int \d E_1 \d E_2\, e^{-\beta_1 E_1-\beta_2 E_2}\langle\rho(E_1)\rho(E_2)\rangle^c~.
\end{equation}
We also collect here a nice fact about orthogonal polynomials, known as the Christoffel-Darboux formula, which states: 
\begin{equation}\label{eq:christoffeldarboux}
   K_L(x,y)=\sum_{j=0}^{L-1}\psi_j(x)\psi_j(y)=\sqrt{\frac{h_L}{h_{L-1}}}\left[\frac{\psi_L(x)\psi_{L-1}(y)-\psi_L(y)\psi_{L-1}(x)}{x-y}\right]
\end{equation}
Applying this to the one-point function, we see that 
\begin{equation}\label{eq:christoffeldarboux1pt}
    \langle\rho(E)\rangle=\sum_{j=0}^{L-1}\psi_j(E)^2=\sqrt{\frac{h_L}{h_{L-1}}}\left[{\psi_L'(E)\psi_{L-1}(E)-\psi_L(E)\psi_{L-1}'(E)}\right]~.
 \end{equation}
 Even simpler, the two-point function has the following density:
 \begin{equation}\label{eq:christoffeldarboux2pt}
    \langle\rho(E_1)\rho(E_2)\rangle^c=-{\frac{h_L}{h_{L-1}}}\left[\frac{\psi_L(E_1)\psi_{L-1}(E_2)-\psi_L(E_2)\psi_{L-1}(E_1)}{E_1-E_2}\right]^2~.
 \end{equation}

 \subsection{The Loop Equations}
 Besides orthogonal polynomials, there is another way to attack the problem, which will indicate how the symmetry preservation/breaking occurs. This technique is known as the loop equations. Let us derive them. We begin with the definition of our matrix ensemble \eqref{eq:defofmatrixensemble} and measure \eqref{eq:defofmeasure}, which we repeat here
  \begin{equation}
   \langle\cdot\rangle=\frac{1}{\mathcal{N}}\int  [\d H]\, \left(\cdot\right)\exp\left\lbrace-\text{Tr} \,V(H)\right\rbrace~,
 \end{equation}
where
\begin{equation}
   [\d H]\equiv \prod_{i=1}^L \d H_{ii}\prod_{i<j}^L \d\text{Re}H_{ij}\,\d\text{Im}H_{ij} ~. 
\end{equation}
If we have done things correctly, the expectation values defined in this way should be are invariant under arbitrary reparametrizations of the matrix $H\rightarrow f(H)$. Let us now consider $f(H)$ infinitesimally close to the identity element: 
\begin{equation}
   f(H)=H + \epsilon\, H^k~. 
\end{equation}
Under this transformation we find
\begin{align}
   \exp\left\lbrace-\text{Tr} \,V(H)\right\rbrace&\longrightarrow \exp\left\lbrace-\text{Tr} \,V(H)\right\rbrace~\left[1-\epsilon \text{Tr}\left(V'(H)\cdot H^k\right)+O\left(\epsilon^2\right)\right]~,\\
   [\d H]&\longrightarrow [\d H]\left[1+\epsilon\sum_{l=0}^{k-1}\text{Tr}\left(H^l\right)\text{Tr}\left(H^{k-1-l}\right)+O\left(\epsilon^2\right)\right]~.
\end{align}
The final answer should not depend on $\epsilon$, implying the following Ward identity: 
\begin{equation}\label{eq:loopeq}
\left\langle \sum_{l=0}^{k-1}\text{Tr}\left(H^l\right)\text{Tr}\left(H^{k-1-l}\right)\right\rangle=\left\langle \text{Tr}\left(V'(H)\cdot H^k\right)\right\rangle
\end{equation}
for any $k$. 
Multiplying \eqref{eq:loopeq} by $E^{-k-1}$ and summing over $k$, we find: 
\begin{equation}
   \left\langle \sum_{k=0}^\infty\sum_{l=0}^{k-1}\frac{\text{Tr}\left(H^l\right)}{E^{l+1}}\frac{\text{Tr}\left(H^{k-1-l}\right)}{E^{k-l}}\right\rangle =\left\langle\sum_{k=0}^\infty\frac{\text{Tr}\left(V'(H)\cdot H^k\right)}{E^{k+1}}\right\rangle~.
\end{equation}
But this is just the power series expansion for the following formal expression: 
\begin{equation}\label{eq:loopR1}
   \left\langle R(E)^2\right\rangle =\left\langle\text{Tr}\left[V'(H)\cdot (E-H)^{-1}\right]\right\rangle~.
\end{equation}
Denoting for the moment $V'(H)=V'(E)+\left(V'(H)-V'(E)\right)$ and defining: 
\begin{equation}
   \langle P(E)\rangle\equiv \left\langle\text{Tr}\left\lbrace\left[V'(E)-V'(H)\right]\cdot (E-H)^{-1}\right\rbrace\right\rangle, 
\end{equation}
The loop equations simplify to: 
\begin{equation}
  \boxed{ \left\langle R(E)^2\right\rangle=V'(E)\left\langle R(E)\right\rangle -\left\langle P(E)\right\rangle}~. 
\end{equation}
Finally, we can inverse Laplace transform the expression \eqref{eq:loopR1} to obtain a constraint on the partition function, rather than on the resolvents. This leads to 
\begin{equation}\label{eq:looppot}
   \boxed{\int_0^\beta \d \beta'\left\langle Z(\beta')Z(\beta-\beta')\right\rangle=-V'(-\partial_\beta)\left\langle Z(\beta)\right\rangle}~.
\end{equation}
Note that these loop equations imply a kind of sewing relation between $n$-point functions and are derived from the infinite reparametrization symmetry of the integral.

\section{The Gaussian Unitary Ensemble}\label{ap:GUE}
\subsection{Orthogonal Polynomials}

The GUE is a specific Hermitian matrix model with the following potential $V(x)=\frac{L}{2\lambda^2} x^2$.\footnote{In this instance $\lambda$ now refers to the width of the distribution and no longer labels an eigenvalue.}  The orthogonal polynomials are well known for this ensemble\cite{Meh2004,Ginsparg:1993is,DiFrancesco:1993cyw,Anninos:2020ccj}:  
\begin{equation}
   P_n(x)=\left(\frac{\lambda}{\sqrt{2L}}\right)^nH_n\left(\frac{x}{\lambda}\sqrt{\frac{L}{2}} \right)~,
\end{equation}
where the $H_n(x)$ are Hermite polynomials.
With this definition, one can check that the largest power of $x$ always comes with a coefficient $1$. The normalizations $h_n$, defined in \eqref{eq:orthpoldef} are found to be: 
\begin{equation}
   h_n= n!\sqrt{2\pi}\left(\frac{\lambda^2}{L}\right)^{n+\tfrac{1}{2}},
\end{equation}
and the wavefunctions are therefore:
\begin{equation}\label{eq:GUEWavefunction}
   \psi_n(x)=\left(\frac{L}{2\pi}\right)^{\frac{1}{4}}\frac{H_n\left(\frac{x}{\lambda}\sqrt{\frac{L}{2}} \right)}{\sqrt{2^n \lambda \,n!}}\,e^{-\frac{L^2}{4\lambda^2}x^2}~.
\end{equation}
 Notice that we already have an explicit factor of $L$ in our distribution. This will allow us to take a large-$L$ limit while keeping $\lambda\sim\mathcal{O}(1)$. In this sense, $\lambda$ is the natural 't Hooft coupling of the GUE.  

Using the Christoffel-Darboux formula \eqref{eq:christoffeldarboux1pt}, we can write the exact one-point density of states for this ensemble: 
\begin{equation}
     \langle\rho(E)\rangle_{\rm GUE}=\sqrt{\frac{L}{2\pi}}\frac{\,e^{-\frac{L}{2\lambda^2} E^2}}{2^L\lambda(L-1)!}\left[H_L\left(\frac{E}{\lambda}\sqrt{\frac{L}{2}} \right)^2-H_{L-1}\left(\frac{E}{\lambda}\sqrt{\frac{L}{2}} \right)H_{L+1}\left(\frac{E}{\lambda}\sqrt{\frac{L}{2}} \right)\right]~.
\end{equation}
As we will see, since we have a non-perturbative definition of this effective density of states, we can use it to find the behavior of the one-point function $\langle Z(\beta)\rangle_{\rm GUE}$ in all parameter regions. 

\subsection{One-point function}\label{ap:GUE1pt}
We are now in a position to compute the one-point function of the GUE. We must evaluate the integral
 \begin{align}
    \langle Z(\beta)\rangle_{\rm GUE} &=\sum_{j=0}^{L-1}\int_{-\infty}^\infty \d E\, e^{-\beta E}\psi_j(E)^2\\
    &=\sqrt{\frac{L}{2\pi}}\sum_{j=0}^{L-1}\frac{1}{2^j \lambda\,j!}\int_{-\infty}^\infty \d E\, e^{-\beta E}H_j\left(\frac{E}{\lambda}\sqrt{\frac{L}{2}}\right)^2e^{-{\frac{L}{2\lambda^2}E^2}}~.
 \end{align}
 Changing variables to $\frac{E}{\lambda}\sqrt{\frac{L}{2}}=x$ and $\beta=\frac{y}{\lambda}\sqrt{2L}$, we can write this as: 
 \begin{equation}\label{eq:relabelhermite}
     \langle Z(\beta)\rangle_{\rm GUE}=\sum_{j=0}^{L-1}\frac{1}{2^j j!\sqrt{\pi}}\int_{-\infty}^\infty \d x\, e^{-x^2-2xy}H_j(x)^2~.
 \end{equation}
 Looking furiously in Gradshteyn and Ryzhik \cite{Gradshteyn:1943cpj},\footnote{BU 148(15), ET II 289(13)a} we find the following integral: 
  \begin{equation}\label{eq:hermiteint}
     \int_{-\infty}^\infty \d x\, e^{-x^2+2xy}H_m(x)H_n(x)=2^n m!\sqrt{\pi}y^{n-m}e^{y^2}L_m^{n-m}(-2y^2)~,\qquad m\leq n~,
 \end{equation}
 where $L_i^\alpha(x)$ is a generalized Laguerre polynomial. This means that we can write down the exact one-point function for this ensemble: 
 \begin{equation}\label{eq:exact1ptsum}
     \langle Z(\beta)\rangle_{\rm GUE}={e^{\frac{(\beta \lambda)^2}{2L}}}\sum_{j=0}^{L-1}L_j\left(-\frac{(\beta\lambda)^2}{L}\right).
 \end{equation}
 We may also use the following Laguerre polynomial identity, which follows from their recurrence relations: 
 \begin{equation}
     L_n^{\alpha+1}=\sum_{i=0}^nL_i^\alpha(x)
 \end{equation}
 which leads to the following exact expression for the one-point function: 
 \begin{equation}\label{eq:onepointexact}
     \boxed{\langle Z(\beta)\rangle_{\rm GUE}={e^{\frac{(\beta \lambda)^2}{2L}}}L_{L-1}^1\left(-\frac{(\beta\lambda)^2}{L}\right).}
 \end{equation}
 Using the fact that $
    L_n^\alpha(0)=\binom{n+\alpha}{n}$, we have that
 \begin{equation}
   \langle Z(0)\rangle_{\rm GUE}= L
 \end{equation}
 as expected since the infinite-temperature partition function computes the dimension of the Hilbert space. 

 \subsubsection{Large-\texorpdfstring{$L$}{L} limit } \label{largeLGUEsec}
 We now focus on computing the large-$L$ limit of the one-point function:  
 \begin{equation}\label{eq:onepointexacta2}
     \langle Z(\beta)\rangle_{\rm GUE}={e^{\frac{(\beta \lambda)^2}{2L}}}L_{L-1}^1\left(-\frac{(\beta\lambda)^2}{L}\right).
 \end{equation}
To do so, we will need to make use of the following asymptotic formula for Laguerre polynomials, which holds for $x\sim \mathcal{O}(1)$: 
 \begin{equation}
     \lim_{L\rightarrow\infty} \frac{L_L^\alpha\left(-\frac{x}{L}\right)}{L^\alpha}\approx e^{-\frac{x}{2L}}\frac{I_\alpha(2\sqrt{x})}{x^{\frac{\alpha}{2}}}~, 
 \end{equation}
 where $I_\alpha(x)$ is a modified Bessel function. This allows us to write down the `equilibrium' one-point function (using the nomenclature from the main text) in the GUE:
 \begin{equation}\label{eq:infLonepoint}
     \boxed{\langle {Z}_{\rm eq}(\beta)\rangle_{\rm GUE}\equiv\lim_{L\rightarrow\infty}\frac{ \langle Z(\beta)\rangle_{\rm GUE}}{L}= \frac{I_1(2\beta\lambda)}{\beta \lambda}~.}
 \end{equation}
 Moreover, there exists an integral expression for the modified Bessel function:
 \begin{equation}
     I_\nu(z)=\frac{\left(\frac{z}{2}\right)^\nu}{\Gamma\left(\nu+\frac{1}{2}\right)\Gamma\left(\frac{1}{2}\right)}\int_{-1}^1\left(1-t^2\right)^{\nu-1/2}e^{-zt}\d t~.
 \end{equation}
  After a simple change of variables, the above expression allows us to write a simple integral expression for the equilibrium one-point function:  
 \begin{equation}
     \langle {Z}_{\rm eq}(\beta)\rangle_{\rm GUE}=\int_{-2\lambda}^{+2\lambda}\frac{\d E}{\pi\lambda}\sqrt{1-\left(\frac{E}{2\lambda}\right)^2}\,e^{-\beta E}~.
 \end{equation}
 From this we can read off the equilibrium density of states
 \begin{equation}\label{eq:rhoinfinitydef}
     \lim_{L\rightarrow\infty}\frac{\langle \rho(E)\rangle_{\rm GUE}}{L}\equiv\langle{\rho}_{\rm eq}(E)\rangle_{\rm GUE}=\frac{1}{\pi\lambda}\sqrt{1-\left(\frac{E}{2\lambda}\right)^2}~. 
 \end{equation}{}

We now take the opportunity to check the large-$L$ limit of the loop equations \eqref{eq:looppot}: 
 \begin{equation}
   L\int_0^\beta \d \beta'\left\langle  {Z}_{\rm eq}(\beta')\right\rangle_{\rm GUE} \left\langle {Z}_{\rm eq}(\beta-\beta')\right\rangle_{\rm GUE}=-V'(-\partial_\beta)\left\langle {Z}_{\rm eq}(\beta)\right\rangle_{\rm GUE}~,
\end{equation}
which can be verified using the following integral found in Gradshteyn and Ryzhik \cite{Gradshteyn:1943cpj}: 
\begin{equation}
   \int_0^z\d x \frac{I_p(x)}{x}\frac{I_q(z-x)}{z-x}=\left(\frac{1}{p}+\frac{1}{q}\right)\frac{I_{p+q}(z)}{z}~.
\end{equation}

\subsubsection{A suggestive expression} \label{qghint}
 Surprisingly, we can recast the non-pertubative, exact, finite-$L$ one-point function as an integral transform of the equilibrium ($L=\infty$) answer. Let us remind the reader of the following integral representation of the Laguerre polynomials: 
 \begin{equation}\label{eq:laguerreintexp}
     L_n^\alpha(-x)=\frac{e^{-x}x^{-\frac{\alpha}{2}}}{n!}\int_0^\infty e^{-t}t^{n+\frac{\alpha}{2}}I_\alpha\left(2\sqrt{xt}\right) \d t~. 
 \end{equation}
Using this in \eqref{eq:onepointexact} we find
\begin{equation}
    \langle Z(\beta)\rangle_{\rm GUE}=\frac{e^{-\frac{(\beta\lambda)^2}{2L}}}{(L-1)!}\frac{\sqrt{L}}{\beta\lambda}\int_0^\infty e^{-t}t^{L-1/2}I_1\left(2\beta\lambda\sqrt\frac{t}{L}\right) \d t~. 
\end{equation}
Defining a new variable $t\equiv L\left(\frac{\tilde\beta}{\beta}\right)^2$ we find that we may write: 
\begin{equation}\label{eq:QG1pt}
   \boxed{ \langle Z(\beta)\rangle_{\rm GUE}=\frac{2 L^{L+2}}{L!}e^{-\frac{(\beta\lambda)^2}{2 L}}\int_0^\infty \frac{\d \tilde{\beta}}{\tilde \beta}e^{-L\left(\frac{\tilde\beta}{\beta}\right)^2+2(L+1)\log\frac{\tilde\beta}{\beta}}\left\langle {Z}_{\rm eq}\left(\tilde\beta\right)\right\rangle_{\rm GUE}}
\end{equation}
where $\left\langle {Z}_{\rm eq}(\tilde\beta)\right\rangle_{\rm GUE}$ is as in \eqref{eq:infLonepoint}. 
This implies that the exact in $L$ result can be viewed as a quantum gravity path integral performed over the Euclidean-time coordinate of the quantum field theory defined at $L=\infty$. The integral over $\tilde\beta$ is highly peaked around $\tilde\beta=\beta$ at large-$L$. The fact that the distribution goes to zero for small $\tilde{\beta}$ implies that this quantum gravity path integral is killing off the high energy states of the putative $L=\infty$ theory. One may compare this to equation (3.23) in \cite{Gross:2019ach}, relating a $T\bar{T}$-deformed quantum mechanics to the undeformed theory. In both instances, the integral transform cuts off the continuum of states above some critical energy defined by $\beta$.

\subsection{Connected two-point function}
\subsubsection{First derivation}\label{ap:firstderivation}
Let us now repeat the task at hand, following equation \eqref{eq:twopointconnecteddef}. To calculate the connected two-point function in the GUE, we need to compute
\begin{equation}\label{eq:conntwopt}
    \langle Z(\beta_1)Z(\beta_2)\rangle^c_{\rm GUE} \equiv\int \d E_1 \d E_2\, e^{-\beta_1 E_1-\beta_2 E_2}\langle\rho(E_1)\rho(E_2)\rangle^c_{\rm GUE}~. 
\end{equation}
We may write this as follows: 
\begin{align}
\langle Z(\beta_1)Z(\beta_2)\rangle^c_{\rm GUE} &=-\sum_{j,k=0}^{L-1}\int \d E_1 \d E_2\, e^{-\beta_1 E_1-\beta_2 E_2}\psi_j(E_1)\psi_j(E_2)\psi_k(E_1)\psi_k(E_2)\\
&=-\sum_{j,k=0}^{L-1}{{\frac{1}{2^{j+k} j!k!\,{\pi}}}}\int \d x_1\,e^{-x_1^2-2x_1y_1}H_j(x_1)H_k(x_1) \int \d x_2\,e^{-x_2^2-2x_2y_2}H_j(x_2)H_k(x_2)~,
\end{align}
where, as in \eqref{eq:relabelhermite}, we have reparametrized the integrals via the change of variables $\frac{E_i}{\lambda}\sqrt{\frac{L}{2}}\equiv x_i$ and $\beta_i\equiv\frac{y_i}{\lambda}\sqrt{2L}$. We can now use \eqref{eq:hermiteint} to evaluate these integrals. We find
\begin{multline}
   \langle Z(\beta_1)Z(\beta_2)\rangle^c_{\rm GUE}=-e^{y_1^2+y_2^2}\Bigg( \sum_{j=0}^{L-1}L_j\left(-2y_1^2\right)L_j\left(-2y_2^2\right)\\
   +2\sum_{k=1}^{L-1}\sum_{j=0}^{k-1}{{\frac{j!}{ k!}}} (2y_1y_2)^{k-j}L_j^{k-j}\left(-2y_1^2\right)L_j^{k-j}\left(-2y_2^2\right)\Bigg) ~. 
\end{multline}
Where the first line comes from terms with $j=k$ and the second line comes from the remaining terms. The sum in the first line can actually be performed exactly using the following identity
\begin{equation}\label{eq:laguerreidentity}
    \sum_{m=0}^n\frac{m!}{\Gamma(m+\alpha+1)}L_m^\alpha(x)L_m^\alpha(y)=\frac{(n+1)!}{\Gamma(n+\alpha+1)(x-y)}\left[L_n^\alpha(x)L_{n+1}^\alpha(y)-L_{n+1}^\alpha(x)L_n^\alpha(y)\right]~.
\end{equation}
Thus we have: 
\begin{multline}\label{eq:exact2ptsum}
   \langle Z(\beta_1)Z(\beta_2)\rangle^c_{\rm GUE}=e^{y_1^2+y_2^2}\Bigg(\frac{L}{2(y_1^2-y_2^2)}\left[L_{L-1}\left(-2y_1^2\right)L_{L}\left(-2y_2^2\right)-L_{L}\left(-2y_1^2\right)L_{L-1}\left(-2y_2^2\right)\right]~\\
   -2\sum_{k=1}^{L-1}\sum_{j=0}^{k-1}{{\frac{j!}{ k!}}} (2y_1y_2)^{k-j}L_j^{k-j}\left(-2y_1^2\right)L_j^{k-j}\left(-2y_2^2\right)\Bigg) ~. 
\end{multline}
This is a nice expression, which can actually be found in \cite{delCampo:2017bzr}. Interestingly, we can further manipulate this expression:
\begin{align}
-2\sum_{k=1}^{L-1}\sum_{j=0}^{k-1}{{\frac{j!}{ k!}}} &(2y_1y_2)^{k-j}L_j^{k-j}\left(-2y_1^2\right)L_j^{k-j}\left(-2y_2^2\right)\nonumber\\&=-2\sum_{\alpha=1}^{L-1}\sum_{j=0}^{L-1-\alpha}{{\frac{j!}{ (j+\alpha)!}}} (2y_1y_2)^{\alpha}L_j^{\alpha}\left(-2y_1^2\right)L_j^{\alpha}\left(-2y_2^2\right)~,\nonumber\\
&=\sum_{\alpha=1}^{L-1}\frac{(L-\alpha)!}{(L-1)!}\frac{(2y_1y_2)^\alpha}{(y_1^2-y_2^2)}\left[L_{L-1-\alpha}^\alpha\left(-2y_1^2\right)L^\alpha_{L-\alpha}\left(-2y_2^2\right)-L^\alpha_{L-\alpha}\left(-2y_1^2\right)L_{L-1-\alpha}^\alpha\left(-2y_2^2\right)\right]~,
\end{align}
where we have again used \eqref{eq:laguerreidentity} in going from the first to the second line. We can now write a the final expression as a single sum:\footnote{The expression has a factor $\frac{\sin(\pi\alpha)}{\pi\alpha}$ in the summand, which we take to be 1 for $\alpha=0$ and zero otherwise.} 
\begin{multline}\label{eq:exact2ptsum2}
   \langle Z(\beta_1)Z(\beta_2)\rangle^c_{\rm GUE}=e^{y_1^2+y_2^2}\Bigg(\sum_{\alpha=0}^{L-1}\frac{1}{1+\frac{\sin(\pi\alpha)}{\pi\alpha}}\frac{(L-\alpha)!}{(L-1)!}\frac{(2y_1y_2)^\alpha}{(y_1^2-y_2^2)}\\\times\left[L_{L-1-\alpha}^\alpha\left(-2y_1^2\right)L^\alpha_{L-\alpha}\left(-2y_2^2\right)-L^\alpha_{L-\alpha}\left(-2y_1^2\right)L_{L-1-\alpha}^\alpha\left(-2y_2^2\right)\right]\Bigg) ~, 
\end{multline}
which can also be found in appendix A of \cite{Okuyama:2018yep}.

None of these sums can be performed explicitly, but as was noted in \cite{Okuyama:2018yep}, if we consider the \emph{derivative} of this function, we find: 
\begin{multline}
   (\partial_{y_1}-\partial_{y_2}) \langle Z(y_1)Z(y_2)\rangle^c_{\rm GUE}=\\\frac{L}{y_1+y_2}e^{y_1^2+y_2^2}\left[L_{L-1}\left(-2y_1^2\right)L_{L}\left(-2y_2^2\right)-L_{L}\left(-2y_1^2\right)L_{L-1}\left(-2y_2^2\right)\right]
\end{multline}
This latter expression will tell us about the slope of the SFF once analytically continued. Let us now focus on real $t$, taking $y_1=i\frac{t\lambda}{\sqrt{2L}}$ and $y_2=-i\frac{t\lambda}{\sqrt{2L}}$, then we have:
\begin{equation}\label{eq:exactderivative}
   \partial_t\langle Z(it)Z(-it)\rangle_{\rm GUE}^c=2t \lambda^2\,e^{-\frac{(t\lambda)^2}{L}}\left[L_{L-1}\left(\frac{(t\lambda)^2}{L}\right)L_{L-1}^1\left(\frac{(t\lambda)^2}{L}\right)-L_{L}\left(\frac{(t\lambda)^2}{L}\right)L_{L-2}^1\left(\frac{(t\lambda)^2}{L}\right)\right]
\end{equation}
so far this is not so illuminating, but it was derived in \cite{Brezin:1995dp,Okuyama:2018yep} that this is the exact eigenvalue density (in $t^2$) for the  Wishart-Laguerre ensemble, such that: 
\begin{equation}
   \lim_{L\rightarrow\infty}\partial_t\langle Z(it)Z(-it)\rangle_{\rm GUE}^c=\frac{2\lambda}{\pi}\sqrt{1-\left(\frac{t\lambda}{2L}\right)^2}~,\qquad\qquad \frac{t\lambda}{L}\text{  fixed.}
\end{equation}
Apparently these results were known in the past, see e.g. \cite{Brezin_1997,Liu:2018hlr}. We can integrate once with respect to $t$, giving the final large-$L$ limit of the spectral form factor:
\begin{equation}\label{eq:nonlinearamp}
 \boxed{  \lim_{L\rightarrow\infty}\tfrac{1}{L}\langle Z(it)Z(-it)\rangle_{\rm GUE}^c=-1+\frac{1}{\pi}\left(\frac{t\lambda}{L}\sqrt{1-\left(\frac{t\lambda}{2L}\right)^2}+2\,\text{arcsin}\left(\frac{t\lambda}{2L}\right)\right)~,\qquad \frac{t\lambda}{L}\text{  fixed.}}
\end{equation}
where the integration constant is determined by comparing with the exact expression. Note, crucially that this final large-$L$ spectral form factor behavior is only linear in $t$ for $t\lambda\ll 2L$. We will argue in the main text that the deviation from linearity is signalling a particular type of symmetry breaking. Haar random ensembles, for example, have an exactly linear spectral form factor. 

\subsubsection{Second derivation} 

We will now present an alternative route to computing the connected part of the two-point function explicitly, this will allow us to derive novel expressions. Recall that we wish to compute the following quantity \eqref{eq:conntwopt}: 
\begin{equation}
    \langle Z(\beta_1)Z(\beta_2)\rangle^c_{\rm GUE} \equiv\int \d E_1 \d E_2\, e^{-\beta_1 E_1-\beta_2 E_2}\langle\rho(E_1)\rho(E_2)\rangle^c_{\rm GUE}~, 
\end{equation}
with 
\begin{equation}
    \langle\rho(E_1)\rho(E_2)\rangle^c_{\rm GUE}\equiv-\sum_{j,k=0}^{L-1}\psi_j(E_1)\psi_j(E_2)\psi_k(E_1)\psi_k(E_2)~,
\end{equation}
with wavefunctions $\psi_j$ defined in \eqref{eq:GUEWavefunction}.
This time we will use the Christoffel-Darboux symbol to write this as: 
\begin{align}
\langle\rho(&E_1)\rho(E_2)\rangle^c_{\rm GUE}=-{\frac{h_L}{h_{L-1}}}\left[\frac{\psi_L(E_1)\psi_{L-1}(E_2)-\psi_L(E_2)\psi_{L-1}(E_1)}{E_1-E_2}\right]^2  ~,\nonumber\\
&=-\frac{e^{-L\frac{E_1^2+E_2^2}{2\lambda^2}}}{4^L\pi((L-1)!)^2} \left[\frac{H_L\left(\frac{E_1}{\lambda}\sqrt{\frac{L}{2}} \right)H_{L-1}\left(\frac{E_2}{\lambda}\sqrt{\frac{L}{2}} \right)-H_L\left(\frac{E_2}{\lambda}\sqrt{\frac{L}{2}} \right)H_{L-1}\left(\frac{E_1}{\lambda}\sqrt{\frac{L}{2}} \right)}{E_1-E_2}\right]^2~.\label{eq:2pthermiteexpr}
\end{align}
We would like to use the nice identity \eqref{eq:hermiteint}, but as we can already anticipate, the denominator will preclude us doing so without some further manipulations. We again proceed by defining $\frac{E_i}{\lambda}\sqrt{\frac{L}{2}}\equiv x_i$ and $\beta_i\equiv\frac{y_i}{\lambda}\sqrt{2L}$. What remains is to perform the following integrals:
\begin{multline}
    \langle Z(\beta_1)Z(\beta_2)\rangle^c_{\rm GUE} =\\-\frac{2^{-2L}}{\pi(L-1)!^2}\int \d x_1 \d  x_2\, e^{-x_1^2-x_2^2-2x_1y_1-2x_2y_2}\left[\frac{H_L(x_1)H_{L-1}(x_2)-H_L(x_2)H_{L-1}(x_1)}{x_1-x_2}\right]^2~. 
\end{multline} 
The trick comes in using the following identity: 
\begin{equation}
    \frac{1}{(x_1-x_2)^2}=-\int_0^\infty \d s\,  s\,e^{-is(x_1-x_2)}~. 
\end{equation}
The factor of $i$ in the exponent is very important, as it allows us to regulate this integral by giving $(x_1-x_2)$ a small imaginary part without worrying about whether $x_1$ is bigger or smaller than $x_2$. And with the above replacement, we can happily use \eqref{eq:hermiteint}. This gives:
\begin{align}\label{eq:connectedcompact}
   \langle Z(\beta_1)&Z(\beta_2)\rangle^c_{\rm GUE}= \frac{L}{2}e^{y_1^2+y_2^2}\int_0^\infty \d s\, s\,e^{-\frac{s^2}{2}+is(y_1-y_2)}\nonumber\\&\Bigg\lbrace L_{L-1}\left(-2\left[y_1+\frac{is}{2}\right]^2\right)L_{L}\left(-2\left[y_2-\frac{is}{2}\right]^2\right)+L_{L-1}\left(-2\left[y_2-\frac{is}{2}\right]^2\right)L_{L}\left(-2\left[y_1+\frac{is}{2}\right]^2\right)\nonumber\\
   &-\frac{(s-2iy_1)(s+2iy_2)}{L}L_{L-1}^1\left(-2\left[y_1+\frac{is}{2}\right]^2\right)L_{L-1}^1\left(-2\left[y_2-\frac{is}{2}\right]^2\right)\Bigg\rbrace~.
\end{align}
This is a nice, compact, and exact expression.  For analytically continued temperatures, $y_1=i\frac{t\lambda}{\sqrt{2L}}$, $y_2=-i\frac{t\lambda}{\sqrt{2L}}$, and again taking $s\rightarrow s\lambda\sqrt{\frac{2}{L}}$ in \eqref{eq:connectedcompact}, we can write the following expression: 
\begin{align*}
   \langle Z(it)Z(-it)\rangle_{\rm GUE}^c=& 2\lambda^2\int_0^\infty \d s\, s\,e^{-\frac{(t+s)^2\lambda^2}{L}}\nonumber\\&\Bigg\lbrace L_{L-1}\left(\frac{(t+s)^2\lambda^2}{L}\right)L_{L}\left(\frac{(t+s)^2\lambda^2}{L}\right)-\frac{(t+s)^2\lambda^2}{L^2}\left[L_{L-1}^1\left(\frac{(t+s)^2\lambda^2}{L}\right)\right]^2\Bigg\rbrace~.
\end{align*} Starting from this integral expression, one can rederive \eqref{eq:exactderivative} by taking a derivative with respect to $t$, rewriting it as a derivative with respect to $s$ inside the integral, then integrating by parts.

We end this appendix by noting that this exact integral expression \eqref{eq:connectedcompact} for the connected GUE two-point function can be brought to a similar form as \eqref{eq:QG1pt}; that is, we can write it as a smearing over temperature of the infinite-$L$ disconnected one-point functions. To do so, we first change integration variables in \eqref{eq:connectedcompact}  $s\rightarrow s\lambda\sqrt{\frac{2}{L}}$, and use the following Laguerre-polynomial identities:
\begin{align}
L_n(x)&=\frac{1}{n+1}\frac{\d }{\d x}\left(x L_n^1(x)\right)~,\\
x L_n^{\alpha+1}(x)&=(n+\alpha)L_{n-1}^{\alpha}(x)-(n-x)L_{n}^\alpha(x) ~,\\
n L_n^{\alpha+1}(x)&=(n-x)L_{n-1}^{\alpha+1}(x)+(n+\alpha)L_{n-1}^\alpha(x) ~,
\end{align}
 which allows us to write the above expression as follows
\begin{align}
   \langle Z(\beta_1)Z(\beta_2)\rangle^c_{\rm GUE}= 
   \lambda^2\int_0^\infty & s \,\d s \frac{\left(\beta_1+is\right)\left(\beta_2-is\right)}{2L^2}\Bigg\lbrace\left[\partial_{\beta_1}+\left(\frac{\beta_1+is}{2}\right)^{-1}\right]\left[\partial_{\beta_2}+\left(\frac{\beta_2-is}{2}\right)^{-1}\right]\nonumber\\ &-2\lambda^2\left[2+\frac{\left(\beta_1+is\right)\left(\beta_2-is\right)\lambda^2}{2L^2}\right]\Bigg\rbrace\langle Z(\beta_1+is)\rangle_{\rm GUE} \langle Z(\beta_2-is)\rangle_{\rm GUE}~.
\end{align}
We have derived, in a precise sense, how the exact connected GUE two-point function can be obtained by correlating two disconnected disc-amplitudes via a particular sewing proceedure.\footnote{In particular this is reminiscent of the loop equations leading up to \eqref{eq:looppot}.} If we wish, using \eqref{eq:QG1pt}, we could interpret this as a particular quantum gravity path integrals over the Euclidean times of the disconnected infinite-$L$ one-point functions, along with an integral over a `wormhole' parameter $s$. We provide this expression here in gory detail: 
\begin{align}\label{eq:QG2pt}
  &\langle Z(\beta_1)Z(\beta_2)\rangle^c_{\rm GUE}=2\lambda^2\left(\frac{2 L^{L+2}}{L!}\right)^2\int_0^\infty s\d s
  \,e^{-\frac{(\beta_1+i s)^2\lambda^2}{2 L}-\frac{(\beta_2-i s)^2\lambda^2}{2 L}}\int_0^\infty \frac{\d \tilde{\beta}_1}{\tilde{\beta}_1}\int_0^\infty \frac{\d \tilde{\beta}_2}{\tilde{\beta_2}}\,\nonumber\\ 
   &\times\left[\left(1-\left(\frac{\tilde{\beta}_1}{\beta_1+is}\right)^2+\frac{(\beta_1+is)^2\lambda^2}{2L^2}\right)\left(1-\left(\frac{\tilde{\beta}_2}{\beta_2-is}\right)^2+\frac{(\beta_2-is)^2\lambda^2}{2L^2}\right)\right.\nonumber\\
   & \qquad \qquad \qquad \qquad\qquad \qquad \qquad \qquad~~\left.-\frac{(\beta_1+is)(\beta_2-is)\lambda^2}{L^2}\left(1+\frac{(\beta_1+is)(\beta_2-is)\lambda^2}{4L^2}\right)\right]\nonumber\\
   &\times e^{-L\left(\frac{\tilde{\beta}_1}{\beta_1+is}\right)^2-L\left(\frac{\tilde{\beta}_2}{\beta_2-is}\right)^2+2(L+1)\left[\log\frac{\tilde{\beta}_1}{\beta_1+is}+\log\frac{\tilde{\beta}_2}{\beta_2-is}\right]}\left\langle {Z}_{\rm eq}\left(\tilde{\beta}_1\right)\right\rangle_{\rm GUE}\left\langle {Z}_{\rm eq}\left(\tilde{\beta}_2\right)\right\rangle_{\rm GUE}\nonumber\\
\end{align}


\subsection{Saddle analysis of one- and two-point functions}

\subsubsection{One-point function}\label{ap:saddleanalysisonepoint}
We now have all the pieces ready to understand all the qualitative features of the one-point function. In this section, we will analytically continue to $\beta=it$ from the get-go. Recall for imaginary temperatures, we have the following expression for the exact one-point function:
\begin{equation}\label{eq:ZHMMexactLor}
  \langle Z(it)\rangle_{\rm GUE}={e^{-\frac{(t \lambda)^2}{2L}}}L_{L-1}^1\left(\frac{(t\lambda)^2}{L}\right),
\end{equation}
and if we take $L\rightarrow\infty$ while keeping $t\lambda$ fixed, we found: 
\begin{equation}\label{eq:powerlawdecay2}
   \lim_{L\rightarrow\infty}\frac{\langle Z(it)\rangle_{\rm GUE}}{L}\equiv \langle {Z}_{\rm eq}(it)\rangle_{\rm GUE}=\frac{J_1(2t\lambda)}{t \lambda}~. 
\end{equation}
Moreoever, using the asymptotic behavior of the Bessel function, we can conclude that this profile decays as: 
\begin{equation}
   \lim_{t\lambda\rightarrow\infty}\langle {Z}_{\rm eq}(it)\rangle_{\rm GUE}\approx\frac{\pi\,\sin\left(2 t\lambda -\frac{\pi}{4}\right)}{(\pi t\lambda)^{3/2}}~,
\end{equation}
as we showed in figure \ref{fig:symbreaking1pt}.
However, we know this polynomial decay can not last forever. In fact, we know that when $t\lambda$ starts becoming sensitive to the $L$-scale, the expression \eqref{eq:ZHMMexactLor} will be best approximated by it's leading polynomial term, that is
\begin{equation}\label{eq:exponentialdecay}
   \lim_{t\rightarrow\infty}\frac{\langle {Z}(it)\rangle_L}{L}\approx \frac{e^{-L-\frac{(t\lambda)^2}{2L}}}{L!}\left(-\frac{(t\lambda)^2}{L}\right)^{L-1}~.
\end{equation}
In a theory where $Z$ is a charged operator, we would claim that this behavior is reminsicent of symmetry restoration. Of course, this physics is non-perturbative, but it would be nice to have a picture for how these two behaviors can emerge. What we should expect to find, is that the polynomial decay (with an oscillatory prefactor) comes from the physics governed by the Wigner semi-circle. The late-time exponential decay tells us that this approximation is bad at $t\lambda \sim L^\gamma$ (for some undetermined $\gamma$), at which point non-perturbative physics kicks in.

Since we can solve this model exactly, let us see how this arises for this ensemble. Recall that the exact density of states is 
\begin{equation}\label{eq:GUEdensity}
     \langle\rho(E)\rangle_{\rm GUE}=
     \frac{1}{\lambda}\frac{e^{-\frac{L}{2}\left(\frac{E}{\lambda}\right)^2}}{2^L(L-1)!}\sqrt{\frac{L}{2\pi}}\left[H_L\left(\frac{E}{\lambda}\sqrt{\frac{L}{2}}\right)^2-H_{L-1}\left(\frac{E}{\lambda}\sqrt{\frac{L}{2}}\right)H_{L+1}\left(\frac{E}{\lambda}\sqrt{\frac{L}{2}}\right)\right]~, 
\end{equation}
and that the quantity we want to compute is: 
\begin{equation}
   \langle Z(it)\rangle_{\rm GUE}=\int_{-\infty}^\infty\d E\, \langle\rho(E)\rangle_{\rm GUE}\, e^{-i E t}~.
\end{equation}
We will also use the integral representation of the Hermite polynomials: 
\begin{equation}
   H_n(x)=\frac{2^n}{\sqrt{\pi}}\int_{-\infty}^\infty \d y(x+i y)^ne^{-y^2}~.
\end{equation}
Using the above integral expression and defining a dimensionless energy $e\equiv \frac{E}{\lambda}$ we are left to compute:\footnote{We have also taken $y_i\rightarrow y_i\sqrt{\frac{L}{2}}$ in the integral definition of the Hermite polynomials and symmetrize the integral representations.} 
\begin{multline}\label{eq:integralexpressionexact1pt}
   \langle Z(it)\rangle_{\rm GUE}=\left(\frac{L}{2\pi}\right)^{3/2}\frac{L^L}{2(L-1)!}\int_{-\infty}^\infty\d e\int_{-\infty}^\infty\d y_1\int_{-\infty}^\infty\d y_2 (e+iy_1)^{L-1}(e+iy_2)^{L-1}\\
   \times\left(y_1-y_2\right)^2 e^{-ie(t\lambda) -\frac{L}{2}\left[e^2+y_1^2+y_2^2\right]}~.
\end{multline}
We are free to take $L\rightarrow\infty$, we can now approximate this integral by saddle point methods. 

\paragraph{Saddle point analysis for $t\lambda\sim 1$:}\mbox{}\\
Let us first assume that $t\lambda\sim 1$. The Large-$L$ saddle point equations obtained from varying with respect to $y_i$ are
\begin{equation}\label{eq:yisaddleeq}
    \qquad y_1=\frac{i}{e+iy_1}~, \qquad y_2=\frac{i}{e+iy_2}~,
\end{equation}
while the saddle point equation that comes from varying with respect to $e$ is
\begin{equation}\label{eq:esaddleeq}
   e=\frac{1}{e+iy_1}+\frac{1}{e+iy_2}~.
\end{equation}
It is straightforward to check that the solutions to \eqref{eq:yisaddleeq} are 
\begin{equation}
   y_1^*=i \frac{e}{2}+\sigma_1\sqrt{1-\left(\frac{e}{2}\right)^2}~,\qquad\qquad y_2^*=i \frac{e}{2}+\sigma_2\sqrt{1-\left(\frac{e}{2}\right)^2}~,\qquad \sigma_i=\pm1~.
\end{equation}
In the end we will be integrating over $e$, so it won't be necessary to also satisfy \eqref{eq:esaddleeq}, however, note that there are two saddles that satisfy this equation, namely the saddles for which $\sigma_1=-\sigma_2$. Crucially, these two solutions with anti-correlated signs are the leading solutions, due to the explicit $(y_1-y_2)^2$ prefactor in the integrand \eqref{eq:integralexpressionexact1pt} which goes to zero when $\sigma_1=\sigma_2$. This is a precise signal of causal symmetry-breaking found in the leading saddle discussed in \cite{Altland:2020ccq}. 

It is easy to see that we must sum over all choices of $\sigma_i$. The two saddles along the $y_1$ (resp. $y_2$) contour have the same imaginary part, but opposite real part (for $|e|\leq 2$), meaning that we can deform the  $y_1$ (resp. $y_2$) contour such that we pick up the contributions from both saddles with each choice of sign. 

Let us consider the leading saddles, namely the ones for which $\sigma_1=-\sigma_2=\pm1$.  Performing the Guassian integrals over the fluctuations and summing over the two saddles, we find: 
\begin{equation}
    \frac{\langle Z(it)\rangle_{\rm GUE}}{L}\approx \frac{1}{\pi\lambda}\int_{-2\lambda}^{2\lambda} \d E\, e^{-iEt}\sqrt{1-\left(\frac{E}{2\lambda}\right)^2}= \frac{J_1(2t\lambda)}{t\lambda}
 \end{equation} 
 as expected from the discussion around \eqref{eq:powerlawdecay}. Note that we returned to the $E$ parametrization for this final step, to make the matching with previous expressions more explicit, and we have restricted the integral over $E$ to the set $[-2\lambda, 2\lambda]$ as this is where our saddle-point analysis is valid. For $E$ outside of this range, the $y_{1,2}$ saddles are purely imaginary, and our above analysis breaks down---not to fret as the contributions from this region are suppressed for small enough $t\lambda$. 

 If we go back and include the two remaining subleading saddles, the ones with \emph{correlated} signs ($\sigma_1=\sigma_2$), we find the precise correction that gives rise to the wiggles on top of the semi-circle, namely: 
 \begin{equation}
    \frac{ \langle Z(it)\rangle_{\rm GUE}}{L}\approx \frac{1}{\pi\lambda}\int_{-2\lambda}^{2\lambda} \d E e^{-iEt}\left\lbrace\sqrt{1-\left(\frac{E}{2\lambda}\right)^2}-\frac{\cos\left(2L\left[\frac{E}{2\lambda}\sqrt{1-\left(\frac{E}{2\lambda}\right)^2}-\cos^{-1}\left(\frac{E}{2\lambda}\right)\right]\right)}{4L\left[1-\left(\frac{E}{2\lambda}\right)^2\right]}\right\rbrace~.
 \end{equation}
Recalling the definition of ${\rho}_{\rm eq}(E)$ from \eqref{eq:rhoinfinitydef}, we can write the answer as 
\begin{equation}
     \frac{\langle Z(it)\rangle_{\rm GUE}}{L}\approx \int_{E_c}^{E_c} \d E e^{-iEt}\left\lbrace\rho_{\rm eq}(E)-\frac{\cos\left(2L\left[\frac{\pi E\rho_{\rm eq}(E)}{2}-\cos^{-1}\left(\frac{E}{2\lambda}\right)\right]\right)}{4L\pi^3\lambda^3\rho_{\rm eq}(E)^2}\right\rbrace~,
 \end{equation}
 with $E_c$ the energy where the term in braces above vanishes. 

 Thus far we have derived the early-time behavior of this one-point function, where the $\langle Z(it)\rangle_{\rm GUE}$ decays polynomially. Once $t\lambda\sim L$, we saw that the behvior of this function changes. In the next section we will analyze the origin for this change of behavior.

\paragraph{Saddle point analysis for $t\lambda\sim L$:}\mbox{}\\
It is instructive to repeat the above analysis for $t\lambda\sim L$ in order to see how the expected exponential decay \eqref{eq:exponentialdecay} arises from a saddle point analysis. To make this explicit, we will now take $t\equiv L \tilde{t}$ and keep $\tilde{t}$ fixed and of order 1. In this regime, the saddle point equations are
\begin{equation}
   e=\frac{1}{e+iy_1}+\frac{1}{e+iy_2}-i\tilde{t}\lambda~, \qquad y_1=\frac{i}{e+iy_1}~, \qquad y_2=\frac{i}{e+iy_2}~.
\end{equation}
with solutions
\begin{equation}
   e^*=\pm i\sqrt{\left(\tilde{t}\lambda\right)^2-4}~, \qquad\qquad y_{1,2}^*=-\frac{1}{2}\left(\tilde{t}\lambda\pm \sqrt{\left(\tilde{t}\lambda\right)^2-4}\right)~,
\end{equation}
and we will assume, from now on, that $\tilde{t}\lambda>2$~. A few things are of note at this stage. Unlike the previous case, here all the signs are correlated, meaning $(y_1-y_2)=0$ on the saddle, and we will have to carefully analyze the fluctuation integral. Secondly, the saddles of the $e$ integral are purely imaginary, so we cannot deform the contour, which sits along the real line, to pass through both of them. In this case we are instructed to choose the $(-)$ saddle since it gives a decaying contribution to the one-point function. 

It is a straightforward, albeit a bit tedious to perform the fluctuation integrals in this regime,\footnote{Explicitly we have Gaussian integral that looks as follows: \begin{equation}
   \int d\vec{y} e^{-\frac{1}{2}\vec{y}\cdot M\cdot \vec{y}}=\text{det}{}^{-1/2}\left(\frac{M}{2\pi}\right)
\end{equation}
with $M$ not diagonal. So performing the fluctuation integral involves finding this matrix $M$, and computing its determinant. This is straightforward, but we skip writing down the explicit steps.} but the end result is
\begin{equation}
  \frac{\langle Z(it)\rangle_{\rm GUE}}{L}\underset{t\lambda\gtrsim 2L}{\approx} \frac{e^{-L-{t\lambda}\sqrt{\left(\frac{t\lambda}{2L}\right)^2-1}}}{L!}\left(L-{t\lambda}\left(\frac{t\lambda}{2L}+\sqrt{\left(\frac{t\lambda}{2L}\right)^2-1}\right)\right)^{L-1}~, 
\end{equation}
which for $t\lambda\gg 2L$ gives: 
\begin{equation}
  \frac{\langle Z(it)\rangle_{\rm GUE}}{L}\underset{t\lambda\gg 2L}{\approx} \frac{e^{-\frac{(t\lambda)^2}{2L}}}{L!}\left(-\frac{(t\lambda)^2}{L}\right)^{L-1}~.\label{eq:GUE1ptlatetimebehavior}
\end{equation}
which is exactly what we saw in figure \ref{fig:symbreaking1pt}. Interestingly, this analysis shows that we expect this behavior to set in when $t\lambda > 2L$. 

\subsubsection{Two-point function}
We can now analyze the two-point function in the same way. Starting from \eqref{eq:2pthermiteexpr} and using the integral representation of the Hermite polynomials, we find
\begin{align}
   \langle Z(it)&Z(-it)\rangle^c_{{\rm GUE}}=-\left(\frac{L}{2\pi}\right)^{3}\left(\frac{L^{L-1}}{(L-1)!}\right)^2\int_{-\infty}^\infty\d e_1\int_{-\infty}^\infty\d e_2\int_{-\infty}^\infty\d y_1\int_{-\infty}^\infty\d y_2 \int_{-\infty}^\infty\d y_3\int_{-\infty}^\infty\d y_4 \nonumber\\
   &\times(e_1+iy_1)^{L-1}(e_1+iy_2)^{L-1}(e_2+iy_3)^{L-1}(e_2+iy_4)^{L-1}e^{-i(t\lambda)(e_1-e_2) -\frac{L}{2}\left[e_1^2+e_2^2+y_1^2+y_2^2+y_3^2+y_4^2\right]}\nonumber\\
   &\times\left(1+i\frac{y_1+y_2-y_3-y_4}{e_1-e_2}+\frac{\tfrac{1}{2}(y_1+y_2)(y_3+y_4)-(y_1y_2+y_3y_4)}{(e_1-e_2)^2}\right)~,
\end{align}
where $e_i\equiv E_i/\lambda$ as before. 
Conceptually, this integral is no different than \eqref{eq:integralexpressionexact1pt}, so it's amenable to the exact same analysis. 
\paragraph{Saddle point analysis for $t\lambda\sim 1$:}\mbox{}\\
As in the case of the one-point function, let us first assume that $t\lambda\sim 1$.
The Large-$L$ saddle point equations from varying with respect to $y_i$ are
\begin{equation}\label{eq:yisaddleeq2pt}
    \qquad y_1=\frac{i}{e_1+iy_1}~, \qquad y_2=\frac{i}{e_1+iy_2}~,\qquad y_3=\frac{i}{e_2+iy_3}~,\qquad y_4=\frac{i}{e_2+iy_4}~.
\end{equation}
As before the solutions to these equations are
\begin{equation}\label{eq:2ptsaddles}
   y_{1,2}^*=i \frac{e_1}{2}+\sigma_{1,2}\sqrt{1-\left(\frac{e_1}{2}\right)^2}~,\qquad\qquad y_{3,4}^*=i \frac{e_2}{2}+\sigma_{3,4}\sqrt{1-\left(\frac{e_2}{2}\right)^2}~,\qquad \sigma_i=\pm1~. 
\end{equation}
The equations that come from varying with respect to $e_{1,2}$ are 
\begin{equation}\label{eq:esaddleeq2pt}
e_1=\frac{1}{e_1+iy_1}+\frac{1}{e_1+iy_2}~,\qquad e_2=\frac{1}{e_2+iy_3}+\frac{1}{e_2+iy_4}~,
\end{equation}
and among the saddle point solutions \eqref{eq:2ptsaddles}, only those with $\sigma_1=-\sigma_2$ and $\sigma_3=-\sigma_4$ satisfy \eqref{eq:esaddleeq2pt}---again evidence of causal symmetry breaking. To simplify notation we will now switch to another parametrization. As a leading approximation, we will take the $e_i$ to be valued in $[-2,2]$ as so happens when the Wigner semi-cirlce appears. We will therefore change variables to 
\begin{equation}
   e_1\equiv2\sin\theta_1~\qquad e_2\equiv2\sin\theta_2~\qquad \theta_i\in\left[-\frac{\pi}{2},\frac{\pi}{2}\right]~.
\end{equation}
Using this parametrization, the saddle points can be expressed as
\begin{equation}\label{eq:2ptsaddlesv2}
   y_{1,2}^*=\sigma_{1,2}\,e^{i\sigma_{1,2}\,\theta_1}~,\qquad\qquad y_{3,4}^*=\sigma_{3,4}\,e^{i\sigma_{3,4}\,\theta_2}~,\qquad \sigma_i=\pm1~, 
\end{equation}
which saves us the trouble of dealing with the cumbersome square-roots. Performing the fluctuation integral around these saddle points, we arrive at a compact expression:
\begin{align}
  \langle Z(it)Z(-it)\rangle_{{\rm GUE}}^c\approx-&\int_{-\frac{\pi}{2}}^{\frac{\pi}{2}}\d\theta_1 \int_{-\frac{\pi}{2}}^{\frac{\pi}{2}}\d\theta_2\,\frac{e^{-2i (t\lambda)(\sin\theta_1-\sin\theta_2)}}{\pi^2(\sin\theta_1-\sin\theta_2)^2}\times\nonumber\\&\left[\sin\left(\frac{\theta_1-\theta_2}{2}\right)\cos(L[\theta_1+\theta_2+\cos\theta_1\sin\theta_1+\cos\theta_2\sin\theta_2])\right.\nonumber\\&\left.-(-1)^L\cos\left(\frac{\theta_1+\theta_2}{2}\right)\sin(L[\theta_1-\theta_2+\cos\theta_1\sin\theta_1-\cos\theta_2\sin\theta_2])\right]^2~.
\end{align}
This is an expression for the connected two-point density of states that is valid for $t\lambda \sim 1$ and $L$ large.  that includes the curvature of the density of states. If we return to the energy representation $2\sin\theta_i\equiv E_i/\lambda$. Now we see that, at the center of the band, the sine-kernel emerges. The above integral can be written as
\begin{equation}
   \langle Z(it)Z(-it)\rangle_{{\rm GUE}}^c\approx\int_{-2\lambda}^{2\lambda}\d E_1 \int_{-2\lambda}^{2\lambda}\d E_2\,{e^{-i t(E_1-E_2)}}\langle\rho(E_1)\rho(E_2)\rangle_{{\rm eq}}^c
\end{equation}
and for $E_i=x_i/L$ and $L\rightarrow\infty$, we see
\begin{equation}
   \langle\rho(E_1)\rho(E_2)\rangle_{{\rm eq}}^c\approx -\frac{L^2\sin^2\left(\frac{x_1-x_2}{\lambda}\right)}{\pi^2(x_1-x_2)^2}~.
\end{equation}
\paragraph{Saddle point analysis for $t\lambda\sim L$:}\mbox{}\\
Now we repeat the analysis for $t\lambda\sim L$, taking $t\equiv L \tilde{t}$ with $\tilde{t}$ or order 1. the saddle point equations are
\begin{equation}
   e_1=\frac{1}{e_1+iy_1}+\frac{1}{e_1+iy_2}-i\tilde{t}\lambda~,\qquad  e_2=\frac{1}{e_2+iy_3}+\frac{1}{e_2+iy_4}+i\tilde{t}\lambda~,  
\end{equation}
and the equations for the $y_i$ are as written in \eqref{eq:yisaddleeq2pt}. Now the saddles are 
\begin{equation}
   e_1^*= i\sigma_1\sqrt{\left(\tilde{t}\lambda\right)^2-4}~,\qquad e_2^*= i\sigma_2\sqrt{\left(\tilde{t}\lambda\right)^2-4}~, 
\end{equation}
and 
\begin{equation}
   y_{1,2}^*=-\frac{1}{2}\left(\tilde{t}\lambda+\sigma_1 \sqrt{\left(\tilde{t}\lambda\right)^2-4}\right)~,\qquad y_{3,4}^*=\frac{1}{2}\left(\tilde{t}\lambda-\sigma_2 \sqrt{\left(\tilde{t}\lambda\right)^2-4}\right)~.
\end{equation}
A straightforward analysis reveals that only saddle that contributes is $(\sigma_1,\sigma_2)=(-1,1)$. It is also straightforward to compute the fluctuation integral about this saddle and we find:
\begin{equation}
  {\langle {Z}(it)Z(-it)\rangle_{\rm GUE}}\underset{t\lambda> 2L}{\approx}- L^{2L}\frac{e^{-2L-2t\lambda
\sqrt{\left(\frac{t\lambda}{2L}\right)^2-1}}}{16 (L!)^2}\frac{\left(\frac{t\lambda}{2L}+\sqrt{\left(\frac{t\lambda}{2L}\right)^2-1}\right)^{4L}}{\frac{t\lambda}{2L}\left(\left(\frac{t\lambda}{2L}\right)^2-1\right)^{3/2}}~.
\end{equation}
For $t\lambda\gg 2L$, this simplifies to
\begin{equation}
{\langle {Z}(it)Z(-it)\rangle_{\rm GUE}}\underset{t\lambda\gg 2L}{\approx} -L^2\frac{e^{-\frac{(t\lambda)^2}{L}}}{(L!)^2}\left(\frac{(t\lambda)^2}{L}\right)^{2(L-1)}
\end{equation}
and note that this implies, given \eqref{eq:GUE1ptlatetimebehavior}, that  
\begin{equation}\label{eq:asymptoticratioGUE}
\langle Z(it)Z(-it)\rangle_{\rm GUE}\underset{t\lambda\gg L}{\approx}-\left(\langle Z(it)\rangle_{\rm GUE}\right)^2~.
\end{equation}
Hence for the exit from the ramp of the GUE, we should expect the same behavior, up to a power of two, as the exponential decay of the one-point function. 

\section{Orthogonal polynomials of the GWW model}\label{ap:GWWorthogonalpolynomials}

Recall that we can reduce the GWW model to an integral over the eigenvalues of $U$: $e^{i\theta_j}$, for $j=1,\dots,L$. We wish to compute expectation values as follows: 
\begin{equation}\label{eq:GWWexpectation}
   \langle\cdot\rangle_{\rm GWW}=\frac{1}{\mathcal{N}_{\rm GWW}}\int_{-\pi}^\pi \left[\prod_{i=1}^L \frac{\d\theta_i}{2\pi}\right]\, \left|\Delta\left(e^{i\theta}\right)\right|^2 (\cdot)\exp\left\lbrace L g\sum_{i=1}^L \cos \theta_i\right\rbrace~,
\end{equation}
where 
\begin{equation}
   \Delta\left(e^{i\theta}\right)\equiv \prod_{k<j}^L \left(e^{i\theta_j}-e^{i\theta_k}\right)=\text{det}\left[e^{i(j-1)\theta_k}\right]~.
 \end{equation}
 To do this, it will be useful to introduce the orthogonal polynomials for this ensemble. These were discovered in \cite{Goldschmidt:1979hq}. Typically orthogonal polynomials are designed such that their leading polynomial has unit coefficient: 
\begin{equation}
P_j(z)=z^j+\sum_{i=1}^{j-1}a_i z^i~.
\end{equation}
 As such, they are not orthonormal, and rather satisfy 
 \begin{equation}
    \int_{-\pi}^\pi \frac{\d \theta}{2\pi}\,e^{L g\, \cos \theta}P_j\left(e^{i \theta}\right)P_k^*\left(e^{i \theta}\right)=h_j\delta_{jk}~,
 \end{equation}
 with the $h_j$ to be determined. 
 To define these orthogonal polynomials, first note the identity:
\begin{equation}\label{eq:besselintegral2}
   I_n(Lg)=\int_{-\pi}^\pi \frac{\d \theta}{2\pi}e^{L g\cos\theta}e^{in\theta}~,
\end{equation}
where $I_n(x)$ is the modified Bessel function.  The orthogonal polynomials for this ensemble are then
 \begin{equation}
    P_n(z)=\frac{1}{c_n}\text{det}\begin{pmatrix}
    1 & z & z^2 &\dots & z^n\\
    I_0(Lg) & I_1(Lg)& I_2(Lg) &\dots & I_n (Lg)\\
    I_1(Lg) & I_0(Lg)& I_1(Lg) &\dots & I_{n-1} (Lg)\\
\vdots & \vdots &\vdots &\ddots & \vdots\\
I_{n-1}(Lg) &\dots & \dots& I_0(Lg) & I_1(Lg)
   \end{pmatrix}~,
 \end{equation}
 and defined such that $P_0(z)$=1 and, similarly, we have defined the coefficients $c_n$ such that
 \begin{equation}\label{eq:cdef}
    c_0=1~, \qquad\qquad c_{n\geq1}=\text{det}[I_{k-l}(Lg)]_{k,l=1,\dots,n}~.
 \end{equation}
The coefficients $h_n$ are then found to be: 
 \begin{equation}
    h_n=c_{n+1}/c_n~. 
 \end{equation}
 These orthogonal polynomials satisfy the following recursion relation
 \begin{equation}
    P_{n+1}(z)=z P_n(z)+(-1)^{n+1}z^n P_n\left(\frac{1}{z}\right)\sqrt{1-\frac{h_{n+1}}{h_n}}
 \end{equation}
 The choice of leading coefficient in $P_j(z)$ allows us to write \cite{Meh2004}: 
 \begin{equation}
    \Delta\left(e^{i\theta}\right)=\text{det}\left[e^{i(j-1)\theta_k}\right]=\text{det}\left[P_{j-1}\left(e^{i\theta_k}\right)\right]~,
 \end{equation}
 \paragraph{Normalization:}
 The above exposition allows us to readily compute the normalization $\mathcal{N}_{\rm GWW}$ (see for example \cref{ap:norm})
 \begin{equation}\label{eq:normalization}
    \mathcal{N}_{\rm GWW}=L!\prod_{n=0}^{L-1} h_n=L!\times c_L~, 
 \end{equation}
 where $c_L$ is defined in \eqref{eq:cdef}. This has the property that $\left.\mathcal{N}_{\rm GWW}\right\rvert_{g=0}=L!$~.
 
\begin{figure}[t!]
\begin{center}
\includegraphics[height=4.1cm]{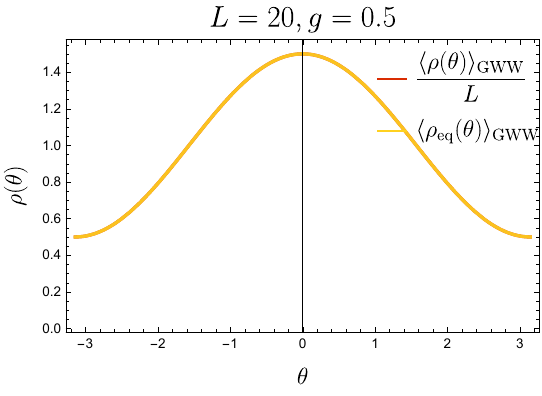}
\includegraphics[height=4.1cm]{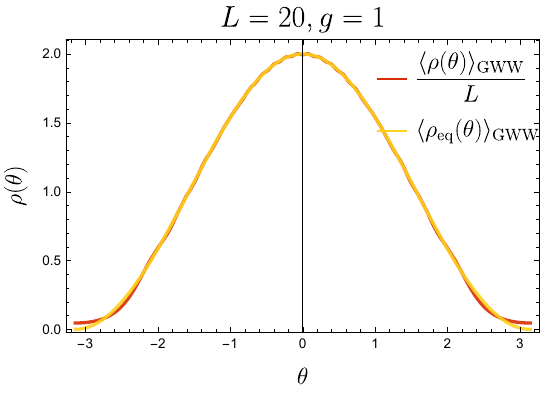}
\includegraphics[height=4.1cm]{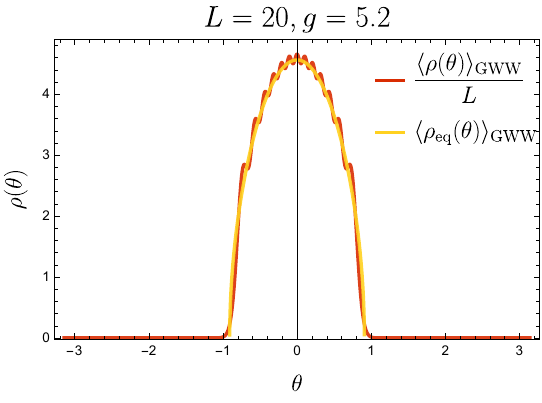}
\end{center}
\caption{Comparison between exact ($L=20$) and equilibrium density of states for various $g$~.}\label{fig:densitycompare}
\end{figure}

\paragraph{One-point function:}
 Defining the `wavefunctions' $\psi_n$: 
 \begin{equation}
    \psi_n(\theta)= \frac{P_n\left(e^{i\theta}\right)}{\sqrt{h_n}}e^{\frac{L g}{2}\cos\theta}~, 
 \end{equation}
 which satisfy
 \begin{equation}
 \int_{-\pi}^\pi \frac{\d \theta}{2\pi}\,\psi_i(\theta)\psi_j^*(\theta)=\delta_{ij}~,
 \end{equation}
 then the exact expression for the Polyakov loop is (see \cref{sec:1pfctgene} for a derivation): 
 \begin{equation}\label{eq:onepointGWW}
    \left\langle \text{Tr}\, U^n\right\rangle_{\rm GWW}=\int_{-\pi}^\pi \frac{\d \theta}{2\pi}e^{i n \theta}\langle\rho(\theta)\rangle_{\rm GWW}~,  
 \end{equation}
 where  
 \begin{equation}
    \langle\rho(\theta)\rangle_{\rm GWW}\equiv \sum_{j=0}^{L-1}\psi_j(\theta)\psi_j^*(\theta)~.
 \end{equation}
 Because the wavefunctions are unit-normalized, we have that
 \begin{equation}
    L=\int_{-\pi}^\pi\frac{\d\theta}{2\pi} \langle\rho(\theta)\rangle_{\rm GWW}~. 
 \end{equation}
 Although it is not obvious, from the derivation, a nice consistency check is that 
 \begin{equation}
 \lim_{L\rightarrow\infty}\frac{\langle\rho(\theta)\rangle_{\rm GWW}}{L}=\langle\rho_{\rm eq}(\theta)\rangle_{\rm GWW}~.
 \end{equation}
with given $\langle\rho_{\rm eq}(\theta)\rangle_{\rm GWW}$ in \eqref{eq:largegedensity}, as can be determined visually in figure \ref{fig:densitycompare}.

\paragraph{Two-point function:} Similarly, for the two-point function we have: 
 \begin{equation}\label{eq:twopointGWW}
     \left\langle \text{Tr}\, U^n\text{Tr}\, U^m\right\rangle_{\rm GWW}=\left\langle \text{Tr}\, U^{n+m}\right\rangle_{\rm GWW}+\int_{-\pi}^\pi \frac{\d \theta_1}{2\pi}\int_{-\pi}^\pi \frac{\d \theta_2}{2\pi}e^{i n \theta_1+im\theta_2}\langle\rho(\theta_1)\rho(\theta_2)\rangle_{\rm GWW}~,  
 \end{equation}
 where
 \begin{equation}\label{eq:rhorhotheta}
   \langle\rho(\theta_1)\rho(\theta_2)\rangle_{\rm GWW} = \det \begin{pmatrix} K_L(\theta_1,\theta_1) & K_L(\theta_1,\theta_2)\\ K_L(\theta_2,\theta_1) & K_L(\theta_2,\theta_2)\end{pmatrix}
\end{equation}
and
\begin{equation}
   K_L(\theta_1,\theta_2)\equiv\sum_{j=0}^{L-1}\psi_j(\theta_1)\psi_j^*(\theta_2)~.
\end{equation}
Expanding out \eqref{eq:rhorhotheta}, we may write
\begin{equation}
  \langle\rho(\theta_1)\rho(\theta_2)\rangle_{\rm GWW} =  \langle\rho(\theta_1)\rangle_{\rm GWW}\langle\rho(\theta_2)\rangle_{\rm GWW}+\langle\rho(\theta_1)\rho(\theta_2)\rangle^{c}_{\rm GWW}~,
\end{equation}
with 
\begin{equation}
   \langle\rho(\theta_1)\rho(\theta_2)\rangle_{\rm GWW}^c\equiv-K_L(\theta_1,\theta_2)K_L(\theta_2,\theta_1)~.
\end{equation}
This implies that we can write
\begin{equation}
 \left\langle \text{Tr}\, U^n\text{Tr}\, U^m\right\rangle_{\rm GWW}=\left\langle \text{Tr}\, U^{n+m}\right\rangle_{\rm GWW}+\left\langle \text{Tr}\, U^{n}\right\rangle_{\rm GWW}\left\langle \text{Tr}\, U^{m}\right\rangle_{\rm GWW}+\left\langle \text{Tr}\, U^n\text{Tr}\, U^m\right\rangle_{\rm GWW}^c~,
\end{equation}
just as we saw in \eqref{eq:GWW2pt} and where we have defined
\begin{equation}
\left\langle \text{Tr}\, U^n\text{Tr}\, U^m\right\rangle_{\rm GWW}^c\equiv\int_{-\pi}^\pi \frac{\d \theta_1}{2\pi}\int_{-\pi}^\pi \frac{\d \theta_2}{2\pi}e^{i n \theta_1+im\theta_2}\langle\rho(\theta_1)\rho(\theta_2)\rangle^c_{\rm GWW}~.
\end{equation}

For the circular ensembles, the orthogonal polynomials $P_j(z)$ admit a Christoffel-Darboux formula \cite{forrester2010log}:
\begin{equation}
\sum_{j=0}^{L-1} P_j(z)P_j^*(\zeta)=\frac{z^L \zeta^L\left(P_L(z)\right)^* P_L\left(\frac{1}{\zeta}\right)-P_L(z)\left( P_L\left(\frac{1}{\zeta}\right)\right)^*}{1-z\zeta}~,
\end{equation}
where we are considering variables on the unit-circle such that $ z^*=1/z$ and $\zeta^*=1/\zeta$~.
This implies
\begin{equation}
   K_L(\theta_1,\theta_2)=e^{i(L-1) \left(\frac{\theta_1-\theta_2}{2}\right)}\frac{e^{-iL \left(\frac{\theta_1-\theta_2}{2}\right)}\psi_L(\theta_1)\psi_L^*(\theta_2)-e^{iL \left(\frac{\theta_1-\theta_2}{2}\right)}\psi_L^*(\theta_1)\psi_L(\theta_2)}{2i\sin\left(\frac{\theta_1-\theta_2}{2}\right)}~.
\end{equation}
From this we can derive that 
\begin{equation}\label{eq:exactdensityGWW}
   \langle\rho(\theta)\rangle_{\rm GWW}=i\left[\psi_L(\theta)\partial_\theta\psi_L^*(\theta)-\psi_L^*(\theta)\partial_\theta\psi_L(\theta)\right]-L\, \psi_L(\theta)\psi_L^*(\theta)~.
\end{equation}

\bibliographystyle{utphys}
\bibliography{ssbrefs.bib}{}

\providecommand{\href}[2]{#2}\begingroup\raggedright\begin{thebibliography}{10}

\bibitem{KitaevTalks}
A.~Kitaev, ``{A simple model of quantum holography},'' {\em KITP strings
  seminar and Entanglement program (Feb. 12, April 7, and May 27,)} (2015) .
  \url{http://online.kitp.ucsb.edu/online/entangled15/}.

\bibitem{Anninos:2013nra}
D.~Anninos, T.~Anous, P.~de~Lange, and G.~Konstantinidis, ``{Conformal quivers
  and melting molecules},''
  \href{http://dx.doi.org/10.1007/JHEP03(2015)066}{{\em JHEP} {\bfseries 03}
  (2015) 066}, \href{http://arxiv.org/abs/1310.7929}{{\ttfamily arXiv:1310.7929
  [hep-th]}}.

\bibitem{Maldacena:2015waa}
J.~Maldacena, S.~H. Shenker, and D.~Stanford, ``{A bound on chaos},''
\href{http://arxiv.org/abs/1503.01409}{{\ttfamily arXiv:1503.01409 [hep-th]}}.

\bibitem{Saad:2019lba}
P.~Saad, S.~H. Shenker, and D.~Stanford, ``{JT gravity as a matrix integral},''
\href{http://arxiv.org/abs/1903.11115}{{\ttfamily arXiv:1903.11115 [hep-th]}}.

\bibitem{Jackiw:1984je}
R.~Jackiw, ``{Lower Dimensional Gravity},''
  \href{http://dx.doi.org/10.1016/0550-3213(85)90448-1}{{\em Nucl. Phys. B}
  {\bfseries 252} (1985) 343--356}.

\bibitem{Teitelboim:1983ux}
C.~Teitelboim, ``{Gravitation and Hamiltonian Structure in Two Space-Time
  Dimensions},'' \href{http://dx.doi.org/10.1016/0370-2693(83)90012-6}{{\em
  Phys. Lett. B} {\bfseries 126} (1983) 41--45}.

\bibitem{mirzakhani2007simple}
M.~Mirzakhani, ``Simple geodesics and weil-petersson volumes of moduli spaces
  of bordered riemann surfaces,'' {\em Inventiones mathematicae} {\bfseries
  167} no.~1, (2007) 179--222.

\bibitem{Eynard:2007fi}
B.~Eynard and N.~Orantin, ``{Weil-Petersson volume of moduli spaces,
  Mirzakhani's recursion and matrix models},''
  \href{http://arxiv.org/abs/0705.3600}{{\ttfamily arXiv:0705.3600 [math-ph]}}.

\bibitem{Klebanov:1991qa}
I.~R. Klebanov, ``{String theory in two-dimensions},'' in {\em {Spring School
  on String Theory and Quantum Gravity (to be followed by Workshop)}}.
\newblock 7, 1991.
\newblock \href{http://arxiv.org/abs/hep-th/9108019}{{\ttfamily
  arXiv:hep-th/9108019}}.

\bibitem{Ginsparg:1993is}
P.~H. Ginsparg and G.~W. Moore, ``{Lectures on 2-D gravity and 2-D string
  theory},'' in {\em {Theoretical Advanced Study Institute (TASI 92): From
  Black Holes and Strings to Particles}}, pp.~277--469.
\newblock 10, 1993.
\newblock \href{http://arxiv.org/abs/hep-th/9304011}{{\ttfamily
  arXiv:hep-th/9304011}}.

\bibitem{Polchinski:1994mb}
J.~Polchinski, ``{What is string theory?},'' in {\em {NATO Advanced Study
  Institute: Les Houches Summer School, Session 62: Fluctuating Geometries in
  Statistical Mechanics and Field Theory}}.
\newblock 11, 1994.
\newblock \href{http://arxiv.org/abs/hep-th/9411028}{{\ttfamily
  arXiv:hep-th/9411028}}.

\bibitem{Maldacena:2004sn}
J.~M. Maldacena, G.~W. Moore, N.~Seiberg, and D.~Shih, ``{Exact vs.
  semiclassical target space of the minimal string},''
  \href{http://dx.doi.org/10.1088/1126-6708/2004/10/020}{{\em JHEP} {\bfseries
  10} (2004) 020}, \href{http://arxiv.org/abs/hep-th/0408039}{{\ttfamily
  arXiv:hep-th/0408039}}.

\bibitem{Altland:2020ccq}
A.~Altland and J.~Sonner, ``{Late time physics of holographic quantum chaos},''
  \href{http://dx.doi.org/10.21468/SciPostPhys.11.2.034}{{\em SciPost Phys.}
  {\bfseries 11} (2021) 034}, \href{http://arxiv.org/abs/2008.02271}{{\ttfamily
  arXiv:2008.02271 [hep-th]}}.

\bibitem{Gaiotto:2014kfa}
D.~Gaiotto, A.~Kapustin, N.~Seiberg, and B.~Willett, ``{Generalized Global
  Symmetries},'' \href{http://dx.doi.org/10.1007/JHEP02(2015)172}{{\em JHEP}
  {\bfseries 02} (2015) 172}, \href{http://arxiv.org/abs/1412.5148}{{\ttfamily
  arXiv:1412.5148 [hep-th]}}.

\bibitem{Hofman:2018lfz}
D.~M. Hofman and N.~Iqbal, ``{Goldstone modes and photonization for higher form
  symmetries},'' \href{http://dx.doi.org/10.21468/SciPostPhys.6.1.006}{{\em
  SciPost Phys.} {\bfseries 6} no.~1, (2019) 006},
  \href{http://arxiv.org/abs/1802.09512}{{\ttfamily arXiv:1802.09512
  [hep-th]}}.

\bibitem{Winer:2023btb}
M.~Winer and B.~Swingle, ``{Reappearance of Thermalization Dynamics in the
  Late-Time Spectral Form Factor},''
  \href{http://arxiv.org/abs/2307.14415}{{\ttfamily arXiv:2307.14415
  [nlin.CD]}}.

\bibitem{Haake2018}
F.~Haake, S.~Gnutzmann, and M.~Ku{\'{s}}, {\em Supersymmetry and Sigma Model
  for Random Matrices},
  \href{http://dx.doi.org/10.1007/978-3-319-97580-1_6}{pp.~205--288}.
\newblock Springer International Publishing, Cham, 2018.
\newblock \url{https://doi.org/10.1007/978-3-319-97580-1_6}.

\bibitem{Dyson:1962es}
F.~J. Dyson, ``{Statistical theory of the energy levels of complex systems.
  I},'' \href{http://dx.doi.org/10.1063/1.1703773}{{\em J. Math. Phys.}
  {\bfseries 3} (1962) 140--156}.

\bibitem{Dyson:1962es2}
F.~J. Dyson, ``{Statistical Theory of the Energy Levels of Complex Systems.
  II},'' \href{http://dx.doi.org/10.1063/1.1703774}{{\em J. Math. Phys.}
  {\bfseries 3} no.~1, (1962) 157–--165}.

\bibitem{Dyson:1962oir}
F.~J. Dyson, ``{Statistical Theory of the Energy Levels of Complex Systems.
  III},'' \href{http://dx.doi.org/10.1063/1.1703775}{{\em J. Math. Phys.}
  {\bfseries 3} no.~1, (1962) 166}.

\bibitem{Dyson:1970tza}
F.~J. Dyson, ``{Correlations between the eigenvalues of a random matrix},''
  \href{http://dx.doi.org/10.1007/BF01646824}{{\em Commun. Math. Phys.}
  {\bfseries 19} no.~3, (1970) 235--250}.

\bibitem{Meh2004}
M.~L. Mehta, {\em Random Matrices}.
\newblock 3rd~ed., 2004.

\bibitem{Alvarez-Gaume:2005dvb}
L.~Alvarez-Gaume, C.~Gomez, H.~Liu, and S.~Wadia, ``{Finite temperature
  effective action, AdS(5) black holes, and 1/N expansion},''
  \href{http://dx.doi.org/10.1103/PhysRevD.71.124023}{{\em Phys. Rev. D}
  {\bfseries 71} (2005) 124023},
  \href{http://arxiv.org/abs/hep-th/0502227}{{\ttfamily arXiv:hep-th/0502227}}.

\bibitem{Mizoguchi:2004ne}
S.~Mizoguchi, ``{On unitary / hermitian duality in matrix models},''
  \href{http://dx.doi.org/10.1016/j.nuclphysb.2005.03.035}{{\em Nucl. Phys. B}
  {\bfseries 716} (2005) 462--486},
  \href{http://arxiv.org/abs/hep-th/0411049}{{\ttfamily arXiv:hep-th/0411049}}.

\bibitem{Aharony:2003sx}
O.~Aharony, J.~Marsano, S.~Minwalla, K.~Papadodimas, and M.~Van~Raamsdonk,
  ``{The Hagedorn - deconfinement phase transition in weakly coupled large N
  gauge theories},'' \href{http://dx.doi.org/10.4310/ATMP.2004.v8.n4.a1}{{\em
  Adv. Theor. Math. Phys.} {\bfseries 8} (2004) 603--696},
  \href{http://arxiv.org/abs/hep-th/0310285}{{\ttfamily arXiv:hep-th/0310285}}.

\bibitem{Liu:2004vy}
H.~Liu, ``{Fine structure of Hagedorn transitions},''
  \href{http://arxiv.org/abs/hep-th/0408001}{{\ttfamily arXiv:hep-th/0408001}}.

\bibitem{Klebanov:1994pv}
I.~R. Klebanov, ``{Touching random surfaces and Liouville gravity},''
  \href{http://dx.doi.org/10.1103/PhysRevD.51.1836}{{\em Phys. Rev. D}
  {\bfseries 51} (1995) 1836--1841},
  \href{http://arxiv.org/abs/hep-th/9407167}{{\ttfamily arXiv:hep-th/9407167}}.

\bibitem{Klebanov:1994kv}
I.~R. Klebanov and A.~Hashimoto, ``{Nonperturbative solution of matrix models
  modified by trace squared terms},''
  \href{http://dx.doi.org/10.1016/0550-3213(94)00518-J}{{\em Nucl. Phys. B}
  {\bfseries 434} (1995) 264--282},
  \href{http://arxiv.org/abs/hep-th/9409064}{{\ttfamily arXiv:hep-th/9409064}}.

\bibitem{Gross:1980he}
D.~J. Gross and E.~Witten, ``{Possible Third Order Phase Transition in the
  Large N Lattice Gauge Theory},''
  \href{http://dx.doi.org/10.1103/PhysRevD.21.446}{{\em Phys. Rev. D}
  {\bfseries 21} (1980) 446--453}.

\bibitem{Wadia:2012fr}
S.~R. Wadia, ``{A Study of U(N) Lattice Gauge Theory in 2-dimensions},''
  \href{http://arxiv.org/abs/1212.2906}{{\ttfamily arXiv:1212.2906 [hep-th]}}.

\bibitem{Marino:2008ya}
M.~Marino, ``{Nonperturbative effects and nonperturbative definitions in matrix
  models and topological strings},''
  \href{http://dx.doi.org/10.1088/1126-6708/2008/12/114}{{\em JHEP} {\bfseries
  12} (2008) 114}, \href{http://arxiv.org/abs/0805.3033}{{\ttfamily
  arXiv:0805.3033 [hep-th]}}.

\bibitem{Eynard:2015aea}
B.~Eynard, T.~Kimura, and S.~Ribault, ``{Random matrices},''
  \href{http://arxiv.org/abs/1510.04430}{{\ttfamily arXiv:1510.04430
  [math-ph]}}.

\bibitem{Okuyama:2017pil}
K.~Okuyama, ``{Wilson loops in unitary matrix models at finite $N$},''
  \href{http://dx.doi.org/10.1007/JHEP07(2017)030}{{\em JHEP} {\bfseries 07}
  (2017) 030}, \href{http://arxiv.org/abs/1705.06542}{{\ttfamily
  arXiv:1705.06542 [hep-th]}}.

\bibitem{Goldschmidt:1979hq}
Y.~Y. Goldschmidt, ``{1/$N$ Expansion in Two-dimensional Lattice Gauge
  Theory},'' \href{http://dx.doi.org/10.1063/1.524600}{{\em J. Math. Phys.}
  {\bfseries 21} (1980) 1842}.

\bibitem{Periwal:1990gf}
V.~Periwal and D.~Shevitz, ``{UNITARY MATRIX MODELS AS EXACTLY SOLVABLE STRING
  THEORIES},'' \href{http://dx.doi.org/10.1103/PhysRevLett.64.1326}{{\em Phys.
  Rev. Lett.} {\bfseries 64} (1990) 1326}.

\bibitem{Bars:1979xb}
I.~Bars and F.~Green, ``{Complete Integration of U ($N$) Lattice Gauge Theory
  in a Large $N$ Limit},''
  \href{http://dx.doi.org/10.1103/PhysRevD.20.3311}{{\em Phys. Rev. D}
  {\bfseries 20} (1979) 3311}.

\bibitem{Rossi:1996hs}
P.~Rossi, M.~Campostrini, and E.~Vicari, ``{The Large N expansion of unitary
  matrix models},'' \href{http://dx.doi.org/10.1016/S0370-1573(98)00003-9}{{\em
  Phys. Rept.} {\bfseries 302} (1998) 143--209},
  \href{http://arxiv.org/abs/hep-lat/9609003}{{\ttfamily
  arXiv:hep-lat/9609003}}.

\bibitem{nistDLMFxA71041}
``{D}{L}{M}{F}:10.41 {A}symptotic {E}xpansions for {L}arge {O}rder; {M}odified
  {B}essel {F}unctions; {C}hapter 10 {B}essel {F}unctions --- dlmf.nist.gov.''
  \url{https://dlmf.nist.gov/10.41}.
\newblock [Accessed 02-07-2024].

\bibitem{Anninos:2020ccj}
D.~Anninos and B.~M\"uhlmann, ``{Notes on matrix models (matrix musings)},''
  \href{http://dx.doi.org/10.1088/1742-5468/aba499}{{\em J. Stat. Mech.}
  {\bfseries 2008} (2020) 083109},
  \href{http://arxiv.org/abs/2004.01171}{{\ttfamily arXiv:2004.01171
  [hep-th]}}.

\bibitem{Okuyama:2018yep}
K.~Okuyama, ``{Spectral form factor and semi-circle law in the time
  direction},'' \href{http://dx.doi.org/10.1007/JHEP02(2019)161}{{\em JHEP}
  {\bfseries 02} (2019) 161}, \href{http://arxiv.org/abs/1811.09988}{{\ttfamily
  arXiv:1811.09988 [hep-th]}}.

\bibitem{Liu:2018hlr}
J.~Liu, ``{Spectral form factors and late time quantum chaos},''
  \href{http://dx.doi.org/10.1103/PhysRevD.98.086026}{{\em Phys. Rev. D}
  {\bfseries 98} no.~8, (2018) 086026},
  \href{http://arxiv.org/abs/1806.05316}{{\ttfamily arXiv:1806.05316
  [hep-th]}}.

\bibitem{Altland:2022xqx}
A.~Altland, B.~Post, J.~Sonner, J.~van~der Heijden, and E.~P. Verlinde,
  ``{Quantum chaos in 2D gravity},''
  \href{http://dx.doi.org/10.21468/SciPostPhys.15.2.064}{{\em SciPost Phys.}
  {\bfseries 15} no.~2, (2023) 064},
  \href{http://arxiv.org/abs/2204.07583}{{\ttfamily arXiv:2204.07583
  [hep-th]}}.

\bibitem{Swingle:2009bg}
B.~Swingle, ``{Entanglement Renormalization and Holography},''
  \href{http://dx.doi.org/10.1103/PhysRevD.86.065007}{{\em Phys. Rev. D}
  {\bfseries 86} (2012) 065007},
  \href{http://arxiv.org/abs/0905.1317}{{\ttfamily arXiv:0905.1317
  [cond-mat.str-el]}}.

\bibitem{Winer:2020gdp}
M.~Winer and B.~Swingle, ``{Hydrodynamic Theory of the Connected Spectral form
  Factor},'' \href{http://dx.doi.org/10.1103/PhysRevX.12.021009}{{\em Phys.
  Rev. X} {\bfseries 12} no.~2, (2022) 021009},
  \href{http://arxiv.org/abs/2012.01436}{{\ttfamily arXiv:2012.01436
  [cond-mat.stat-mech]}}.

\bibitem{forrester2010log}
P.~J. Forrester, {\em Log-gases and random matrices (LMS-34)}.
\newblock Princeton university press, 2010.

\bibitem{DiFrancesco:1993cyw}
P.~Di~Francesco, P.~H. Ginsparg, and J.~Zinn-Justin, ``{2-D Gravity and random
  matrices},'' \href{http://dx.doi.org/10.1016/0370-1573(94)00084-G}{{\em Phys.
  Rept.} {\bfseries 254} (1995) 1--133},
  \href{http://arxiv.org/abs/hep-th/9306153}{{\ttfamily arXiv:hep-th/9306153}}.

\bibitem{Anninos:2020geh}
D.~Anninos and B.~M\"uhlmann, ``{Matrix integrals \& finite holography},''
  \href{http://dx.doi.org/10.1007/JHEP06(2021)120}{{\em JHEP} {\bfseries 06}
  (2021) 120}, \href{http://arxiv.org/abs/2012.05224}{{\ttfamily
  arXiv:2012.05224 [hep-th]}}.

\bibitem{Gradshteyn:1943cpj}
I.~S. Gradshteyn and I.~M. Ryzhik, {\em {Table of Integrals, Series, and
  Products}}.
\newblock 1943.

\bibitem{Gross:2019ach}
D.~J. Gross, J.~Kruthoff, A.~Rolph, and E.~Shaghoulian, ``{$T\overline{T}$ in
  AdS$_2$ and Quantum Mechanics},''
  \href{http://dx.doi.org/10.1103/PhysRevD.101.026011}{{\em Phys. Rev. D}
  {\bfseries 101} no.~2, (2020) 026011},
  \href{http://arxiv.org/abs/1907.04873}{{\ttfamily arXiv:1907.04873
  [hep-th]}}.

\bibitem{delCampo:2017bzr}
A.~del Campo, J.~Molina-Vilaplana, and J.~Sonner, ``{Scrambling the spectral
  form factor: unitarity constraints and exact results},''
  \href{http://dx.doi.org/10.1103/PhysRevD.95.126008}{{\em Phys. Rev. D}
  {\bfseries 95} no.~12, (2017) 126008},
  \href{http://arxiv.org/abs/1702.04350}{{\ttfamily arXiv:1702.04350
  [hep-th]}}.

\bibitem{Brezin:1995dp}
E.~Br\'{e}zin, S.~Hikami, and A.~Zee, ``{Oscillating density of states near
  zero energy for matrices made of blocks with possible application to the
  random flux problem},''
  \href{http://dx.doi.org/10.1016/0550-3213(96)00063-6}{{\em Nucl. Phys. B}
  {\bfseries 464} (1996) 411--448},
  \href{http://arxiv.org/abs/cond-mat/9511104}{{\ttfamily
  arXiv:cond-mat/9511104}}.

\bibitem{Brezin_1997}
E.~Br\'{e}zin and S.~Hikami, ``Spectral form factor in a random matrix
  theory,'' \href{http://dx.doi.org/10.1103/physreve.55.4067}{{\em Physical
  Review E} {\bfseries 55} no.~4, (Apr, 1997) 4067--4083}.
  \url{https://doi.org/10.1103%2Fphysreve.55.4067}.

\end{thebibliography}\endgroup

\end{document}